%% file: thesis.tex
\documentclass[a4paper,10pt]{report}

\usepackage{thesis}

\newcommand{\shortdoctitle}{Master's Thesis}
\newcommand{\doctitle}{Standards-based modeling and deployment of machine learning workflows via BPMN on FaaS}
\newcommand{\docsubtitle}{Master Thesis}

\newcommand{\me}{Laurens Martin Tetzlaff}
\newcommand{\keywords}{keyword1, keyword2, keyword3}

\author{\me}
\def\thesistitle{BPMN4sML: A BPMN Extension for Serverless Machine Learning}
\def\docsubtitle{Technology Independent and Interoperable Modeling of Machine Learning Workflows and their Serverless Deployment Orchestration}

\def\yourname{Laurens Martin Tetzlaff}

\def\yourprogramme{Data Science and Entrepreneurship}
\def\yourstudentnumbertilburg{2048299}
\def\yourstudentnumbereindhoven{1525786}
\def\finalmonth{May}
\def\finalyear{2022}
\def\supervisorone{Dr. Indika P.K. Weeransingha Dewage (Supervisor)}
\def\supervisortwo{Prof. Dr. Willem-Jan van den Heuvel (Supervisor)}

\hypersetup
{
    pdfauthor={\me},
    pdftitle={\shortdoctitle},
    pdfsubject={\doctitle},
    pdfkeywords={\keywords}
}

\begin{document}

\pagenumbering{roman}
\include{titlepage}

\normalsize

\clearemptydoublepage




\chapter*{Abstract}\label{chapter:abstract}
\input{chapters/abstract}

\clearemptydoublepage



\chapter*{Preface}\label{chapter:preface}
\input{chapters/preface}

\clearemptydoublepage

\addtocontents{toc}{\protect\setcounter{tocdepth}{-1}}
\tableofcontents
\addtocontents{toc}{\protect\setcounter{tocdepth}{2}}

\clearemptydoublepage

\listoftables

\clearemptydoublepage

\listoffigures

\clearemptydoublepage



\chapter{Introduction}\label{chapter:introduction}
\setcounter{page}{0}
\pagenumbering{arabic}

\input{chapters/1.introduction}

\clearemptydoublepage
\chapter{Literature Review}\label{chapter:related_work}
\input{chapters/2.related_work}

\clearemptydoublepage

\chapter{Methodology}\label{chapter:methodology}
\input{chapters/3.methodology}

\clearemptydoublepage

\chapter{Requirements for Standard based Conceptual Modeling for Serverless ML Workflows}\label{chapter:requirement_analyses}
\input{chapters/4.requirement_analyses}

\clearemptydoublepage

\chapter{BPMN4sML: BPMN for serverless Machine Learning}\label{chapter:results}
\input{chapters/5.results}

\clearemptydoublepage
\chapter{Conceptual Mapping for BPMN4sML to TOSCA Conversion}\label{chapter:implementation}
\input{chapters/6.Implementation}

\clearemptydoublepage

\chapter{Validation}\label{chapter:IllustrativeUseCase}
\input{chapters/7.IllustrativeUseCase}

\clearemptydoublepage

\chapter{Discussion and Limitations}\label{chapter:discussion}
\input{chapters/8.discussion}

\clearemptydoublepage

\chapter{Conclusion}\label{chapter:conclusions}
\input{chapters/9.conclusions}

\clearemptydoublepage

\bibliographystyle{unsrt}
\bibliography{references}

\clearemptydoublepage

\appendix
\addcontentsline{toc}{chapter}{Appendix}
\input{appendices/main}

\end{document}

%% file: titlepage.tex
\begin{titlepage}
%
%
%
%
%
%
%
%
\begin{figure}[!htb]
\minipage{0.32\textwidth}
  \includegraphics[width=\linewidth]{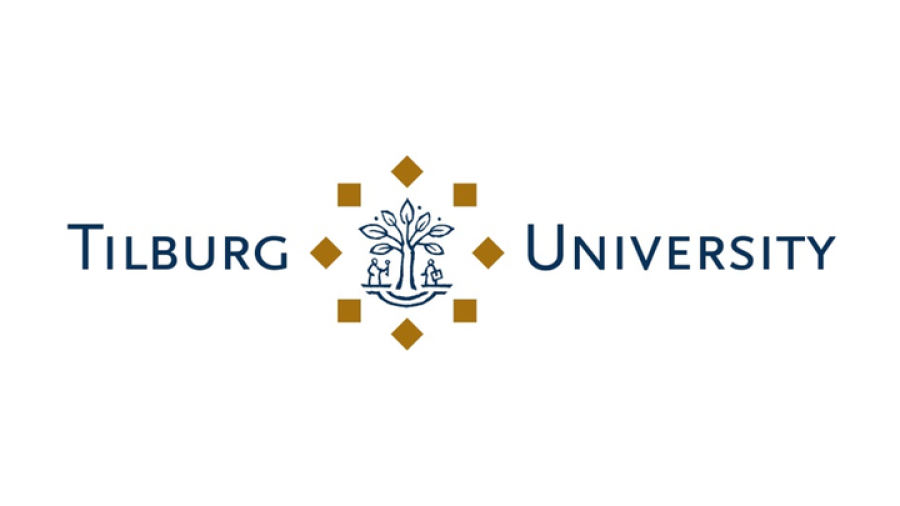}
\endminipage\hfill
\minipage{0.32\textwidth}
  \includegraphics[width=\linewidth]{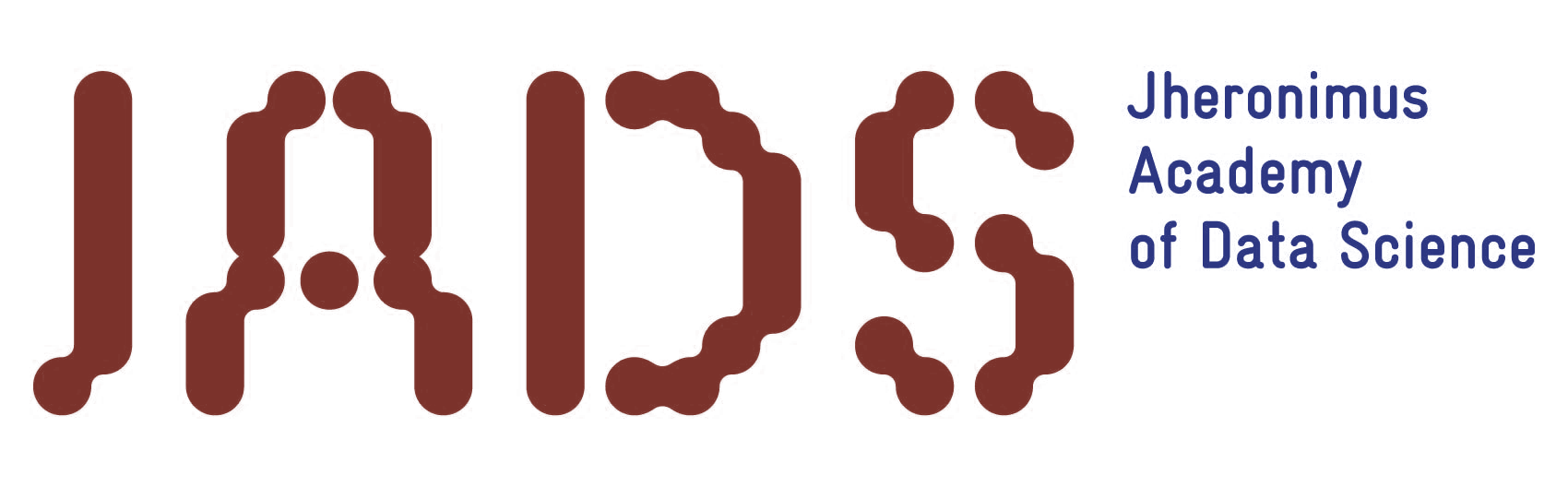}
\endminipage\hfill
\minipage{0.32\textwidth}%
  \includegraphics[width=\linewidth]{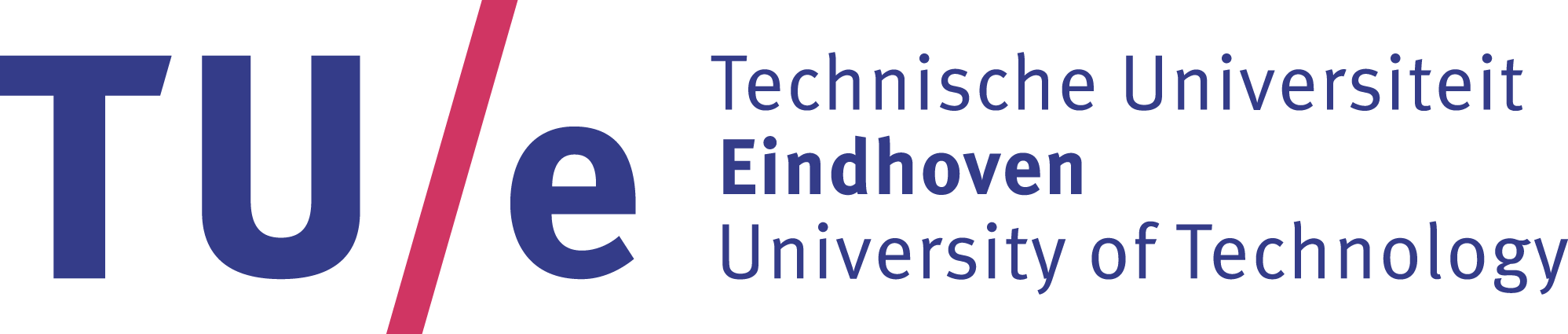}
\endminipage
\end{figure}

\begin{center}

\huge{\textbf{\thesistitle{}}} \\[0.5cm]
\Large{{\docsubtitle{}}}\\[1.2cm]
\normalsize{\yourname{}} \\
\normalsize{\textsc{Student number Tilburg}: \yourstudentnumbertilburg{}} \\
\normalsize{\textsc{Student number Eindhoven}: \yourstudentnumbereindhoven{}} \\ [1cm]
\normalsize{\textsc{Thesis submitted in partial fulfillment}} \\
\normalsize{\textsc{of the requirements for the degree of}} \\
\normalsize{\textsc{Master of Science in \yourprogramme{}}}\\
\normalsize{\textsc{Jheronimus Academy of Data Science}} \\[1.5cm]
\normalsize{Thesis committee:} \\[0.2cm]
\normalsize{\supervisorone{}} \\
\normalsize{\supervisortwo{}} \\
\vfill
\normalsize{Tilburg University, Eindhoven University of Technology} \\
\normalsize{Jheronimus Academy of Data Science} \\
\normalsize{'s-Hertogenbosch, The Netherlands} \\
\normalsize{\finalmonth{} \finalyear{}} \\

\end{center}

\end{titlepage} 

%% file: chapters/abstract.tex

\textit{Context}: Machine learning (ML) continues to permeate all layers of academia, industry and society.  Despite its successes, mental frameworks to capture and represent machine learning workflows in a consistent and coherent manner are lacking. For instance, the de facto process modeling standard, Business Process Model and Notation (BPMN), managed by the Object Management Group, is widely accepted and applied. However, it is short of specific support to represent machine learning workflows.\\
Further, the number of heterogeneous tools for deployment of machine learning solutions can easily overwhelm practitioners. Research is needed to align the process from modeling to deploying ML workflows.\\

\noindent \textit{Objective}: Confronting the shortcomings with respect to consistent and coherent modeling of ML workflows in a technology independent and interoperable manner, we extend BPMN and introduce BPMN4sML (BPMN for serverless machine learning). We further address the heterogeneity in deployment by proposing a conceptual mapping to convert BPMN4sML models to corresponding deployment models using TOSCA.\\

\noindent \textit{Method}: We first analyze requirements for standard based conceptual modeling for machine learning workflows and their serverless deployment. Building on the requirements, we extend BPMN's Meta-Object Facility (MOF) metamodel and the corresponding notation. Our extension BPMN4sML follows the same outline referenced by the Object Management Group (OMG) for BPMN. We then take the extended metamodel elements and relate them to corresponding TOSCA elements to identify a conceptual mapping.\\

\noindent \textit{Results}: BPMN4sML allows technology independent and interoperable modeling of machine learning workflows of various granularity and complexity across the entire machine learning lifecycle. It aids in arriving at a shared and standardized language to communicate ML solutions. Moreover, it takes the first steps toward enabling the conversion of ML workflow model diagrams to corresponding deployment models for serverless deployment via TOSCA. However, orchestrating the deployment models requires further research. The capabilities and advantages of using BPMN4sML are illustrated by real-life examples.\\

\noindent \textit{Conclusion}: BPMN4sML extends the standard BPMN 2.0 with additional modeling capabilities to support (serverless) machine learning workflows, thereby functioning as a consistent and coherent mental framework. The conceptual mapping illustrates the potential of leveraging BPMN4sML workflow models to derive corresponding serverless deployment models.

%% file: chapters/preface.tex
This thesis synthesizes the work of the past few months which truly has been a rollercoaster of ups and downs, dead-ends and new solutions. Certainly, conceptualizing a domain (or even parts of a domain) as vast as machine learning can seem like an overwhelming task, but once the first step is taken, another one follows and soon enough the mountain is climbed. To me, this work represents a great learning experience, both personally and academically, starting with conceptual modeling over to serverless computing, model-driven engineering and general design science research.

Through it all, my supervisor Dr. Indika P.K. Weerasingha Dewage has been a tremendous help. You gave me unlimited support and confidence when things did not go as planned, pushed me to reach beyond what I thought possible, remained critical whenever necessary and ensured to still enjoy the research. You provided me with incredibly helpful resources and advice and stood patiently next to me (virtually) without hesitation. For that I am sincerely grateful.

I would also like to express my sincere gratitude to my supervisor Prof. Dr. Willem-Jan van den Heuvel. You supported me and provided me with great opportunities and learning experiences both within and outside the scope of this thesis.

Finally, I want to thank my friends and my family, for being patient, for their relentless support and for answering countless phone calls.

I hope that you, the reader, will benefit from this work as much as I did.

\rightline{\textit{Laurens Martin Tetzlaff}}

\rightline{\textit{'s-Hertogenbosch, May 2022}}

%% file: chapters/1.introduction.tex
\textbf{"Data is food for AI"} says Andrew Ng~\cite{AndrewNg} stimulating the search for mental frameworks to better capture and represent the lifecycle of machine learning (ML). 

The domain is at the forefront of methods to analyse and understand large volumes of data for value creation. It is thus a vast and resourceful solution for any organization to turn to. Not just novel applications refer to machine learning for 'intelligent' solutions, also regular business processes are being overlaid with or amplified by ML, interconnecting digitized services with fully automated \textit{intelligent} functionalities~\cite{McKinseyMLOperationalization}.
Nonetheless, keeping track of machine learning workflows and integrating them with existing processes is a cumbersome endeavour not least due to the lack of a shared mental framework and language that allows to communicate what an ML solution entails~\cite{BeyondHypeAnalytics}. General approaches to alleviate such challenges of complexity include the adoption of standardized modeling languages to help conceptualizing, visualizing and operationalizing a domain in a straightforward manner.
For instance, process modeling as part of general business process management practices or deployment modeling embodied in model-driven engineering help in streamlining the identification and representation of individual components and modeling their interaction for ultimately both non-technical and technical individuals to grasp logic, workflow and the overall process or application.

Accounting for the intricacies of machine learning by means of a conceptual model, let alone modeling and deploying ML workflows that are derived from it, remains difficult with the functionality provided by existing standards.

\section{Problem Indication}

\textbf{Why is ML workflow modeling relevant?}\, Machine learning comes with a large variety of heterogeneous solutions spanning from a myriad of potential methods, algorithms and tools over to ever-increasing options of deploying them. Different machine learning approaches, i.e. supervised, unsupervised, reinforcement learning etc., as well as sub-categories each pose their own requirements~\cite{ManagementModelingLifecycleinDL,DMLSurvey1}. This leads to lengthy and complicated development processes. Generalization of the underlying concepts is challenging as machine learning workflows can be extraordinarily complex and sophisticated. 

The large number of methods and algorithms requires practitioners to have a profound technical understanding on how to properly design and embed ML solutions within new applications or the existing process ecosystem of an organization. 
Further, a majority of current ML services is still being developed in siloed machine learning pipelines targeting one specific use case, making re-use of trained models or parts of the ML pipeline difficult to achieve. Allowing for seamless integration or combination of (parts of) existing ML services into new solutions across departments and domains of an organization is necessary to fully leverage the benefits ML offers~\cite{CMLEnterprise,CMLFuture}. To realize this, however, an all-encompassing picture of the machine learning service, its workflow as well as associated processes is required. Lukyanenko et al.~\cite{ConceptualModelingML} formulate this problem as a challenge that can potentially be alleviated by leveraging conceptual modeling. Applying conceptual modeling to machine learning workflows can be advantageous to: 1)
increase business understanding, 2) enhance effectiveness of the ML solution and 3) improve overall comprehensibility, understandability and documentation of the ML workflow. \\

\noindent \textbf{How does serverless computing relate to ML workflows?}\, Similar to general cloud computing, serverless computing abstracts from the underlying hardware infrastructure and associated management and operational activities. Serverless computing takes the abstraction to a new extreme by further decoupling business and application logic from resource management activities. Function-as-a-Service (FaaS) represents a popular realization of serverless computing. Instead of referring to virtualized machines or containers for services that are running on the cloud's infrastructure, a developer can directly concentrate on business logic encapsulated by a serverless function that is managed by a FaaS platform on the cloud. Such functions operate in an event-driven paradigm and enable granular-billing, i.e. only invoking cost when actually running, thus promising a fraction of the cost of most common microservice solutions. 

The development towards modeling services as functions which can be independently triggered, invoked and chained goes well with the sequential nature of machine learning. Challenges are however the increase in complexity regarding function composition to represent ML workflows. Additionally, cloud providers differ heavily with respect to how serverless workflows can be defined~\cite{FaaSBPMNOrchestration,FaaSComparison}. Consequently, next to understanding the machine learning domain and the possible ways it relates to associated processes, practitioners also have to master provider specific tools and language to build serverless ML workflows.\\

\noindent \textbf{What are modeling and deployment challenges with current standards and tools?}
Process modeling languages such as FlowChart or the Business Process Model and Notation (BPMN) offer a variety of functionalities to represent workflows~\cite{OMGBPMN}. 
Available extensions to BPMN cover recently emerged fields like ubiquitous computing technologies or quantum applications~\cite{uBPMN,bpmnt,BPMNExQuant}, meanwhile still neglecting specific requirements to allow for appropriate representation of machine learning concepts.

Several solutions to help realize and manage machine learning workflows exist, commonly referred to as machine learning in operations (MLOps). However, MLOps frameworks come with their own language and functionality making it difficult for ML practitioners to master all of them. Provider differences regarding serverless workflows further intensify the heterogeneity. First steps are taken to decrease this burden on the developer~\cite{FaaSBPMNOrchestration,SWSA}.\\

To facilitate and standardize the process of designing and communicating machine learning workflows irrespective of the selected cloud provider and technology, new modeling constructs are required. Moreover, to bridge the gap between modeling of machine learning workflows and their serverless deployment, the workflow model needs to be relatable to a deployable artefact.

\section{Research Questions}
\label{RQs}

To address the indicated research problem, the conducted study is formalized alongside the following main research question:
\newline
\newline
 \textbf{How can ML engineers be supported to model ML workflows and their serverless deployment orchestrations in a technology independent and interoperable manner?}
\newline
\newline
In order to better answer the main research question, it is operationalized into sub-questions. 
First, a knowledge base needs to be established on which the remaining work can build up ensuring that it draws from the current state-of-the-art. Thus, the first set of questions is directed towards the main domains this work references, them being 1) technology independent and interoperable modeling of processes and application deployment, 2) machine learning and its processes and 3) serverless computing. This results in the following questions:
\begin{enumerate}[label=1.\arabic*]
\item What constitutes standard-based modeling and deployment?
\item What constitutes machine learning and its processes?
\item What constitutes serverless computing?
\end{enumerate}

With an established understanding of the domains, the underlying concepts and requirements for modeling machine learning workflows considering the goal of their subsequent serverless deployment orchestration need to be identified. Further, established requirements are to be associated with capabilities of existing process modeling solutions to discover potential shortcomings and need for extension. As the Business Process Model and Notation is the de-facto standard in process modeling, it is chosen as the language of comparison.
This results in the following set of questions:
\begin{enumerate}[label=2.\arabic*]
\item What are the underlying concepts of a machine learning workflow, i.e. operations, artefacts, events?
\item What are the relevant concepts for FaaS-based deployment of machine learning workflows?
\item What are the requirements for and shortcomings of BPMN as a modeling language standard to represent serverless machine learning workflows?
\end{enumerate}

Answering this set of questions lays the foundation to address the guiding research question - through application of the identified concepts and requirements the creation of artefacts that enables ML engineers and other stakeholders to model ML workflows in a technology independent and interoperable manner is possible. To further structure and realize this process, the following set of sub-questions is derived:
\begin{enumerate}[label=3.\arabic*]
\item How can a modeling language standard be extended to allow for modeling of serverless machine learning workflows? 
\begin{enumerate}
    \item How can serverless machine learning workflows be conceptually captured?
    \item How can serverless machine learning workflows be visually represented?
\end{enumerate}
\end{enumerate}

In answering these questions, the core artefacts of this work can be developed. To further address the remaining aspects of the guiding research question, i.e. the modeling of serverless deployment orchestration of designed ML workflows, a conceptual mapping is required that enables derivation of a deployment model while continuously adhering to the requirements of technology independence and interoperability. This leads to the last sub-question:
\begin{enumerate}[label=4.\arabic*]
\item How can a machine learning workflow model be conceptually transformed into a deployment model for serverless deployment orchestration in a technology-independent and interoperable manner?
\end{enumerate}

In answering this question, a method proposition can be derived to support the modeling of ML workflows and their serverless deployment orchestration that ML engineers and other stakeholders may draw from. It further facilitates communication and analysis of designed ML workflows in a standardized and structured manner.

\section{Research Scope and Relevance}
The scope of this work is to 1) analyse machine learning concepts (supervised, unsupervised and selected methodologies) that form the ML workflow and the overarching ML lifecycle. Leveraging this, we 2) identify the conceptual modeling requirements of machine learning workflows and their serverless deployment orchestration. We map the requirements against the current functionality of a process modeling standard such that 3) a minimal and sufficient set of modeling elements (Tasks, Events, Data Objects and other artefacts) can be extracted that allow modeling (serverless) ML workflows. To that end, this work 5) extends the Business Process and Model Notation to capture and represent the respective ML concepts. Referencing the extension, an initial conceptual mapping is proposed to 6) translate ML workflow model diagrams into corresponding (TOSCA) deployment models such that their serverless deployment orchestration is facilitated. 

\subsection{Theoretical Relevance}
This work contributes to existing studies by combining machine learning research with the fields of serverless computing as well as process and deployment modeling. Currently, no extension exists to specifically support modeling machine learning workflows in a standardized and technology agnostic manner. Thus, the research joins related publications on extending the Business Process Model and Notation standard~\cite{uBPMN,bpmnt,BPMNExQuant,BPMNExclinical}. By operationalizing conceptual modeling for machine learning, this study addresses the call for investigation and the research gap identified by Lukyanenko et al.~\cite{ConceptualModelingML}. It moreover contributes to existing work on technology independent and interoperable modeling of serverless workflows by leveraging both BPMN and TOSCA while focusing specifically on ML workflows~\cite{FaaSBPMNOrchestration,TOSCAServerless}.


\subsection{Practical Relevance}
Organizations and business continue to struggle with integrating machine learning services in their processes, not least due to the inherent complexity, technical challenges and lack of a shared understanding~\cite{AiBMInnovation,McKinseyMLOperationalization}. The BPMN extension for machine learning allows to leverage existing business process management practices to model ML processes transparently and communicate them both among technical and managerial people. By depicting a machine learning service as a process, existing process modeling knowledge can be leveraged to better incorporate new ML solutions into the existing process ecosystem of an organization.

Machine learning practitioners, e.g. data scientists and ML engineers, can benefit from a unified notation and semantic to ideate on and model ML workflows in a technology independent and interoperable manner. Further, they can draw from the methodology applied in this work to derive deployment models for the corresponding serverless deployment orchestration.

\section{Reading Guide}
This chapter introduced the domain of machine learning, the problem of modeling machine learning workflows as well as their serverless deployment orchestration in a technology independent and interoperable manner and outlined the approach taken towards addressing this challenge. The following chapter~\ref{chapter:related_work} presents a literature review establishing the underlying topics of this work: an in-depth overview of machine learning and its lifecycle, serverless computing, conceptual and process modeling alongside business process management as well as deployment modeling and general model-driven engineering. Relevant related work is discussed. Chapter~\ref{chapter:methodology} describes design science as chosen methodology referenced throughout this thesis and tailors it to the challenge of extending BPMN for serverless ML workflows and their subsequent deployment orchestration. Following, we analyse the requirements for conceptual modeling of machine learning workflows and identify the constraints created by their serverless deployment in chapter~\ref{chapter:requirement_analyses}. The chapter concludes with a requirement synthesis and an equivalence check to existing BPMN functionality. Based on the identified elements, in chapter~\ref{chapter:results}, we present the BPMN extension for (serverless) machine learning. We propose new metamodel elements to the standard and define their semantic alongside the created notation. To bridge the gap to serverless deployment orchestration, chapter~\ref{chapter:implementation} describes a conceptual mapping from the proposed BPMN extension to corresponding TOSCA elements. Further, the end-to-end modeling approach of machine learning workflows and their deployment is consolidated and visualized as an initial method proposal. The created artefacts are then reviewed by means of illustrative use cases in chapter~\ref{chapter:IllustrativeUseCase}. We discuss and scrutinize our findings in chapter~\ref{chapter:discussion}. In closing, with chapter~\ref{chapter:conclusions} we summarize the answers we arrived at as well as their academic and practical implications and provide directions for future work.

%% file: chapters/2.related_work.tex
This chapter presents the dominant themes and concepts of importance to the research questions in section \ref{RQs} thereby establishing a knowledge base and the current state-of-the-art upon which the remaining work builds. Section~\ref{MachineLearning} introduces an overview to the domain of machine learning. Following, section~\ref{MLLifecycle} explicates the machine learning lifecycle while accounting for regular machine learning as well as for federated learning workflows. Additionally, a brief introduction to machine learning in operations is given. Subsequently, a primer on serverless computing establishes the recent paradigm in section~\ref{Serverless}. As an upcoming service model, section~\ref{FaaS} puts forward Function-as-a-Service and delineates possible ways of architecting FaaS-based solutions in event-driven or orchestrated fashions.  By drawing from business process management, its lifecycle and associated activities, the field of business process modeling is introduced in section~\ref{BusinessProcessManagement}. As a solution to business process modeling, the Business Process Modeling Notation is presented alongside its extension mechanism as the de-facto standard in section~\ref{BPMN}. Following, section~\ref{ModelDrivenEngineering} describes different ways of deployment modeling as part of the overall model-driven engineering methodology which builds the connection to the deployment modeling standard \textit{Topology and Orchestration Specification of Cloud Applications} that is elaborated on in section~\ref{TOSCASection}. Finally, section~\ref{relatedWork} concludes this literature review by synthesizing a selection of related studies.

\section{Machine Learning }
\label{MachineLearning}
The fundamental challenge of identifying and extracting patterns, trends and relationships in data has been a part of many aspects in science, industry and society. Machine learning concerns itself with automatically discovering recurrent patterns, i.e. regularities, in data by applying computational statistics in form of algorithms to ultimately map or approximate a mathematical function representing the unknown distribution inherent to the considered dataset~\cite[p.~1ff.]{bishop2006pattern}.  In other terms, the statistical model that has been adapted to the data estimates some output based on some input~\cite[p.~9ff., p.~486ff.]{hastie2009elements}. Thus, machine learning can be seen as a part of artificial intelligence that facilitates understanding of and learning from data in order to make sound inferences and predictions based on the uncovered patterns. The data itself consist of observations for various features (variables) that contain values describing instances of the studied domain. These can be of different modality, generally such as of a textual, numeric, audio or graphical nature. The features ideally inform on the problem at hand - they contain a signal which can be picked up by the algorithm.

\subsection{Machine Learning Methods}
\label{MLMethods}
While the field of Machine Learning is quickly developing and growing with new methodologies being continuously researched and introduced to practice, the domain can still be divided into three main categories that focus on learning problems that require different kinds of feedback - information returned to the algorithm for gradual improvement of its quality: \textit{supervised learning}, \textit{unsupervised learning} and \textit{reinforcement learning}. Supervised learning tackles challenges for which a correct output, that is to be predicted, exists. Hence, based on labeled input-output pairs the algorithm searches for a mapping from input instances $x$ to output instances $y$. Thus, it can be trained on a ground truth through a loss function which measures and minimizes the offset or distance between the predicted value and the correct one. In contrast, unsupervised learning problems address situations in which the measurement of the outcome $y$ is absent and only input instances $x$ are given. In that case, rather than predicting a target variable the unsupervised algorithm focuses on organizing or clustering the data which can be later used to for instance group new observations with existing ones, thereby informing on their context and relationships in the overall dataset. In this learning problem, it is not explicitly stated what patterns shall be looked for and clearly defined error metrics as applied in supervised learning are missing~\cite[p.~2ff., p.~9ff.]{murphy2012machine}. A combination of the two, semi-supervised learning, addresses problems for which a small number of labeled observations is available next to a larger subset of unlabeled data. To improve upon the performance of the supervised learner, information in the unlabeled instances is exploited through unsupervised algorithms. Reinforcement learning focuses on identifying and performing the correct action in a situation - learning what to do in a given context. Hence in this field, algorithms map a situation to the corresponding action such that a numerical reward signal can be maximized that is computed through a reward function. In this problem context, actions taken by the algorithm can influence later inputs (situations). Further, the action in each situation ultimately yielding the highest reward is unknown to the algorithm and needs to be discovered through trial and error~\cite[p.~15ff.]{sutton2018reinforcement}. In the context of this work, supervised and unsupervised learning concepts are considered while aspects of reinforcement learning are left to future research as characteristics and processes specific to it differ and are therefore out of scope.

Apart from the introduced sub-categories, various overarching machine learning methods exist - too many to allow for exhaustive presentation. The methodologies, most of which pertaining to supervised learning, address various challenges inherent to ML such as learner performance, making the statistical model more robust to unseen data, dealing with limited computational resources or decentralized datasets to list a few. In the following, an overview of a popular subset is given. Nonetheless, each topic can be elaborated upon much more in depth, hence, the interested reader may consult the indicated references for further information.

Aside from semi-supervised ML, other hybrid methodologies have been brought forward one of which is self-supervised learning. Similar to semi-supervised learning, it combines unsupervised learning problems with supervised learning techniques to circumvent the bottleneck of labeled datasets. By creating an artificial supervisory signal, i.e. output $y$, an unsupervised learning task can be reformulated as a supervised learning problem for which a supervised algorithm is able to pick up co-occurring relationships inherent to the dataset~\cite{Self_supervised1,self_supervised3, self_supervised2}.

As another method, transfer learning can be applied in order to leverage possibilities of re-use and build up on already well-performing ML models from different but related domains with respect to the current ML problem. To do so, given one or several source domains, a target domain and corresponding ML tasks, the related information from the already solved ML source task(s) is used to improve the predictive function that is applied to the ML task of the target domain. As a consequence, dependence on data from the target domain needed to create the target learner can be greatly reduced~\cite{transferLearning1,Transferlearning2,DeepTransferLearningExample}.

Next to differentiating on the learning problem, some machine learning techniques not limited by it are applicable for this work. Online learning addresses challenges inherent to streaming data, observations that become available over time. In contrast to conventional machine learning practices in which an algorithm is applied to batch data offline, online learning structures the ML task in such a way that the algorithm continuously learns from observations one by one or in grouped intervals at a time. Thereby, the online learner can leverage more and more knowledge on ground truths that previously have been prediction tasks~\cite{OnlineLearningSurvey1}. 

In some scenarios, one ML model may not be sufficient to adequately solve a machine learning problem in which case the application of the model to unseen data results in poor learner performance. By considering several different learners applied to the same problem and combining them, one learner may compensate for the other's errors and limitations.  Known as ensemble learning, multiple ML algorithms are applied to the ML task, producing inferior results on their own. However, when fusing the results via a voting mechanism, better performance can be achieved. Various techniques exist to create the set of distinct learners such as iterative permutation of the dataset used for training the algorithms or applying the same algorithm with changing constraints and parameters. Further, several output fusion approaches have been brought forward. For an overview of ensemble learning the reader is referred to~\cite{EnsembleDL,EnsembleLearningSurvey1,EnsembleLearningSurvey2}.

Inherent challenges to machine learning are computational bottlenecks resulting in the origin of distributed ML. Demands for data processing for large ML models exceed development progress in computing power. Therefore, various solutions to transfer the machine learning task and share the associated workload over multiple machines gained popularity. To accelerate a workload, conventional distributed computing differentiates between vertical scaling, adding more resources to one single machine, and horizontal scaling, adding more machines (worker nodes) to the system. Distributed ML considers a system made up of several machines, thus the ML problem can be solved by parallelizing the data, data-parallel approach, or the model, model-parallel approach, or both. When applying the data-parallel approach, the dataset is partitioned equally over all worker nodes, thereby creating distinct data subsets on which the same algorithm is then trained. A single coherent result, output $y$, can emerge since all machines have access to the same model. In contrast, the model-parallel approach is more constrained. Each machine receives an exact copy of the full dataset while it operates on a distinct part of the  algorithm. The final ML model is then an aggregate of all ML model parts. Nonetheless, distributing the ML algorithm across multiple machines has limits since most times the algorithm parameters cannot be easily split up~\cite{DMLSurvey1}. Consequently, most distributed ML approaches refer to a data-parallel solution. Next to parallelising the algorithm training also other steps occurring during a machine learning workflow can benefit from distribution over multiple worker nodes such as trying out different hyperparameter settings, one setting per worker node. A closer look at other steps involved in the machine learning workflow is taken in the subsequent section~\ref{MLLifecycle}. To take load off of the data scientist or engineer, recent solutions have been brought forward that allow offloading the ML task to a more suitable machine learning platform that potentially applies distributed techniques hidden behind the service interface. Leveraging such a solution abstracts the low-level implementation and management intricacies of sophisticated techniques.

Not only computational limits but also restricted access to decentralized proprietary data sources are a challenge in machine learning research. An upcoming solution is federated learning, a term first coined by Google~\cite{FedLearnGoogle}, that facilitates collaboratively solving a ML problem across a heterogeneous set of devices while retaining data ownership and adhering to legislation and other restrictions of the location where data is stored - albeit with limitations. Federated learning can differ in terms of workflow and life-cycle operations that are presented in section~\ref{MLLifecycle}. Consequently, to clarify differences section~\ref{FL} elaborates on it.

Other ongoing streams of ML research are methods such as active learning or multi-task learning, see~\cite{ActiveDLsurvey,activelearning,multitasklearning}. 

When considering a general machine learning workflow, activities involved may differ significantly depending on the method applied. Most of the established tools in practice support an end-to-end ML workflow particularly covering regular supervised learning. It is one of the best researched and well-established disciplines. Accordingly, the focus of this study lies predominantly on supervised learning workflows. To account for upcoming fields, an look is taken at federated learning as an instance of distributed learning however with a limited scope.

\section{Machine Learning Lifecycle}
\label{MLLifecycle}
The development of a machine learning model is a complex and iterative process. Integrating it into an organization's business processes and gaining value from it is ever more so. Consequently, various efforts in research and industry have been made to establish a shared understanding of the steps and activities involved, commonly referred to as the \textit{machine learning lifecycle}. Definitions can vary on what constitutes the ML lifecycle and how it differs from related concepts such as a \textit{machine learning workflow} or a \textit{machine learning pipeline}, sometimes applying these terms interchangeably. As an alleviation, this work defines the machine learning lifecycle as the global process used for developing, integrating and maintaining ML models in a full fledged system, sharing the definition by Ashmore et al.~\cite{AssuringMLLifecycle}. In contrast, the ML pipeline or ML workflow refer in this thesis to a more focused subset of activities that are required to create the ML model, albeit with the term \textit{ML workflow} being used sometimes interchangeably with the term ML lifecycle or the term ML pipeline in related works. A closer look at what constitutes them is taken in the subsequent sections, presenting a minimal, industry-compliant collection of phases and actions involved.

The ML lifecycle shares ideas defined and introduced in widely accepted data science and data mining frameworks such as the CRoss Industry Standard Process for Data Mining (CRISP-DM)~\cite{CRISP-DM}, the Team Data Science Process (TDSP)~\cite{TDSP} or the process of Knowledge Discovery in Databases (KDD)~\cite{KDD}. Commonalities include a data-centric focus of process activities, re-occuring feedback loops as well as the integration of development activities with business context analysis and corresponding requirements to better inform decision-making during ML model development. 

\begin{figure}
    \centering
    \includegraphics[width=1\linewidth]{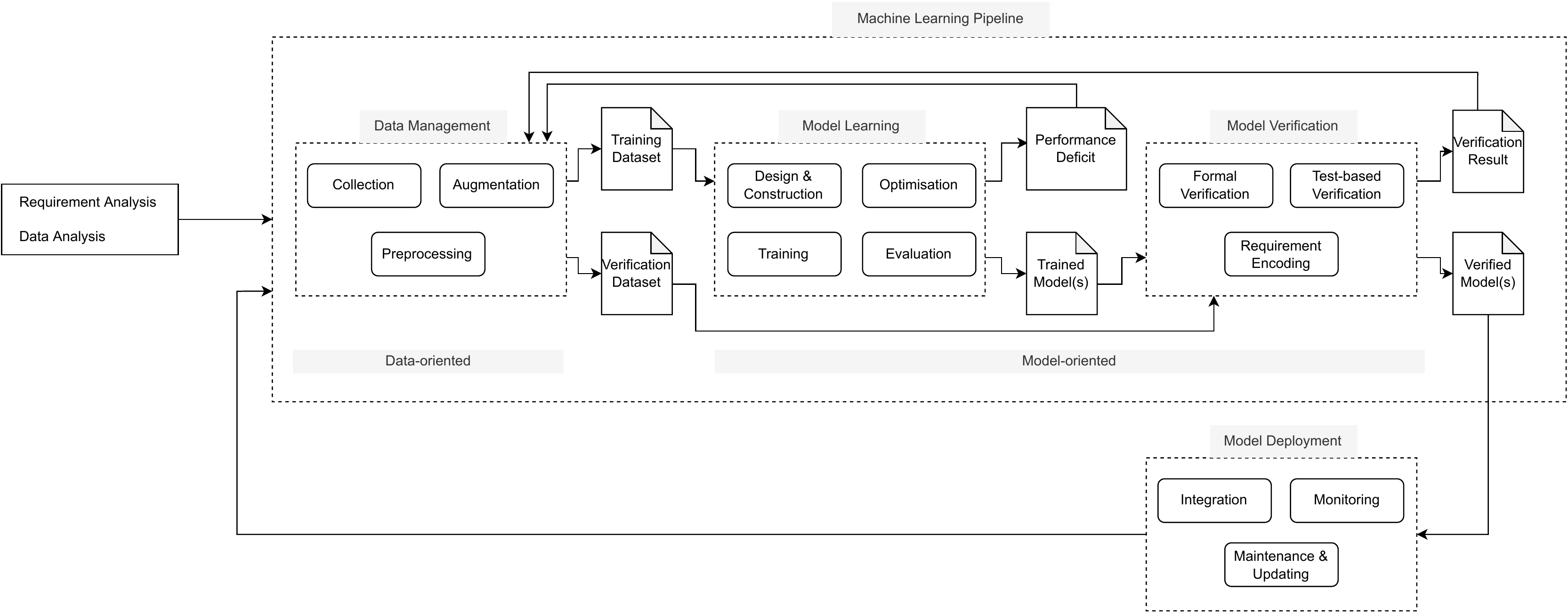}
    \caption[Generic Machine Learning Lifecycle]%
    {Generic Machine Learning Lifecycle \par \small The main phases and respective stages of the machine learning lifecycle (highlighted by headlines in grey bars above the respective square). Requirement analysis can be considered a preceding phase. Data-oriented stages (i.e. analysis, collection, pre-processing and augmentation) are followed by model-oriented ones (i.e. design, training, evaluation, optimisation, verification). Some steps are not exclusive to data or model focus, i.e. pre-processing and augmentation may change w.r.t. the requirements of a selected model. For readability, sequence flow within the phases is omitted, the main feedback loops are indicated by arcs (i.e. model learning / model verification may loop back to any prior stages in e.g. data management; model deployment can loop back to any previous step).}
    \label{fig:MLLifecylce}
\end{figure}

According to Ashmore et al., the ML lifecycle can be differentiated into four key phases: 1) \textit{Data Management}, 2) \textit{Model Learning}, 3) \textit{Model Verification} and 4) \textit{Model Deployment}~\cite{AssuringMLLifecycle}. Idowu et al.~\cite{AssetML} abstract further from the actual stages or activities involved, merging  \textit{Model Learning} and \textit{Model Verification} into the so-called \textit{Model-oriented} phase and introducing \textit{Requirements Analysis} as a phase prior to \textit{Data Management}. Model deployment is understood as a set of activities contained within the \textit{DevOps Works} phase. The actual machine learning workflow (respectively machine learning pipeline) can be considered to encompass the phases directly involved with producing the ML model, thus, data- and model-oriented works, corresponding to \textit{Data Management}, \textit{Model Learning} and \textit{Model Verification}. An abstracted depiction of the main phases is presented in Figure~\ref{fig:MLLifecylce}.  Similar research overall agrees on the discussed breakdown of phases~\cite{ManagementModelingLifecycleinDL,MicrosoftMLWorkflow,SEperspectiveML}.  
Each phase is presented below while elaborating on key activities and particularities.

\subsection{Requirement Analysis}
\label{MLLRA}
The \textit{requirement analysis} accounts for foregoing business related activities as well as functional and non-functional baselines that the ML service needs to adhere to in order to properly address the ML problem. As a result, first decisions can be made regarding applicability and consideration of relevant machine learning methods and algorithms given the ML problem~\cite{MicrosoftMLWorkflow}. For instance, a 'blackbox' ML service such as a complex deep learning solution that provides little transparency and limited explanations for its predictive outcomes is less likely to be accepted in data sensitive situations affecting the life of a customer, e.g. loan approval. Understanding the status quo and related constraints can further inform on the subsequent phases. Among others, access to available data sources and the modality of data may limit the range of potential solution candidates as for example an image or video analysis task is best solved by convolutional neural nets which in turn have higher demands for computing and processing power. Further, in an organization the requirement analysis phase can be closely connected to related business processes that the ML service will be interacting with. Hence, it benefits from established business process management practices which can provide more transparency and information. To further establish requirements, exploratory data analysis can be conducted. Related literature accounts such data analysis to various phases, i.e. to requirement analysis, as part of data management or both~\cite{AssuringMLLifecycle,AssetML,MicrosoftMLWorkflow}. In this work, such analytical tasks are assigned to the requirement phase to improve differentiation with the subsequent operational ML pipeline that can be deployed~\cite{googleML}.

\subsection{Data Management}
\label{MLLDM}
High quality data sources are fundamental to the success of a machine learning solution which is why a multitude of methods and techniques has been introduced in research and industry to optimize data-oriented activities prior to applying an ML method. Some of them can be accounted to categories such as \textit{data collection}, \textit{data pre-processing} and \textit{data augmentation}~\cite{AssetML, AssuringMLLifecycle, MicrosoftMLWorkflow, ManagementModelingLifecycleinDL}. The core results of the \textit{data management} phase are a training dataset and a verification dataset. This is, however, subject to change w.r.t. the ML method applied. If for instance only unsupervised learning algorithms are considered, a verification dataset is less applicable due to the lack of evaluation and verification metrics. Data collection refers to activities involved in the process of sourcing available data samples that shall be fed to the ML algorithm. If existing data samples are scarce, new observations can be gathered, e.g. by accessing previously siloed databases in an organization or by measuring and recording the underlying phenomenon of the ML problem. Moreover, augmentation can be considered to e.g. synthetically create observations or re-structure and transform existing ones if the available feature space is too small, contains too much noise or has other shortcomings~\cite{preprocessing}. 
Most ML algorithms make strict demands on datasets which are addressed by various pre-processing and preparation activities to improve consistency and quality. For instance, data can be cleaned to filter out noise and various scaling and normalization techniques help in preventing algorithms susceptible to e.g. heterogeneous numeric ranges or data formats from misinterpreting the given information. Imputation techniques help mitigating challenges with missing data. 
Data preparation activities, especially cleaning datasets or feature engineering, often go hand-in-hand with the model learning phase to further increase the ML model's performance and are highly dependent on the ML algorithm applied - deep learning techniques as an example require little feature engineering~\cite{ManagementModelingLifecycleinDL}. The interested reader is referred to Zheng et al. for an introduction to pre-processing techniques~\cite{featureeng}.

\subsection{Model Learning and Verification}
\label{MLLMLV}
During the \textit{model learning} phase one or several ML algorithms are selected and applied to the output of the previous phase, the training dataset. The phase typically starts with a selection, i.e. \textit{design and construction}, of algorithms to consider alongside with the potential specification of the loss-function that is to be minimized by adjusting the algorithm's parameters when \textit{training} it on the given dataset~\cite{AssuringMLLifecycle}. Prior to training, the dataset can be divided into a training and validation set, the validation set being used for \textit{evaluating} the 'learned' ML model applying pre-defined evaluation metrics that ideally are already informed by results of the requirement analysis. In more complex scenarios, hyperparameter tuning can be considered to \textit{optimize} the parameter settings of an algorithm on the given training dataset and further improve its prospective performance on new unseen observations. In these cases, the algorithm can be trained several times on variations of the dataset that are obtained through resampling strategies, popular ones being cross-validation and bootstrapping~\cite[p.62ff]{kuhn2013applied}. Applying these resampling strategies and splitting the training data into training and validation data is often directly incorporated in the tuning strategy. Tuning the hyperparameters of ML models facilitates adjustment of common challenges such as the bias-variance trade-off (over- and underfitting) and model complexity to achieve better generalization, see Hastie et al. for more information on the topic~\cite[p.~223ff]{hastie2009elements}. The stages of model learning are inherently iterative in that various hyperparameter settings are experimented with during tuning before deciding on a final configuration for an ML model that is passed on to the next phase in case sufficient performance is achieved. If not, reiteration over the data management phase is necessary to further improve data quality, for instance by applying different pre-processing techniques or adding more data~\cite{MicrosoftMLWorkflow,AssuringMLLifecycle}.

\textit{Model Verification} centrally ensures that the final trained model(s) comply with all previously established requirements to determine their readiness for deployment~\cite{uber}. In case of supervised learning, the generalization performance of the proposed ML solution is tested on another set of unseen data, the \textit{verification dataset}, which has been exempt from model training to avoid data leakage, i.e. \textit{test-based verification}. Moreover, additional constraints may be defined through \textit{requirement encoding}, for instance in case of a supervised classification problem that deals with data imbalance and attributes more importance to the correct classification of one particular class. \textit{Formal verification} may be applicable in case a set of formal properties encoding core requirements needs to be adhered to~\cite{AssuringMLLifecycle}. Ultimately, the model verification phase collects evidence that justifies the ML solution's deployment, for instance if it outperforms an already deployed model~\cite{googleML}. In case the proposed ML solution does not meet the model verification criteria, re-iteration over data management or model learning activities is necessary which is informed by the \textit{verification result}~\cite{AssuringMLLifecycle}.

\subsection{Model Deployment and Monitoring}
\label{MLLMDM}
With a passed verification check the ML solution is prepared, compiled and deployed as a \textit{verified model} in the given infrastructure. The verified model components are then \textit{integrated} with other system components. Deployment can take various forms apart from direct provisioning to external access - for instance, a verified model can be deployed in a shadowing phase or A/B testing manner, to further gather evidence if it should replace an existing model~\cite{googleML}. Next to integrating the model, continuous \textit{monitoring} takes place~\cite{ManagementModelingLifecycleinDL} to identify execution errors and potential risks such as data, concept or model drift, i.e. a change in the properties of the dataset, the target variable or the ML model's performance. The monitoring activities can in turn inform on necessary \textit{updates} such as re-training the ML model on the latest dataset available, either offline as part of maintenance or through online learning when new training data become available~\cite{AssuringMLLifecycle}.

\subsection{Federated Learning Workflow}
\label{FL}
\begin{figure}
    \centering
    \includegraphics[width=0.6\linewidth]{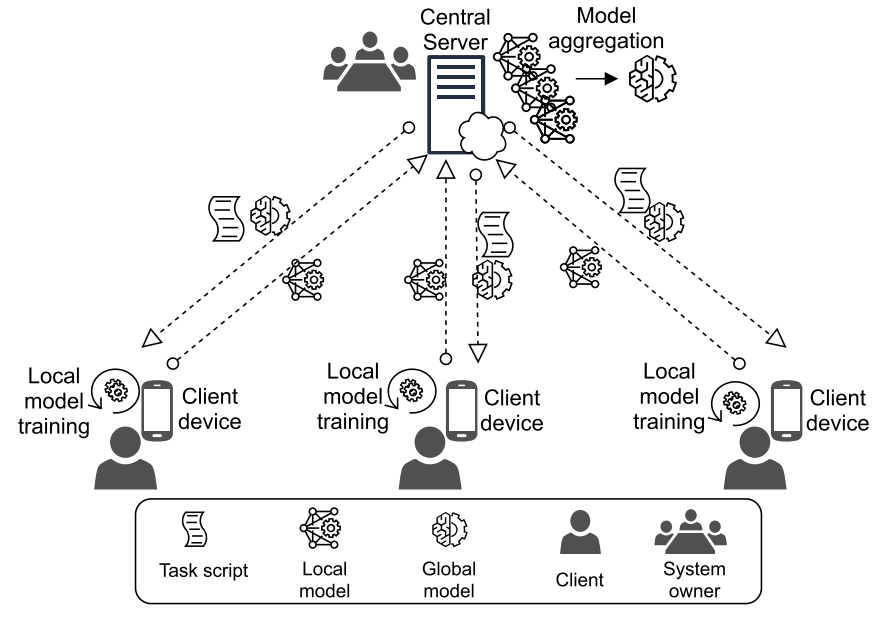}
    \caption[Generic overview of federated learning]%
    {Generic overview of federated learning \par \small High-level picture on interaction between client and central server alongside core data artefacts that are propagated. Drawn from Lo et al.~\cite{FLArch}.}
    \label{fig:FLArch}
\end{figure}

As explored earlier, federated learning (FL) offers techniques to collaboratively learn a machine learning model across various distributed devices and data that is not necessarily independent and identically distributed (iid).
The ML problem that is solved by the distributed nodes (e.g. mobile devices) is coordinated by a central aggregator. The key difference to other distributed ML solutions is the concept of not only using a ML model locally on the distributed node but also training it only with the data available in place while simultaneously benefiting from a shared global model that has been informed by all local model instances through transmission of local model parameters. The parameters are uploaded to the aggregator, creating a global model which can in turn be shared again with the participants (distributed nodes) to improve the solution of the ML task~\cite{FedLearnSurv1,FedLearnOverviewChallenges,FedLearnEdgeComp,dist2fed}. The two main parties in a federated learning workflow are represented by the mentioned \textit{clients} and the \textit{central server}~\cite{FLRA}, see Figure~\ref{fig:FLArch} for an overview. The notion of what a client constitutes may differ depending on the scenario - in case of cross-silo FL a client may be an organization or institution that runs the client-side operations on cloud or on their own data centres; in cross-device FL a client would constitute an edge device~\cite{FedLess}.

In comparison to the presented machine learning lifecycle, federated learning can differ in that the data management, model learning and verification and model deployment phases are more interconnected and can occur at different times on different devices thereby also changing the location and environment of executed tasks and generated artefacts. While data management activities such as a data collection are mostly equivalent in their logic to the conventional ML lifecycle, especially the subsequent stages, i.e. model training, evaluation, deployment and monitoring can vary, not least due to the federated nature of the ML workflow. The model learning phase in particular encapsulates a set of activities that interact between federated clients and the central server, i.e. local model training, model uploading, global model collection and aggregation as well as potential model evaluation and broadcasting back to the clients to iterate over the phase. 

\begin{figure}
    \centering
    \includegraphics[width=1\linewidth]{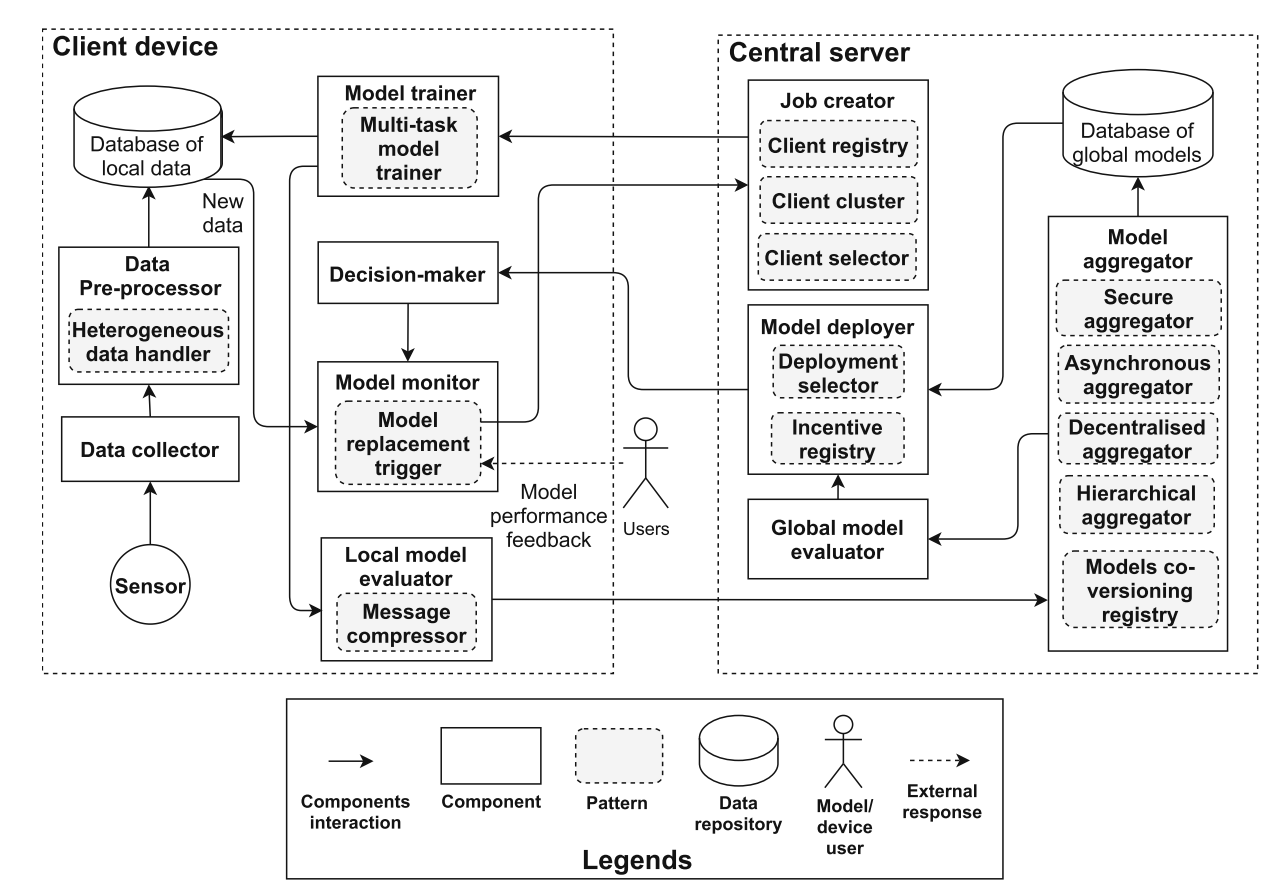}
    \caption[Reference architecture of federated learning]%
    {Reference architecture of federated learning \par \small Reference architecture on activities involved in a federated learning workflow. Components are required to realize a federated learning application whereas patterns can be sourced from and are optional. Drawn from Lo et al.~\cite{FLRA}.}
    \label{fig:FLrefenceArch}
\end{figure}

Since federated learning has only just recently been introduced to the research community and industry, its applications and processes involved are still debated and remain inconsistent~\cite{FLRA,FedLearnSurv1,FedLearnOverviewChallenges,doorbell,FedLess}. To mitigate this, the present work leverages the reference architecture proposed by Lo et al.~\cite{FLRA} to establish an overview and common ground regarding the federated learning workflow, see Figure~\ref{fig:FLrefenceArch}. Note that for an in-depth discussion of various federated learning architectures and intricacies, the reader is referred to the following works~\cite{FLRA,FLArch,FedLearnSurv1,FedLearnOverviewChallenges,FedLearnEdgeComp}.

To condense the intricacies and possibilities of federated learning workflows and focus on the elements studied within this work, only the main components, i.e. activities, are elaborated upon. For further explanations on patterns, we refer to the original paper by Lo et al.~\cite{FLRA}. 

Federated learning does not replace all phases and activities that are relevant to the machine learning lifecycle previously explained. Instead, it intermingles them.  In most cases, a federated learning process starts with a \textit{job creation} executed by the central server. This implicates the requirement analysis, potential data analysis (if possible) and other configuration activities prior to initialization of the workflow in order to define its goal, constraints and set-up, similar to the ML lifecycle. Also data management activities that are possible on a central level can be executed. Once a job is created, a first version of a global ML model is generated. Thus, the initialization of a federated learning process can be closely associated with the model learning phase of the ML lifecycle. After job creation and its propagation to the respective clients, local data collection and processing activities are executed that only differ from conventional operations in terms of the entity that manages them, i.e. the client side. Following, the propagated model is trained on the local dataset and evaluated. In case evaluation complies with previously set requirements and constraints the now local ML model is sent back to the central server for aggregation. The aggregation considers the parameters of all models received thus far, resulting in a new updated global model. As previously mentioned this process can be iterative in that the now aggregated model can be evaluated by the central server and in case of malperformance sent back to other clients to train and update it again. Once a sufficiently good global model is trained, it can be deployed on the client side to allow for its inference services - represented in Figure~\ref{fig:FLrefenceArch} by the Decision-Maker component. These activities are in line with the previously covered model verification and model deployment phase. The ML model monitoring can be executed on a client level to investigate local model performance and potentially demand a local re-training or initialization of a new global training process.

Note that optional activities and artefacts represented by \textit{Patterns} in Figure~\ref{fig:FLrefenceArch} do not explicitly include data privacy related implementations. Further, patterns such as a client registry or deployment selector are needed to fully specify the clients which are to be included in the federated training or deployment.

\subsection{Machine Learning in Operations}
\label{MLOps}
With the machine learning lifecycle being extraordinarily complex, managing it and integrating its phases and activities in the existing IT infrastructure of an organisation through regular software and data engineering practices poses challenges due to the nature of machine learning itself. Machine learning in operations (MLOps) is thus a solution that naturally evolved from related software engineering practices, e.g. development operations (DevOps) and data operations (DataOps), to provide answers~\cite{dataOps}. Since the present study researches potential BPMN-based solutions to technology-agnostic modeling and FaaS-based methods of deployment of ML pipelines, certain aspects, components and artefacts of MLOps systems become relevant, warranting a brief introduction to the field. Fundamentally, MLOps describes a 'set of standardized processes and technology capabilities for building, deploying and operationalizing ML systems rapidly and reliably'~\cite[p.3]{googleML}. To do so, practices of continuous development, continuous integration and continuous deployment that originate from DevOps and DataOps principles are extended with continuous training as an additional core pillar. Moreover, also the existing practices require adjustment due to the inherent complexity and characteristics of machine learning. As explained in previous sections, various components can change and dynamically influence each other throughout an ML process, such as 1) the business, legal and ethical requirements, 2) distribution or format of input data, 3) the ML model itself and consequently also 4) the underlying codebase which in most cases is part of a sophisticated cloud-based architecture~\cite{SustainableMLOps}. The leading cloud providers, e.g. Google Cloud, Amazon Web Services, Microsoft Azure, support open-source or proprietary MLOps technology, for instance Kubeflow\footnote{https://www.kubeflow.org/} or Amazon SageMaker\footnote{https://aws.amazon.com/sagemaker/}, which are more often than not largely aligned with the rest of the providers' offerings and differ in terms of degree of component support. With regards to this study, mostly shared functional artefacts that are required to enable or support the ML lifecycle, particularly the actual ML pipeline, are of interest. 

\section{A Primer on Serverless}
\label{Serverless}
Modern IT architectures become more and more complex by being reliant on new systems, functionalities or technologies to meet increasing expectations. For an organization to cope with this development, two main solutions can be taken - 1) strengthen the level of automation or 2) increase the degree of outsourcing to third parties. Both demands are met by providers that offer cloud-based solutions. A shared characteristic of these solutions is the \textit{as-a-service} type of offering information technology, which entails that the customer can consume the solution in a flexible 'pay-as-you-go' manner allowing for more freedom of choice and opportunity to adapt to latest demands without having to worry about complex challenges such as configuration, management or maintenance of components~\cite{TOSCA}. Some offerings are for instance \textit{infrastructure-as-a-service} or \textit{software-as-a-service}~\cite{ServerlessTaxonomy}.

\textit{Serverless} computing extends the idea of cloud computing as it further abstracts away large portions of the hardware and software stack, such as server or resource allocation, required to operate application systems to effectively enable the development and deployment of cloud applications through a high-level application programming model. Responsibilities regarding operational aspects, e.g. fault tolerance, scaling or communication resources, are assigned to cloud providers who further offer utilization-based billing that only takes into account the actually consumed computing resources which is enabled by autoscaling, adding or removing resources automatically~\cite[p.90]{ServerlessComputingReport}. Consequently, cloud computing becomes more fine-grained, accessible and affordable. Serverless computing follows the paradigm of miniaturizing microservice-based architectures in order to operationalize compact self-contained execution units as \textit{containers} or similar and integrate those in service models~\cite[p.35~ff]{ServerlessComputingReport}. 
With it come certain characteristics unique to the serverless paradigm which are however still actively discussed among researchers and practitioners~\cite[p.78]{ServerlessComputingReport}\cite{serverlessTrendsProblems}:
\begin{itemize}
\item utilization-based billing, i.e. paying solely for actually used resources which can be enabled by scaling them to zero when idle
\item limited control (NoOps), i.e. abstraction of management and maintenance away from the user
\item event-driven, i.e. changes in states, so called events, e.g. a pub-sub message, function as triggers to invoke (request) a serverless execution unit when needed.
\end{itemize}
A popular service model that is used almost equivalently with serverless computing is \textit{Function-as-a-Service} (FaaS). Since this thesis particularly focuses on FaaS as a means for serverless computing and uses it as the underlying mode of deployment for a modeled ML pipeline, a primer on FaaS is in order.  


\subsection{Functions-as-a-Service}
\label{FaaS}
Function-as-a-Service describes another level of explicit abstraction, away from virtual machines, containers and micro-services towards event-driven code snippets, functions, that incorporate only the program's logic and are managed and scaled automatically by providers (as is typical in serverless computing). The cloud providers take care of the entire deployment stack that is required for hosting the function\footnote{FaaS functions are referred to as \textit{functions}, \textit{serverless functions}, \textit{cloud functions} or \textit{FaaS functions} throughout this thesis.}~\cite{FaaSBPMNOrchestration}. In contrast, even with container solutions such as Docker the developer is still required to consider and construct deployment of environments. 

Fundamentally, the idea behind FaaS directly resembles the concept of functions in mathematics or traditional functional programming. A function is created mapping inputs to outputs. A  composition of several functions would then result in a program. Transferred to the concept of cloud computing, a developer should then be able to 'program the cloud' by registering serverless functions on it and combining them into a program~\cite{ServerlessBackwards}.

What makes FaaS not only attractive for users but also for cloud providers are promises that can be enforced through this paradigm. For instance setting a hard upper limit on function running time allows to better predict and allocate resource utilization as well as basing the billing on actual running time of a function. Moreover, statelessness enables the provider to safely remove or relocate all information, i.e. states, of the function after its execution providing more possibilities to utilize their computing resources~\cite{ServerlessComputingReport}. Next to the traits typical to generic serverless computing, additional characteristics can therefore be defined for FaaS:
\begin{itemize}
\item statelessness, i.e. pure-function like behaviour of an invocation of an execution unit
\item fixed memory, i.e. limited amount of memory an invocation can be allocated with
\item short running time, i.e. a limited time window for execution completion.
\end{itemize}
FaaS functions are stateless in that they are ephemeral. They are deployed, run through and the instance is then deleted which also deletes eventual variables or states that were created during function invocation inside the environment of that function. This may not always be the case, as potential solutions are investigated to make the FaaS concept more performant by for example reducing the impact of so called cold-starts, i.e. having to redeploy a function environment from scratch instead of reusing existing ones. However, providers do not guarantee that any state is still available in the next invocation, thus making it stateless. An ongoing effort in research and industry is put into relaxing such characteristics for example by enabling stateful serverless applications\footnote{https://docs.microsoft.com/en-us/azure/azure-functions/durable/durable-functions-overview}, facilitating state propagation, expanding the memory a function can use or extending possible function lifetimes~\cite{StatefulServerlessWorkflows,DurableServerlessFunction}. While these particularities of FaaS simplify the general understanding of the overall concept it comes with a certain slew of complications. As Hellerstein et al.~\cite{ServerlessBackwards} postulate while FaaS enables to harvest some benefits of cloud computing such as pay-as-you-go, at least at the moment of writing it still limits realising other potentials such as unlimited data storage or distributed computing power. The reason for that essentially lies in some of the conceived benefits. Limited lifetimes, isolation of the function from data and statelessness slows down communication between components, for instance by making it necessary to write produced data artefacts to persistent storage locations such as AWS S3 or Google cloud storage in between function calls. Novel solutions to this are continuously being introduced, e.g. to speed up data querying and caching, and yet to be fully tested for various use cases. A full discussion on topics such as choosing the ideal serverless architecture and products for certain applications such as heavy machine learning is out of the scope of this work given the abundance of potential solutions and vendor differences (one only needs to look for example at AWS storage options\footnote{https://docs.aws.amazon.com/whitepapers/latest/aws-overview/storage-services.html} and related services or Google cloud versions\footnote{https://cloud.google.com/products/storage} of it).

Regardless, due to the computational limitations of FaaS, composition of several functions often is necessary to cover the entire logic required for representing the requested program or functionality, positing an ongoing area of research~\cite{FaaSBPMNOrchestration}. In the context of this work, the FaaS paradigm is of most interest from a user, e.g. developer, perspective. Therefore in the following sections, design patterns to compose functions in order to model workflows or full application systems are introduced.   

\subsubsection{FaaS Choreography}
\label{FaaSChoreography}
\begin{figure}
    \centering
    \includegraphics[width=1\linewidth]{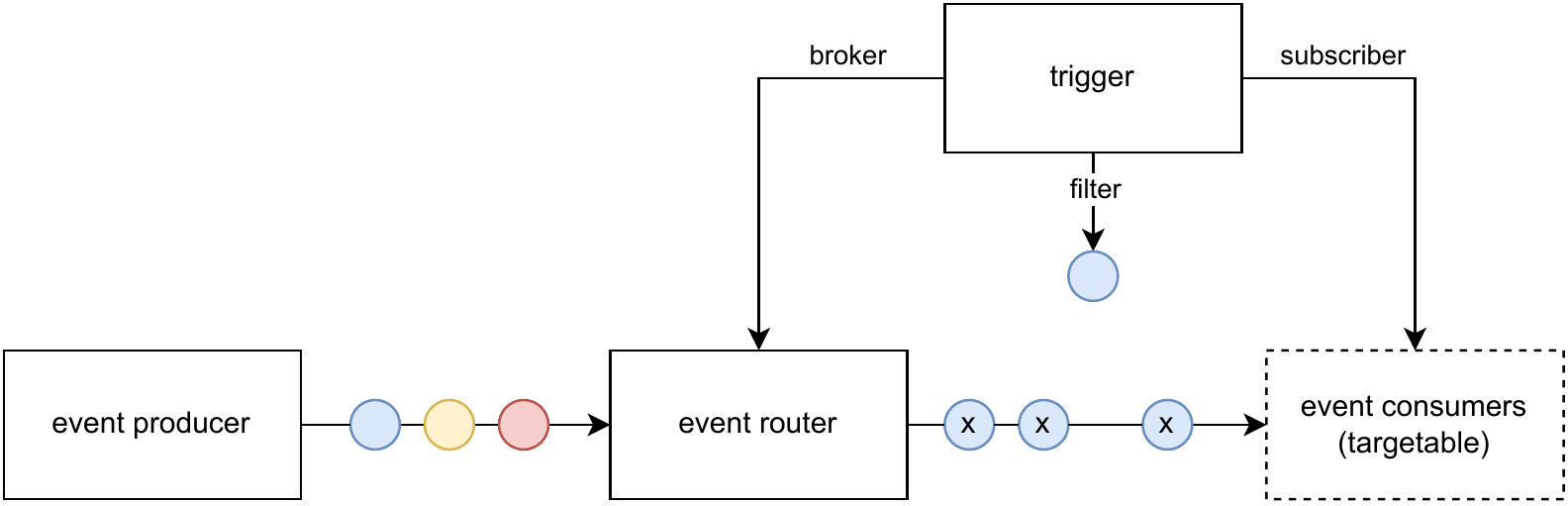}
    \caption[Example snippet of event-driven logic]%
    {Example snippet of event-driven logic \par \small Producers generate events, the broker receives, filters and sends it to the corresponding consumers. Drawn from Google Cloud~\cite{EventDrivenGoogle}.}
    \label{fig:event_driven}
\end{figure}

FaaS choreography is based on the typical serverless paradigm of event-driven architectures~\cite{EventProcessing}. When applied to composing multiple related functions, event-driven paradigms are applied to chain them together through data dependencies that allow to pass along outputs and inputs. Data dependencies arise when functions manipulate states in some way that can be tracked by queueing systems or object stores which in turn inform the subsequent function and results in its invocation~\cite{ServerlessBackwards}. For instance, a component responsible for data pre-processing may lead to producing an event once the pre-processing is finished and the dataset is written to an object store.  These so called \textit{event producers} are responsible for generating events. A broker, i.e. \textit{event router} then ingests and filters incoming events and maps them to the respective \textit{event consumer}. Finding the right event consumer is defined beforehand by \textit{triggers}\footnote{Exact terminology may differ between providers and among researchers. We make use of Google Cloud's vocabulary~\cite{EventDrivenGoogle}}. The event router is responsible for connecting the different components and serves as the medium through which messages are propagated by executing a response to the originally produced event and sending it out to the corresponding consumers. Regarding our ML example, an event consumer may be another component that reads the processed dataset for model training or tuning, see Figure~\ref{fig:event_driven}. If several different models are to be trained on this data set, multiple model training components can be set as subscribers to this event. 
By creating logic around these three artefacts (producers, router, consumers) loose-coupling can be achieved~\cite{EventDrivenGoogle}. Each component can independently be deployed, updated or scaled depending on changing requirements. 

Some challenges with exclusively event-driven approaches, particularly with regards to complex potentially long-running workflows such as machine learning pipelines, lie in properly tracking state and individual execution failures and, as a result, also triggering recovery actions from errors. While the representation of typical workflow patterns, see section~\ref{PMCM}, is also supported in event-driven architectures~\cite{FaaSPatterns}, providers claim to enable better and more consistent support and alleviation of other challenges through function orchestration engines. To that end, explicit function orchestration (function orchestration in short)  gained popularity as an approach to composing multiple severless functions together with other applications while supporting common control flow patterns, e.g. branching constructs, and integration with other internal and external services and actors.

\subsubsection{FaaS Orchestration}
\label{FaaSOrchestration}
Function orchestrations draw from workflow research to represent for example business processes or more generic applications in a serverless manner. Most providers support some degree of workflow patterns, most commonly patterns such as sequences of a set of consecutive functions, branching, e.g. parallel execution of different sets of functions based on the same input, or conditional process flows. Next to that, also other characteristics common to business processes can be supported by some, for instance integrating events such as waiting timers to pause the workflow for a specified duration. Leveraging function orchestration engines, better state and execution control is possible which also simplifies troubleshooting and helps with recovery actions through explicit error handling for each step in the workflow. A closer look at these functionalities is taken in section~\ref{FAASCDReq} when considering inherent characteristics that create requirements for the BPMN modeling extension. 
To create a serverless function orchestration at least two different modeling concepts have to be considered~\cite{FaaSBPMNOrchestration}:
\begin{itemize}
\item modeling the control flow of the orchestration, which would represent the workflow, covering multiple functions
\item modeling the deployment of the overall serverless application that surrounds the function orchestrations.
\end{itemize}
When modeling the control flow, also business logic aspects need to be considered and implemented with respect to constraints set by the respective orchestration engine and serverless function provider. For actually modeling the function orchestration, either vendor-specific modeling languages, such as Amazon's ASL (Amazon State Language), or proprietary technology or regular programming languages such as Python can be used. This heterogeneity leads not only to lock-in effects but also to challenges when trying to even semantically transfer the function orchestration logic~\cite{FaaSBPMNOrchestration, FaaSComparison}. These characteristics are however also typical for the level of maturity of function orchestration technologies. Most providers offer their own function orchestration engines with varying levels of functionality such as Amazon's AWS Step Functions~\cite{AWSStepFunctions}, Google's Workflows~\cite{GWorkflows} or Azure's Durable Functions~\cite{AzureDurable} which are ready to be used as-a-service. Finally deploying the modeled function orchestration also requires considerations regarding which vendor or open-source technology and respective services as well as deployment modeling language to use. This leads to two levels of expertise a developer requires to make use of function orchestrations, 1) the function orchestration modeling language and 2) the deployment modeling language~\cite{FaaSBPMNOrchestration,FaaSComparison}. 

%

\section{Business Process Management}
\label{BusinessProcessManagement}
A business process (BP) can be understood as a set of structured and coordinated activities, events and decisions involving various actors and objects and executed in order to generate a product or service which ultimately helps in achieving an organization's objective~\cite{SurveyBPMNChallengesSolutions}. When formally representing those activities, their context and associated information, the represented process can be properly designed, orchestrated, monitored and improved which eventually allows for a sophisticated management practice of all workflows within an organization~\cite{bpmnt}. Business process management (BPM), thus, focuses on systematically keeping track of and improving operational business workflows to ultimately optimise their efficiency and effectiveness~\cite{SurveyBPMNChallengesSolutions}. Optimization in this case refers to the entire sequence of activities, decisions and events, i.e. the process as a whole, for it to add more value to the organization and its environment and not just to each respective activity in isolation~\cite{BPMBookCh1}. Therefore, business process management covers both, operational aspects such as process automation and analysis, as well as more strategic concepts like operations management and overall organization of work, ultimately leading to a more process-centred organization~\cite{BPMNSurveyvanDerAalst}. 

Business process management can be facilitated and automated by implementing BPM systems to source together information from existing systems and related software applications in an organization, for instance an enterprise resource planning system such as the one provided by \textit{SAP}~\cite{BPMNSurveyvanDerAalst}. The data provided by the applications then give information on past and running processes as well as on their performance which in turn can inform specified key performance indicators (KPI).  Tracking processes as such can lead to interesting synergies with overall ML workflows and their management~\cite{BPMImproveTheProcess}. It is thus of interest to treat a machine learning workflow as a business process to leverage BPM practices and facilitate integration of ML services into the present process infrastructure of an organization. The BPM practices can be categorized into several iterative phases, mainly 1) process (re-) design, 2) process configuration and implementation and 3) process running and adjustment where process analysis, monitoring and improvement can be part of each respective phase~\cite{BPLM_SocialTech,BPMNSurveyvanDerAalst}. Together theses practices build the business process management life-cycle.



\subsection{Business Process Management Lifecycle}
\label{BPMLifecycle}
The business process management life-cycle addresses an approach for systematic management of processes in an organization in an iterative fashion such that continuous improvement of existing and ideal implementation of novel processes can be achieved. Depending on the domain and point of focus, various phases have been identified, starting with an abstract representation towards a more fine-grained categorization of the stages~\cite{BPMNSurveyvanDerAalst,SurveyBPMNChallengesSolutions,BPLM_SocialTech}. This work considers the latest definition by van Aalst as an overarching frame in which the life-cycle comprises three phases, see Figure~\ref{fig:BPMLifecycle}, and elaborates on some in more detail according to Durmas et al.~\cite{BPMBookCh1}:

\begin{figure}
    \centering
    \includegraphics[width=0.70\linewidth]{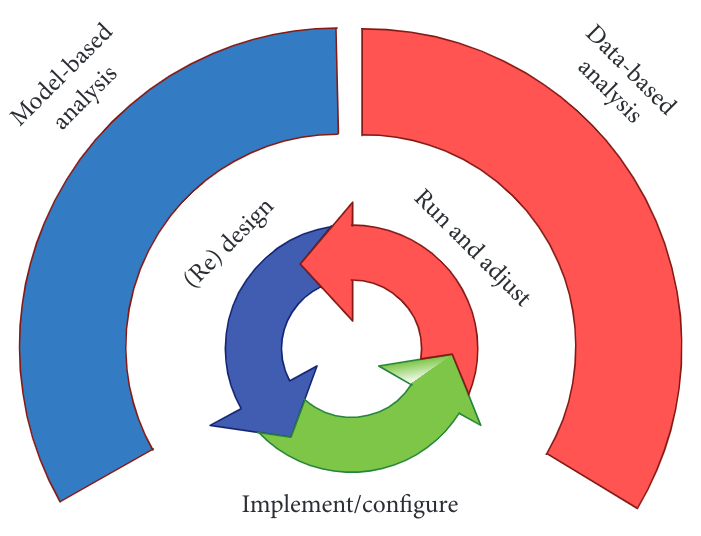}
    \caption[Business Process Management life-cycle]%
    {Business Process Management life-cycle \par \small The BPM life-cycle as introduced by van Aalst \cite{BPMNSurveyvanDerAalst}. The three main phases (re-)design, implement / configure and run and adjust build the life-cycle's core. Overarching activities include model-based and data-based analysis with the former being applied during process model creation and the latter while analysing running processes.}
    \label{fig:BPMLifecycle}
\end{figure}

\begin{enumerate}
\item (Re-) design: A process model is designed. To arrive at a process model, the process needs to be 1) identified w.r.t. the problem under study and related to potentially existing processes, 2) discovered in detail and documented in its as-is state and if re-designed also 3) analysed for limitations and improved accordingly, thereby generating a to-be state of the process. 
\item Implementation/configuration: The process model is transformed into a deployable or running system. Depending on the level of sophistication, the process model can be automatically transformed and deployed or it requires manual intervention and further codifying of rules and logic. Notably, implementing and configuring a process solely reaps benefits if the process in itself is sound and effective to achieve the overarching goal as it otherwise only enhances ineffective activities. 
\item Run \& adjust: The implemented process is put into practice, monitored and adjusted using predefined controls if required. No overall re-design or implementation take place. In case the process no longer fulfils its purpose, the cycle needs to be restarted.
\end{enumerate}

The presented research mainly addresses the first two phases, (re-)design and implementation/configuration in the context of machine learning workflows. To design a processes model, the Business Process Modeling Notation standard is considered, while implementation and configuration are realized severlessly as FaaS functions once a coherent process model is generated. The BPM life-cycle is relevant to this research to the extent that the study addresses the limitations of current process modeling - one of the core activities of BPM - with respect to machine learning. The BPM life-cycle thus serves as a more general orientation, putting process modeling into perspective and highlighting its importance in the context of business process management, particularly regarding design and analysis of the process as well as its representation and communication among various stakeholders. Thus, subsequently process modeling is presented as a part of general conceptual modeling.

\subsection{Process Modeling as an instance of Conceptual Modeling}
\label{PMCM}
Conceptual modeling experienced wide-spread adoption in information system development and database design~\cite{ConceptualModelingML}. As its overarching goal, it aims to facilitate the creation of models that formally represent certain aspects of the physical, digital and social world for a better understanding and communication~\cite{ConceptualModelling}. Characterized by three properties such a model has 1) a \textit{mapping}, 2) a level of \textit{abstraction} and 3) a \textit{purpose}. The mapping refers to the real-world phenomenon that is described by the model, each element of it thus relates back to its original counterpart. By neglecting irrelevant details of the subject under study, the model abstracts from unnecessary information and solely documents pertinent aspects. The model's purpose guides which aspects of the phenomenon can be omitted~\cite{BPMBookCh3}. The act of \textit{modeling} can thus be understood as decomposing the phenomenon into distinct lower-level aspects, each addressing a part of the whole. The term \textit{model} is referenced in multiple areas of science, albeit with slightly varying connotations to it. Important in this work is the difference between a machine learning model, referring to the algorithm that has been trained on data - i.e. statistical model also known as \textit{Learner}, and a conceptual model. For clarity's sake a machine learning model will also be referenced as an ML model in case of a lack of differentiating context throughout this thesis. Different types of conceptual models exist that often take the form of diagrams made up of graphics with descriptive text which ideally represents a particular domain or use case such as a data model, a model of an IT system's architecture or a process model. These models address three basic needs - i) to better cope with complexity by focusing on relevant aspects and ordering requirements, ii) to arrive at a shared understanding by interpreting the conceptual model as a boundary object and iii) to eventually solve the problem being tackled through domain analysis and designing of solutions~\cite{CMbenefits}. Ultimately, a conceptual model can guide development activities in their respective domain~\cite{ConceptualModelingML}.


As a version of conceptual modeling, process modeling derives a conceptualization specialized in largely representing a workflow, i.e. a flow of activities, decisions and events that are interdependent. A large body of research has been spent to better understand workflows, their patterns, functionality, potentials and drawbacks. The fundamental element, a task, can, when combined with other elements, be arranged to describe complex patterns, i.e. \textit{workflow patterns}, the most fundamental ones being~\cite{workflowPatterns}:
\begin{itemize}
\item \textit{sequence}, i.e. one task is dependent on the completion of the preceding one
\item \textit{parallel}, i.e. several tasks can be realized independent of each other
\item \textit{distribution} (e.g. fan-out), i.e. $m$ tasks directly follow the completion of a single preceding task  
\item \textit{aggregation}, i.e. a dependency relation between one task and several directly preceding ones.
\end{itemize}
For a more elaborate discussion on workflow patterns see Russel et al.~\cite[p.~105ff.]{AalstWOrkflowPatterns}.

Once a process model depicting a workflow is generated, in operations an instance of it can represent the running process in its actual form. The practice of process modeling consequently aims to abstractly express the process at its current state or to-be state through a clear and standardized graphical notation and formalized schema language thereby reducing ambiguity, risks of misinterpretation and allowing for backtracing of process instances against the designed model to identify areas of improvement. Process modeling can be applied for two main purposes, namely 1) \textit{organizational design} which leads to conceptual models to improve understanding, communication and facilitate process (re-)design and analysis and 2) \textit{application system design} which concentrates on IT-oriented aspects such as process automation and implementation details represented in an executable process model~\cite{BPMBookCh3}.

The level of detail that a process model should contain depends on its purpose. Forr documentation means, a high-level model alongside text annotations is sufficient. When quantitative analysis on process performance should be performed, more fine-grained information in the process model is required such as time taken to fulfill a task. Executable process models that are to be deployed require even more granular information on the process itself and any information associated with or required by it, for instance inputs and outputs of the tasks. The presented work focuses on the latter level of detail when extending the modeling language as the process model needs to contain the required information for it to enable conversion to TOSCA and deployment of the ML pipeline. Various process modeling languages exist that allow to design a process diagram. Most are based on two basic kinds of nodes - \textit{control nodes} and \textit{activity nodes} with activity nodes representing a unit of work that is to be performed by a person, a software or other agent and control nodes regulating the execution flow among activities. Further, \textit{event nodes} are supported by some languages as a third major category to indicate an event taking place that requires a corresponding reaction during the process~\cite{BPMBookCh1}. 
More generally, a modeling language is compromised of four aspects - 1) vocabulary, 2) syntax, 3) semantics and 4) notation. The modeling elements are given by the vocabulary which are in turn constrained by the syntax to enforce the rules of the language and describe the relationship between the elements available. Their precise meaning in the domain context is bound by the language's semantics and mostly given via textual descriptions~\cite{theofoundations}. Finally, the notation provides graphical symbols to visualize the elements when modeling the concept at hand~\cite{BPMBookCh3, Friedenstab2012}.

A \textit{flowchart}, one of the oldest process modeling languages, is a graphical representation of formalised structures such as a software logic sequence or a work process~\cite{Flowchart_def}. By using symbols the process is described as a sequence of actions in an easy-to-understand fashion while providing a high degree of flexibility and low adoption threshold. With it, however, come several weaknesses such as a blurred boundary of the actual process and a rapid increase in size of the model due to it having no differentiation between main and sub-activities. Further, responsibilities and performers cannot be easily represented. The language suits itself well for explaining processes with a high level of detail while providing an overview of the process itself falls short. Also numerous other process modeling techniques have been brought forward over the years originating from various methodologies such as Gantt charts or petri nets each having their own characteristics and weaknesses~\cite{BP_modeling_Flowchart_UML,petrinets}.

\begin{figure}
    \centering
    \includegraphics[width=0.85\linewidth]{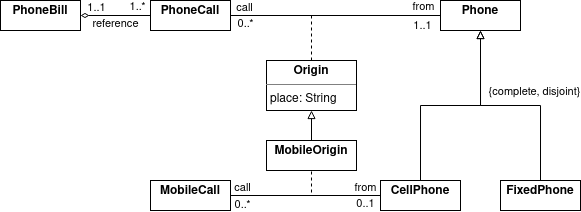}
    \caption[UML class diagram notation example]%
    {UML class diagram notation example \par \small A UML class diagram depicting a conceptual model of phone calls, their originating devices and resulting phone bills. Adapted from \cite{UML_reasoning}.}
    \label{fig:umlExample}
\end{figure}

A noteworthy modeling technique that took inspiration from flowcharts are UML diagrams. The \textit{Unified Modeling Language} (UML), an object-oriented method, can model a process which is represented by objects that are transformed by the process activities over the process lifetime. UML, further, provides nine different diagram types addressing various aspects of a system that is to be modeled~\cite{uml}. All of them build up on three main concepts - 1) \textit{objects} that represent the entity under study by incorporating the data structure, i.e. attributes, and its behaviour, i.e. supported operations, 2) \textit{state} a condition that the object may be in with its attributes taking on specific values, and 3) \textit{behaviour} which are the actions and reactions according to the operations the object supports and can perform and that lead to state changes. As Aguilar-Savén constitutes UML aides in 'specifying, visualizing, constructing and documenting the artefacts of [...] systems as well as [in] business modeling'~\cite{BP_modeling_Flowchart_UML}. 
In this work, the UML class diagram notation is applied to conceptualize the meta-model artefacts of the BPMN extension for ML workflows. A \textit{class} refers to a set of objects that have similar properties. Classes can be \textit{associated} with one another to depict relations between instances of them, e.g. an aggregation to explain a part-whole relationship or multiplicities to elaborate on various forms of cardinality. Additional features are supported to facilitate enrichment of the conceptual models such as generalizations and other forms of hierarchies~\cite{UML_reasoning}. A minimal example applying the class diagram notation to arrive at a conceptual model is shown in Figure~\ref{fig:umlExample}.
The Unified Modeling Language established itself as a standard among object-oriented modeling techniques. While UML allows for internal consistency checking and to directly build software off of it, it is not withstanding shortcomings. Modeling UML diagrams is a time-consuming and complex process. Additionally, users may quickly be overwhelmed with excessively large models and fragmented information.

Among the available process modeling languages the Business Process Modeling Notation established itself as a standard answering the need for a language that is both expressive and formal while still being understandable and comprehensive to take into account the various stakeholders, technical and non-technical, affected by a process and contributing to the overall business process management practices~\cite{BPMroles, BPMNintrostandard, BPMNenough}. Consequently, throughout the rest of this thesis, we place our focus on BPMN.

\section{Business Process Model and Notation}
\label{BPMN}
\begin{figure}
    \centering
    \includegraphics[width=0.90\linewidth]{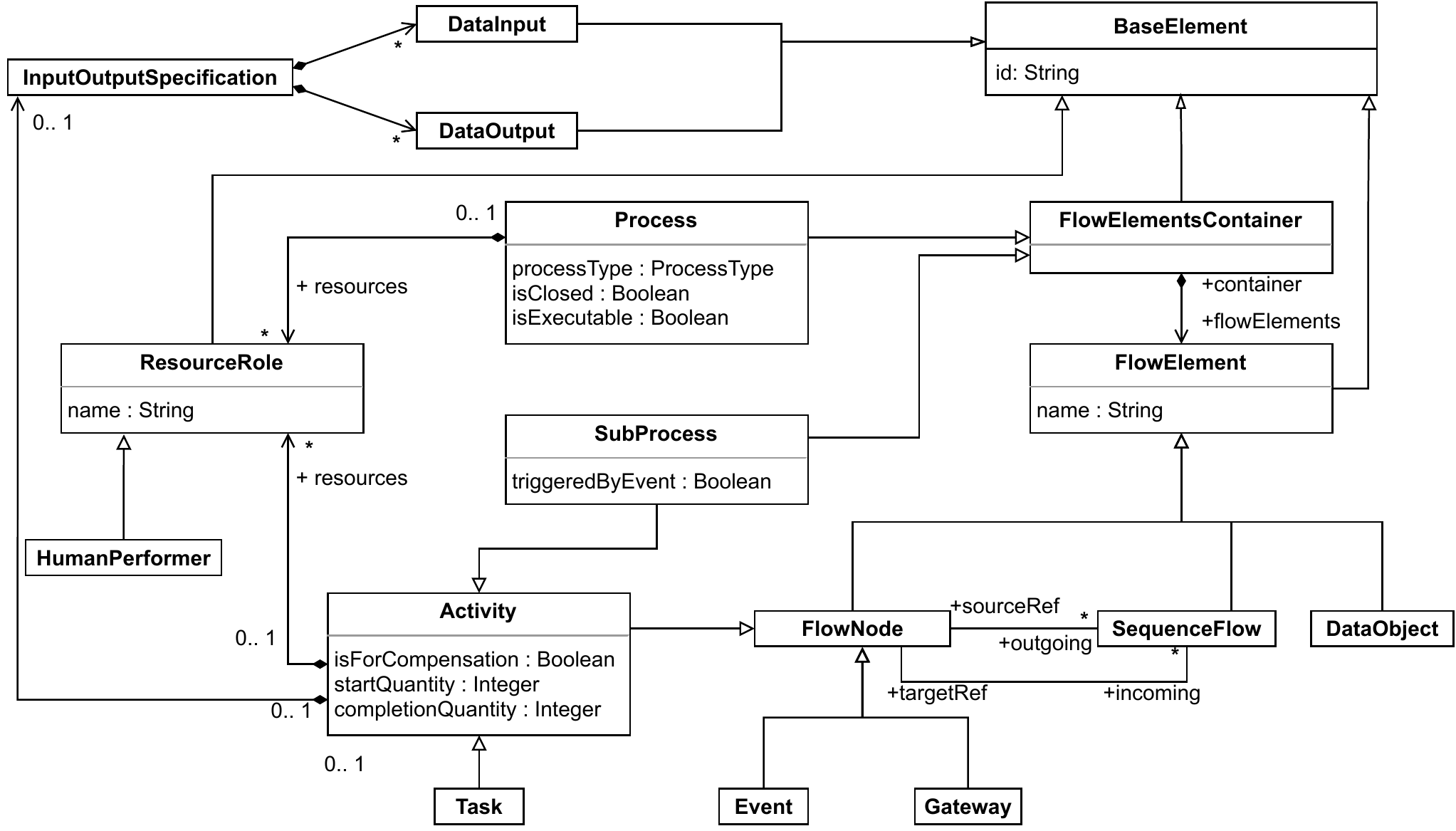}
    \caption[BPMN 2.0 core metamodel]%
    {BPMN 2.0 core metamodel \par \small A condensed metamodel of core BPMN elements, properties and relationships. Adapted from~\cite{bpmnt}.}
    \label{fig:BPMN_meta_model_overview}
\end{figure}

The Business Process Model and Notation standard provides an elaborately formalized and graphically expressive way of representing various processes, from organizational ones to workflows of systems and logical procedures in software applications, particularly aiding in domain analysis and high-level system design~\cite{BPMNenough}. With its functionality, the language covers most application domains a modeling technique may be used for, sometimes in combination with additional supporting languages - purely describing the process, simulating it for analysis or allowing for execution. The graphical notation can be easily understood by the various process participants such as business and process analysts or software engineers with business processes themselves represented as Business Process Diagrams (BPD) \cite{BPMNintrostandard}. Graphical elements follow flowcharting notations for better readability and flexibility while execution semantics are completely formalized~\cite{OMGBPMN}. As a standard, the language has been adopted and is being actively maintained by the Object Management Group (OMG), at the moment of writing under version 2.0.2. As such it can be placed within the four-layered metadata architecture OMG uses to define languages through their corresponding metamodels. A \textit{metamodel} formalizes statements about a class of models, such as their syntax, semantics and other constructs, in its own further abstracted model~\cite{metamodeldef}. At its highest level the OMG four-layered metadata architecture consists of a meta-metamodel layer to define a language for specifying a metamodel, represented by the Meta Object Facility (MOF)~\cite{MOF}. The second layer specifies a language to define a model, at OMG the Unified Modeling Language is applied to do so. At the model layer, specific models from the metamodel can be created by a user. Finally, run-time instances of the model elements specified in the model are contained in the forth layer of the architecture, the data layer~\cite{metamodelarchitecture}.  Derived from Pillat et al.~\cite{bpmnt} an aggregated version of the main BPMN 2.0 meta-classes of process elements and objects alongside their relations is presented in Figure~\ref{fig:BPMN_meta_model_overview} using UML notation. Next to the metamodel used to standardize the Business Process Modeling Notation, interchange formats for BPMN models are supported through the Extensible Markup Language (XML) specified in XML schema documents (XSD). Further, the interchange formats can be used as blueprints to derive executable specifications that can directly be deployed in workflow systems while also enabling validation and verification control of the process model~\cite{BPMNintrostandard,bpmnt}.

\begin{figure}
    \centering
    \includegraphics[width=0.75\linewidth]{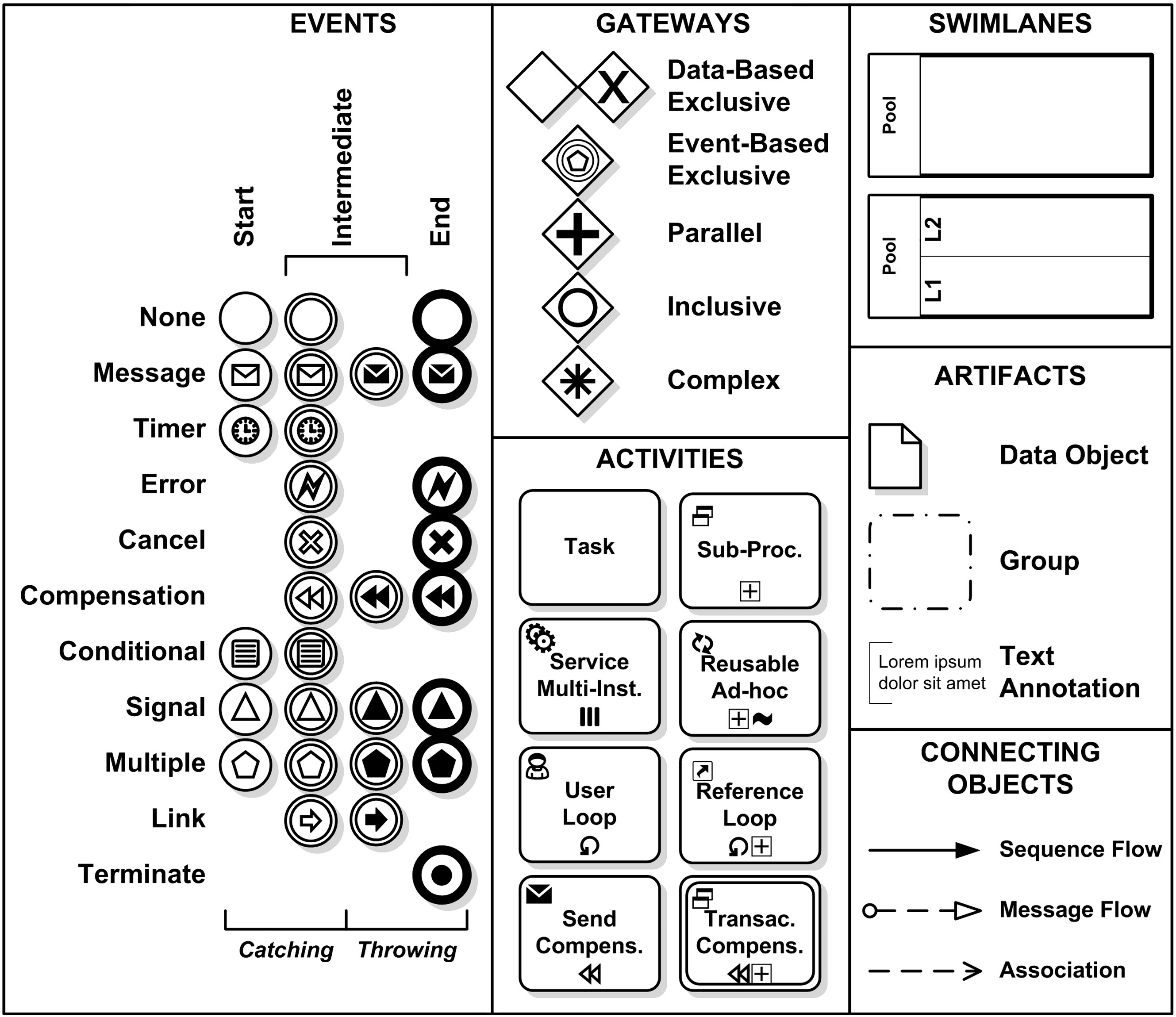}
    \caption[Main BPMN 2.0 notation elements]%
    {Main BPMN 2.0 notation elements \par \small The main BPMN 2.0 notation elements separated into events, gateways, activities, swimlanes, artifacts and connecting objects. Adapted from~\cite{BPMNintrostandard}.}
    \label{fig:BPMN_elements}
\end{figure}

Most elements of the BPMN vocabulary can be grouped into activities, events, gateways, connecting objects, artifacts and swimlanes~\cite{BPMNintrostandard,BPMBookCh3}. Figure~\ref{fig:BPMN_elements} provides a summary of the standard's visual notation - graphical elements to represent the vocabulary. Activities are presented as rounded rectangles. More specifically, an activity can be a task - an atomic unit of work - or a composite of several tasks, i.e. a sub-process. Gateways, BPMN's control nodes, are indicated in diamond shapes. Both elements can be structured via sequence flows, arcs that are part of connecting objects and graphically depicted by an arrow with a full arrow-head~\cite{BPMBookCh1}. Events are described by circles. They can mark the start or end of a process as well as intermediate situations happening instantaneously during execution and can be further classified into catching and throwing events. Circles with a thin (thick) border describe start (end) events while a double layered border indicates an intermediate event. Catching events are markers with no fill whereas throwing events are markers with a dark fill~\cite{BPMBookCh3}. BPMN further accounts for abstract artefacts such as data related objects, for instance a data store or data in- and outputs. Additional information can be provided via text annotations which however complicate process validation or verification as the textual data contained within is ill-defined~\cite{uBPMN}. Process resources such as process participants are described by pools or lanes. Resources can be active and passive - the former capable of autonomously performing an activity while the latter is only somewhat involved with the activity's performance. A complete overview of the BPMN vocabulary, syntax, semantics and entire range of elements is available in the latest documentation of BPMN~\cite{OMGBPMN}.
Making use and extending BPMN in the course of this research comes with two benefits - 1) leveraging widely known and properly defined semantics, syntax and logic and 2) building up on existing state-of-the-art and thereby improving on relevance and validity of the to be created artefact.

\subsection{BPMN Extension}
\label{BPMNExtensionMechanism}
Several BPMN extensions have been proposed over the years to incorporate more functionality and address various shortcomings for specific domains to which the language is applied. These rank from fully business related aspects such as business activity monitoring or communications to more recent technological developments such as ubiquitous computing or quantum applications that need to be properly represented in the modeling standard, see among others~\cite{BPMNExComms, BPMNExBAM, BPMNExKM,BPMNExQuant, bpmnt,uBPMN,BPMNExclinical,BPMNExDataQuality,BPMNExIoT,BPMNExWirelessSensorNetworks}.

In general, the extensions enable the representation of domain-specific concepts in BPMN~\cite{BPMN_extension_guide}. Similarly, extending the language may allow transformation into executable files or code for specific purposes. To enrich the general BPMN standard with domain-specific concepts one can apply extension by addition, i.e. attaching novel elements to the pre-defined existing ones of the language. When extending BPMN, three pillars can be considered - 1) the MOF meta-model defining abstract BPMN objects in UML, 2) the XML schema documents which are derived from the meta-model and represent the structure in a machine readable format and 3) the graphical notation that is being used by the modeler. A semantically correct extension primarily focuses on the first pillar which in turn requires the designer to consider low level implementation challenges and particularities of the respective domain. Based on the metamodel, the XML schema documents may be extended. A standardized approach for visualizing the graphical representation of the extended elements has not been fully formalized, thus most extensions stick close to existing BPMN elements incorporating slight changes via context-representative icons~\cite{BPMNExComms,BPMNExQuant,uBPMN}. The latest BPMN version provides a guiding extension mechanism that focuses on the addition of elements and attributes to the BPMN metamodel via four components - 1) \textit{ExtensionDefinition} to group new attributes under a new concept name, 2) \textit{ExtensionAttributeDefinition} to represent the respective attribute, 3) \textit{ExtensionAttributeValue} to store the attribute's value and 4) \textit{Extension} to bind the new concept to the BPMN model. Notably, this is a means of guidance instead of hard-written rules to align with the BPMN core semantics and various approaches referenced in literature exist.
Building up on the BPMN metamodel representation, new extensions can be precisely ideated, defined and explained \cite{bpmnt, BPMNExKM}.

Depending on the purpose of the extension, the introduced BPMN additions in related research focus on different process modeling perspectives - be it to for instance conceptualize and visualize the domain-specific process in a more articulate fashion to process participants or to enrich the data contained in the process model, e.g. the data flow with its corresponding inputs and outputs, such that it contains the required information to enable execution (transformation towards executable models) in the respective application domain. As the focal point of the proposed extension artefact is of a technical nature - the ML pipeline design and implementation -, the latter existing extensions are of interest considering potential re-use of proposed elements or as a stepping stone to build on top of. The BPMN extension artefact presented in this thesis is based on a metamodel extension.

\section{Model-driven Engineering}
\label{ModelDrivenEngineering}
Model-driven engineering leverages the characteristics of models to decompose complex phenomena and  explain them more abstractly in a structured organisation of distinct yet related sub-components. By doing so, the inherent complexity of developing and maintaining large software systems can be better dealt with. Through the combination of two main technologies - 1) a modeling language that properly represents domain-specific constructs and 2) a transformation engine and generator that can analyse these constructs and synthesizes the corresponding artefacts - an automated transformation process from model to artefact, e.g. source code or deployment instructions, can be achieved. Next to the domain concepts, the domain-specific modeling language can define associated semantics and constraints in the form of a metamodel which in turn also facilitates automated model checking and constraint enforcement. Further facilitating communication, this approach helps in opening up the designed system to a larger range of stakeholders~\cite{MDEngineering}. By structuring the deployment descriptions derived through the transformation engine as deployment models, deployment automation can be achieved.  

\subsection{Deployment Modeling}
Particularly in cloud-based, serverless environments management of services and components becomes complex which in turn requires deployment processes that take care of the continuous automated delivery over the components' entire lifetime, i.e. installing , starting, stopping or terminating them~\cite{DeploymentMetamodel}. In the context of machine learning, meeting this need becomes more important with increasing size of the ML system and number of artefacts. Two main deployment modeling concepts exist, i.e. \textit{declarative deployment modeling} and \textit{imperative deployment modeling}. The former one leverages declarative deployment models which contain descriptions about structure, components, configuration and relationships of the to be deployed application. These descriptions are then interpreted by a declarative deployment system which fetches the technical actions that need to be taken to deploy the modeled application and enacts them subsequently. Consequently, the deployment system takes care of the actual deployment logic whereas the deployment model only describes the components alongside their context which need to be deployed. In contrast, imperative deployment models also contain the actual procedural deployment logic, i.e. the process steps such as technical deployment tasks that need to be executed alongside their order and data flow~\cite{DeclarativeVSImperative}. While both deployment concepts find application in various contexts, declarative deployment modeling is largely accepted as the most appropriate approach~\cite{DeploymentMetamodel}. An OASIS open-source industry standard supporting both deployment modeling concepts in a technology-agnostic manner is the \textit{Topology and Orchestration Specification for Cloud Applications}~\cite{TOSCAOasis} (TOSCA) - a standard which is referenced in this work as the go-to choice for deploying the modeled machine learning workflow.  Related work has shown that standardized TOSCA modeling constructs are well suited for integrating heterogeneous technologies and serverless architectures due to the focus on portability as well as interoperability, thereby justifying the choice~\cite{TOSCAServerless,wettinger2014unified,TOSCA4QC}. 


\subsection{Topology and Orchestration Specification for Cloud Applications}
\label{TOSCASection}
\begin{figure}
    \centering
    \includegraphics[width=0.8\linewidth]{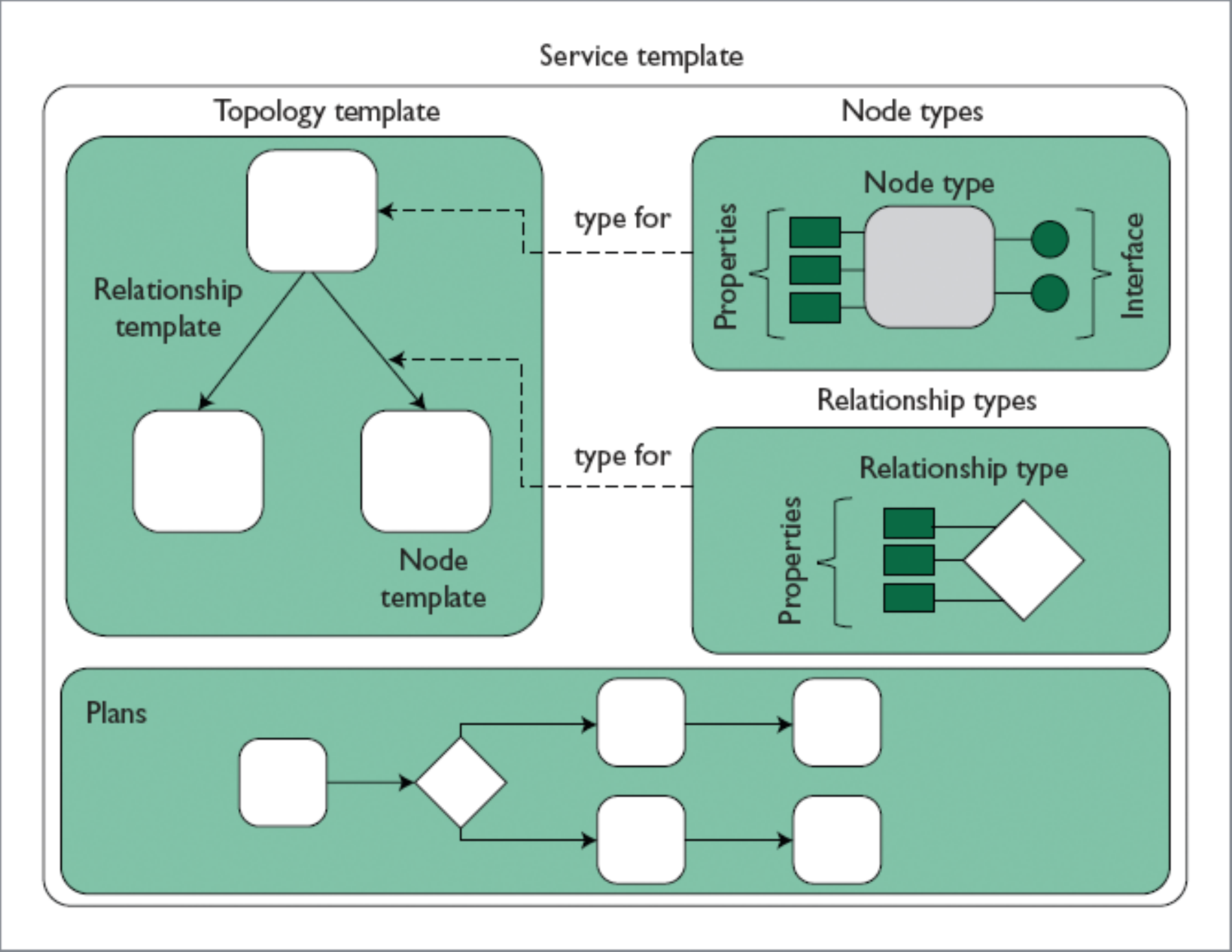}
    \caption[TOSCA Service Template]%
    {TOSCA Service Template \par \small Application components are represented by nodes and structured by relationships into a topology. The cloud's service operational aspects are captured by plans. Drawn from Binz et al.~\cite{TOSCACloudServices}.}
    \label{fig:toscaServiceTemplate}
\end{figure}

The \textit{Topology and Orchestration Specification for Cloud Applications} enables creation, automated deployment and management of portable cloud applications by specifying an XML-based modeling language to formalize the structural description of an application as a \textit{typology graph} and capture the related management tasks as plans, see Figure~\ref{fig:toscaServiceTemplate}.

Next to this, software files are responsible for instantiating the topology~\cite{TOSCA}. The typology graph can be understood as a directed acyclic graph comprised of nodes that describe the application's components and of edges specifying the relationships between them. The structure is then defined in the form of a \textit{typology template} that at its core is comprised of \textit{node templates} and \textit{relationship templates} as building blocks which also include application-specific middleware and infrastructure components. Reusable \textit{node types} and \textit{relationship types} are referenced to define the characteristics, i.e. the semantics, of the corresponding node and relationship templates. This way, type hierarchies can be encoded. A set of normative types is specified by the TOSCA standard, e.g. 'hostedOn' to define dependency relationships or 'connectsTo' to define communicative relations~\cite{TOSCA4QC}. To connect nodes through a certain relationship a \textit{requirement} can be introduced in the source node type and linked to a target node by adding the corresponding \textit{capability}~\cite{FaaSBPMNOrchestration}.

\begin{figure}[h!]
    \centering
    \includegraphics[width=0.9\linewidth]{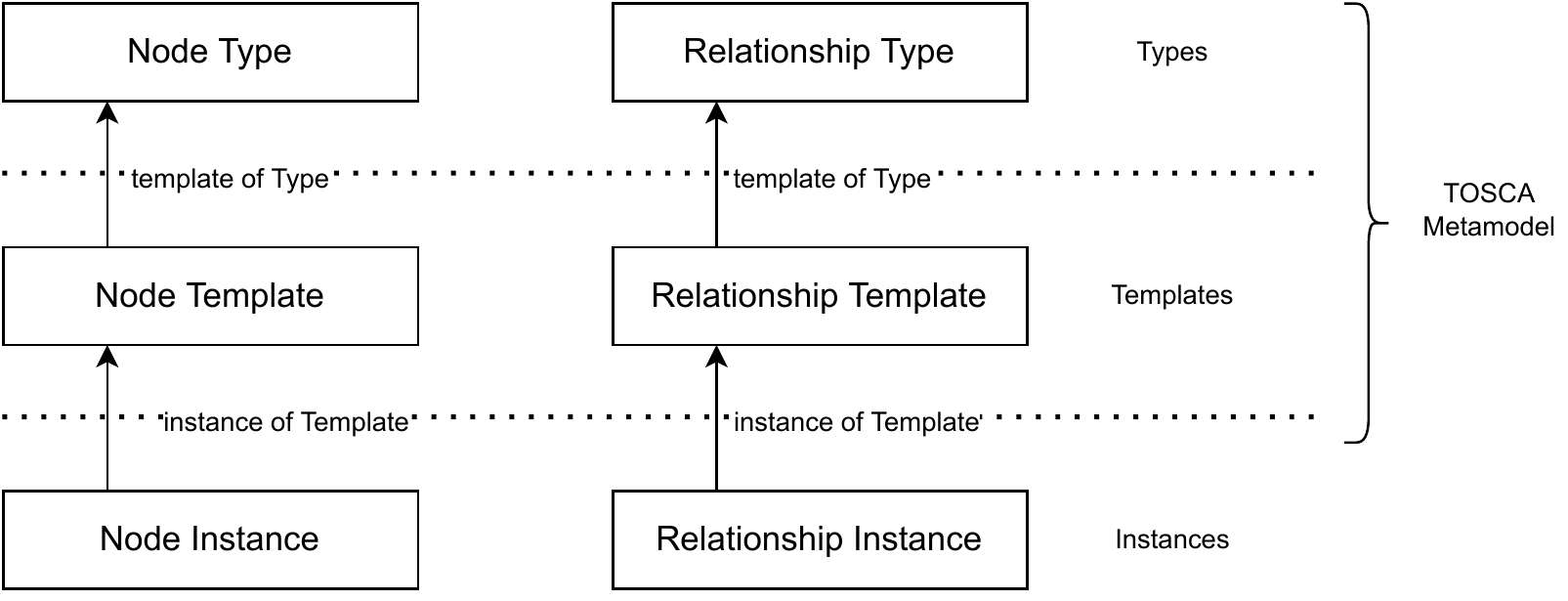}
    \caption[Conceptual layers of TOSCA nodes and relationships]%
    {Conceptual layers of TOSCA nodes and relationships \par \small Drawn from Binz et al.~\cite{TOSCA}.}
    \label{fig:toscaMeta}
\end{figure}

Node and relationship types allow to abstract one layer from the more concrete node and relationship templates, see Figure~\ref{fig:toscaMeta}. An instance of the template then functions as a real existing and instantiated component or relationship~\cite{TOSCA}. 

Similar to what Yussopov et al.~\cite{FaaSBPMNOrchestration} propose, more customized and domain-specific types can be derived from the existing normative one. In order to better configure node and relationship templates, \textit{properties} can be specified such as a port number for communication or an identifier defining in which region of a cloud provider a node should be set up.  In addition, nodes and relationships can offer defined \textit{interface} operations to specify actual deployment and management details which enable TOSCA-compliant provisioning engines to trigger the correct lifecycle operation. Again, normative lifecycle operations are pre-defined, i.e. \textit{create}, \textit{configure}, \textit{start}, \textit{stop}, \textit{delete}, and can be further customized~\cite{TOSCA4QC,TOSCAServerless}. To provide the actual logic performing the required operation \textit{implementation artefacts} (IA) can be assigned to the correct node or relationship type through a \textit{node type implementation} or respective \textit{relationship type implementation}.  The actual business logic, i.e. what work a node should perform, can be attached as \textit{deployment artefact} (DA) to the corresponding node template, e.g. as a .zip archive of the component's code~\cite{FaaSBPMNOrchestration}. 

The combination of components of structure and behaviour information as well as other metadata and attached artefacts are referred to as a \textit{service template} and represent a complete, deployment-ready application~\cite{TOSCAServerless}. The TOSCA application model can then be packaged into a \textit{cloud service archive} (CSAR) that groups all required information into one file. The CSAR enables TOSCA-compliant deployment technologies to consume all necessary artefacts, enact their logic and ultimately deploy the application.

\section{Related Work}
\label{relatedWork}
This study sources from and synthesizes four overall streams of research - machine learning, serverless computing as well as process and deployment modeling. A large body of publications addresses various areas of these fields. Of particular relevance to this work are studies researching ML workflows and lifecycles with different areas of focus, serverless deployment of such workflows as well as potential process and deployment modeling solutions to represent them. A large amount of publications has already been referenced throughout this chapter, thus this section highlights selected relevant but different publications that the current work builds up on or contributes to.

\subsection{Machine Learning Workflows and Lifecycle}
Lukyanenko et al.~\cite{ConceptualModelingML} formulate first ideas towards the use of conceptual modeling as a standard activity to support machine learning in order to address challenges around interpretation of ML models, execution of ML algorithms as well as their overall integration in organizational processes. While no direct solutions are provided, the authors propose future research directions structured along the six phases of the CRISP-DM framework, involving improved 1) business and 2) data understanding, 3) data preparation, 4) modeling, 5) evaluation and 6) deployment.

To address their call for research, a thorough understanding of ML workflows and ML artefacts is required. A large body systematic literature surveys and related studies on this topic exist. Among others Ashmore et al. study the ML lifecycle with respect to assurance of ML applications in safety-critical systems~\cite{AssuringMLLifecycle}, Gharibi et al. review the ML lifecycle by focusing on deep learning in particular~\cite{ManagementModelingLifecycleinDL}, Qian et al. investigate it with respect to IoT applications~\cite{IotMLLifecycle}, Giray examines it through a software engineering perspective~\cite{SEperspectiveML} and others such as Lo et al. address it by focusing on ML methodologies such as federated learning or overall distributes ML~\cite{FLRA,FLArch,DMLSurvey1}. Also industry publications address the topic of ML workflows or more applied derivations thereof such as MLOps. Amershi et al. share a Microsoft view on the topic~\cite{MicrosoftMLWorkflow}, Sun et al. provide insights into solutions at Uber~\cite{uber} and Salama et al. place machine learning workflows in the bigger picture of MLOps at Google~\cite{googleML}. Moreover, other work exist, both academic and industrial, that investigates in particular data artefacts and metadata that occur throughout the ML lifecycle. A conceptualization of ML assets is described by Idowu et al.~\cite{AssetML}. Schelter propose Amazon's understanding of ML metadata~\cite{MLOpsMetadata} and Renggli et al. consider data-quality related aspects~\cite{MLOpsDataQuality}. 

This work differs from the mentioned publications as it aggregates the large body of research into a formalized set of phases, activities, events and data artefacts that compose machine learning workflows and the ML lifecycle and allow its representation.

\subsection{Serverless Machine Learning}
With serverless computing gaining popularity to realize machine learning, first efforts in streamlining the combination of the two domains are proposed. Most studies focus on ML sub-domains such as deep learning or federated learning to leverage serverless functions. Consequently, several custom-made frameworks are introduced.

Carreira et al. describe a Python framework to define ML workflows that solely relies on AWS serverless functions and storage solutions~\cite{ServerlessCirrus}. Minchen et al. introduce a framework to automatically partition and serve deep learning networks via serverless functions~\cite{Yu2021GillisSL}. Grafberger et al. propose a serverless framework to realize federated learning in which both clients and server run FaaS functions~\cite{FedLess}. An architecture proposal based on AWS services to realize serverless deep learning workflows is presented by Chahal et al~\cite{ServerlessDL}. Kurz outlines a Python prototype to implement serverless double machine learning via AWS Lambda functions~\cite{Kurz2021}.

In another vein, Jiang et al. compare the solution performance between serverless and 'serverful' machine learning~\cite{DemystifyingSML}. To circumvent the constraints of purely serverless ML, Bac et al. propose leveraging edge computing layers~\cite{ServerlessEdge}. As a general discussion on serverless computing, the proceedings of the Dagstuhl Seminar address limitations that among others hinder the adoption of serverless functions in the context of ML and suggest hybrid solutions to offload certain machine learning jobs to more appropriate platforms or hardware~\cite{ServerlessComputingReport}.

This work differs from the mentioned publications as it does not aim to develop yet another custom-made framework to realize serverless machine learning but instead proposes a standardized, technology independent and interoperable way to model ML workflows and the subsequent deployment while accounting for serverless intricacies.

\subsection{Serverless Application and Workflow Modeling}
To design serverless applications and workflows, currently architecture diagrams are used which are then implemented 
as code and infrastructure as code definitions~\cite[p.76]{ServerlessComputingReport}. Consequently, active research is conducted towards an intermediate language or respective modeling tool that can bridge the gap between the complex combination of code and infrastructure as code and non-standardized architecture diagrams. In the field of provider-agnostic modeling of serverless applications and workflows, various recent solutions have been published.

Wurster et al.~\cite{TOSCAServerless} offer an initial answer to technology agnostic modeling and deployment of serverless applications via TOSCA. However, their main focus lies on mixed infrastructures, that consist of both serverless and traditional applications. While their solution supports event-driven approaches to configure and relate cloud services with functions and corresponding events, function orchestrations are not covered. Ristov et al. present AFCL, an abstract function choreography language, to describe serverless workflows. They further introduce a custom-made system to execute the AFCL workflow models on different FaaS providers~\cite{Ristov2021AFCLAA}. The Serverless Workflow project represents efforts to create a workflow modeling language standard that is composed of a custom-made domain specific language (DSL) as well as a software development kit (SDK) and other language-specific tools to define function orchestrations and execute them by means of a workflow engine that corresponds to the DSL~\cite{SWSA}. Yussupov et al. introduce a standardized set of constructs for the provider independent modeling of FaaS functions as BPMN generic Task equivalents and subsequent deployment orchestration via TOSCA, albeit with limitations. Their work is one of the initial studies towards adopting existing standards for modeling, deploying and executing function orchestration models on available function orchestration engines.

This work draws from the efforts in leveraging TOSCA to derive deployment models for function choreographies as well as serverless and serverful applications. It differs by extending BPMN to properly model serverless workflows for the domain of machine learning.

\subsection{BPMN Extensions}
Several publications have been introduced to propose BPMN extensions for specific domains, however the field of machine learning has been left untouched. Nevertheless, constructs proposed by existing related extensions can potentially be leveraged and are therefore reviewed.

Yousfi et al.~\cite{uBPMN} address the development in technology towards ubiquitous computing - one subject (e.g. an individual) accessing, interacting and operating multiple computers anytime anywhere. The resulting ubiquitous business processes require better representation in BPMN to properly incorporate novel data input and collection technologies. The BPMN 2.0 metamodel definition is extended to account for sensing, reading and other data collection technologies. Next to defining corresponding task elements, a \textit{SmartObject} is introduced to enhance usage possibilities of regular BPMN \textit{DataInputs} for processes. Further, various \textit{EventDefinitions} carrying data are added allowing for modeling of situations triggered by ubicomp technologies. The extension focuses mainly on input technologies for any business process and is not directed towards ML workflows. The concept of smart sensor events which are triggered in the presence of new data (SmartObjects) can however also be considered in ML pipelines, albeit on a more abstract level - for instance in case raw data is sourced from smart sensors such events may trigger data sourcing or processing activities and subsequent runs of a ML pipeline. Furthermore, the reasoning of data input events used as task triggers can be translated to similar ML lifecycle events such as the arrival of new data input for the respective ML repositories (e.g. feature repository, dataset repository etc.). Additionally, this concept directly corresponds to the event-driven paradigm of serverless applications (for example an AWS S3 data input event).

Next to Yousfi et al.~\cite{uBPMN} also Tranquillini et al.~\cite{BPMNExWirelessSensorNetworks} and Meyer et al.~\cite{BPMNExIoT} study possibilities to better integrate edge devices in BPMN process flows, thereby shifting focus away from human and IT-system centered business processes towards technology-dependent digital workflows. Tranquillini et al.~\cite{BPMNExWirelessSensorNetworks} further experiment on BPMN-based deployment of sensor networks, directly addressing the importance that data flow and code artefacts play throughout the process. Similarly, machine learning data and code artefacts need to be taken into consideration.

Pillat et al.~\cite{bpmnt} present the concept of formally tailoring a standard software process to individual needs of an organization. Proposing a tailoring mechanism for BPMN-based software processes, the researchers extend the BPMN syntax and semantic to facilitate removal, replacement or addition of elements in an operational enterprise process according to their described procedure. The idea of facilitating and standardizing removal, replacement or addition of process elements can be related to the context of ML workflows for which operations need to be updated. However, the suggested process tailoring mechanism is not transferable to the context of this work.

Weder et al.~\cite{BPMNExQuant} propose an approach to incorporate the modeling of quantum computing applications in workflow languages to further facilitate orchestration of quantum circuits and classical applications. The researchers particularly focus on the alignment with existing workflow execution engines to avoid having to extend the execution engine itself. As such, the proposed quantum modeling extension remains language independent and, in a second step, requires a mapping towards the native constructs of the desired workflow language that is used to model the process and subsequently execute it. Similar to previous research, a BPMN extension is showcased subsuming the extended quantum modeling extension with BPMN native elements. New \textit{tasks} types are introduced whose attributes contain relevant information for the execution of the quantum application. Additionally, two \textit{DataObjects} are added to clearly define domain-specific data and properties that are passed along the quantum circuit. While the presented extension elements cannot be applied to a machine learning workflow, equivalent elements can be created such as different \textit{ResultObjects}, \textit{ResultDocuments} or an adapted \textit{DataPreparationTask} representing operations during the data management phase.

The BPMN extension by Braun et al~\cite{BPMNExclinical} focuses on medical pathways, processes in the medical field. Additional \textit{Task} and \textit{Gateway} types are defined next to new \textit{DataObjects}, e.g. a \textit{DocumentObject} or a  \textit{CPG Reference},  that allow to better incorporate domain-specific values in the process. The new elements are introduced as they allow to represent peculiarities and specific information inherent to the medical field thereby answering the need to lower the level of abstraction of ordinary BPMN elements such as deriving a \textit{DocumentObject} with additional of \textit{properties} from a \textit{DataObject}. A similar reasoning for lowering, or rather adapting, the level of abstraction of certain BPMN elements applies to the machine learning domain since ML workflows produce their own data artefacts and information documents. Thus, such concepts need to be provided by the BPMN extension of this work.

In summary, current depictions of the machine learning lifecycle and workflows lack a common language and formalization of what constitutes each phase in terms of activities, events and generated or referenced data artefacts. While several extensive surveys have been presented by related works, heterogeneity still exists. Further, serverless machine learning is just experiencing its advent and experimentation on how to best leverage serverless computing for ML is ongoing. Meanwhile, publications on this topic tend to introduce custom-made frameworks and solutions constrained to a specific cloud provider. Further, the solutions are based on different interpretations of what constitutes a machine learning workflow thereby limiting the extend to which they can be leveraged or generalized. Ongoing efforts are made to consolidate the serverless offerings of different cloud providers within one technology agnostic tool. While this study does not aim to provide new solutions on how to orchestrate serverless workflows, it allows to abstract from provider specific workflow modeling by leveraging and building on top of BPMN and TOSCA as two established standards. With respect to the machine learning domain, currently no formalized modeling extension exists to represent ML workflows in a coherent and standardized vocabulary, semantic and notation. Therefore, a technology independent and interoperable solution to 1) model machine learning workflows and 2) their serverless deployment orchestration contributes to the discussed areas of research. It further addresses the initial call for work on conceptual modeling for machine learning by Lukyanenko et al.~\cite{ConceptualModelingML}.

%% file: chapters/3.methodology.tex
This chapter introduces the research methodology followed throughout this thesis. Since the present work ties in with information systems and software engineering research, design science research as a prospective methodology can be considered and is discussed in section~\ref{DesignScience}. Further, with the main goal of this work being the creation of artefacts, the Design Science Research Methodology process can be applied. Thus, next to introducing design science, its mandated steps to follow while creating the artefacts are explained alongside presenting the research design of this work in section~\ref{RMD}.

\section{Design Science}
\label{DesignScience}
As Hevner et al. postulate, research in information systems takes to the largest extent from either of two paradigms - 1) behavioural science and 2) design science~\cite{Hevner2004}. The former focuses on the establishment and verification of theories to understand and anticipate human as well as organizational behaviour. To that end, behavioural science aims to leverage the developed understanding on the interplay of organizations, technology and people to improve the efficiency and effectiveness of information systems. It stems from natural science research methods and can be complementary to design science~\cite{MARCH1995251}. The latter's roots lie in engineering and the sciences of the artificial~\cite{Hevner2004,SciencesOfTheArtificial}. Its focal point is the creation and evaluation of novel and innovative solutions, i.e. viable artefacts, to stretch human and organizational capabilities. Such artefacts must be differentiated via identification and highlighting of a clear contribution to for instance the foundational knowledge base or methodologies while being relevant and important to existing business problems. Among others, such artefacts can be associated with constructs, models, methods and instantiations~\cite{Peffers2007,Hevner2004}. The present work aligns with these categories as its goal is the creation of an extension to the BPMN modeling standard via 1) a conceptual model of (serverless) machine learning workflows and 2) corresponding development of modeling constructs (i.e. notations) and semantics. Further, the proposition of a method in the form of a conceptual mapping to convert an instance of a ML workflow model to a deployment model for serverless deployment is pursued.

Next to calling for rigorous development, Hevner et al. further stipulate that the proposed artefacts are rigorously evaluated in terms of their utility, quality and efficacy by providing evidence that justifies the artefacts' creation and existence~\cite{Hevner2004}. The overall contribution shall be verifiable or falsifiable. Procedures to collect such evidence are suggested by Peffers et al. and will be elaborated upon in the subsequent presentation of this study's research design~\cite{Peffers2007}. 

Within design science, different stipulations on how to concretely structure the research design exist~\cite{Hevner2004,Peffers2007}. Derived from concepts of both behavioural science as well as design science, Hevner et al. propose seven guidelines as means of orientation: 1) design as an artefact, 2) problem relevance, 3) design evaluation, 4) research contributions, 5) research rigor, 6) design as a search process and 7) communication of research. While these guidelines provide a general outline of concepts on what and what not to do, a concretization of the exact process and activities involved is lacking. Peffers et al. close this gap for the case of information systems by extrapolating the guidelines and related work into a Design Science Research Methodology which involves six activities structured as a process in nominal sequence which can be iterated upon. The research design of this work follows the Design Science Research Methodology.

\section{Research Method and Design}
\label{RMD}
The Design Science Research Methodology (DSRM) process prescribes six key activities in order to ultimately generate knowledge that can be referenced by practitioners in the domain to create specific solutions for their problem instance. The activities are:  1) problem identification and motivation, 2) definition of the objectives for a solution, 3) design and development, 4) demonstration, 5) evaluation and 6) communication, see Figure~\ref{fig:DSRM}~\cite{Peffers2007}. These steps comprise the underlying methodology and will be elaborated upon in the context of this research:

\begin{figure}
    \centering
    \includegraphics[width=1\linewidth]{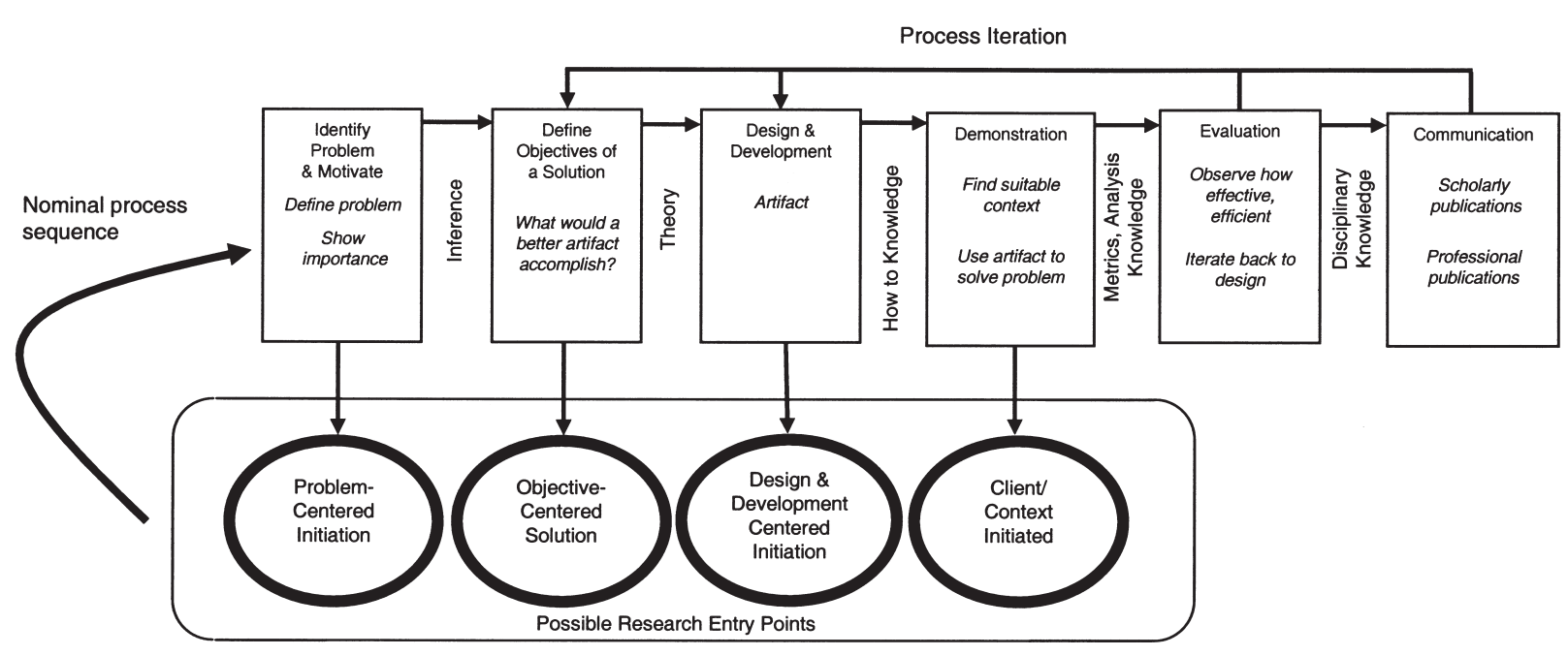}
    \caption[Design Science Research Methodology process]%
    {Design Science Research Methodology process \par \small Figure drawn from Peffers et al.~\cite{Peffers2007}.}
    \label{fig:DSRM}
\end{figure}

\subsubsection{1) Problem identification and motivation}
Refers to the specific research problem that is addressed and legitimizes a potential solution. The problem definition will be referenced throughout the development of the artefact. 

Machine learning, particularly serverless ML, is still largely seen as obscure workflow that is hard to integrate with existing business processes of an organization and difficult to communicate among technical and non-technical stakeholders. Further, a gap in research and practice has been identified regarding the requirements of modeling machine learning workflows as well as their serverless deployment orchestrations in respect to the provided capabilities of state-of-the-art practices such as a process modeling language standard, i.e. BPMN. Among others, the gap in research and limitations of current practical solutions makes modeling, communication, analysis as well as deployment of ML workflows an unstructured and error-prone process and further inhibits overall adoption of machine learning in organizations.

\subsubsection{2) Definition of objectives for the solution}
Refers to the quantitative and / or qualitative goals that are rationally derived from the problem definition in consideration of what is possible and feasible and takes shape in form of an innovative artefact.

The main objective is the extension of the Business Process Modeling Notation to enable modeling, communication and analysis of (serverless) machine learning workflows in a standardized, technology independent and interoperable manner. It is sub-divided into 1) the creation of a conceptual model for serverless ML workflows based on a requirement analysis and 2) the corresponding development of modeling constructs and semantics. The focus of analysis lies on supervised and unsupervised ML as machine learning categories alongside several ML methodologies and the general ML lifecycle. The solution should allow to model various scenarios and complexities of serverless ML workflows. 

A second-order goal is to propose a conceptual mapping to convert a ML workflow model instance based on the proposed extension to a deployment model of a technology-agnostic standard for serverless deployment. Consolidating the core artefact, the BPMN extension, with the conceptual mapping shall allow for the proposition of an end-to-end method to design ML workflows via the BPMN extension and subsequently derive their serverless deployment orchestration.

\subsubsection{3) Design and development}
Refers to the core activities of design science, i.e. the creation of the artefact by 'determining the artefact’s desired functionality and [...] architecture and then creating the actual artefact'~\cite[p.55]{Peffers2007}. 

The primary artefact takes the form of a model extension and associated constructs. The model extension is realized through meta-models that conceptually define data artefacts, tasks and events that are created and occur throughout (serverless) machine learning workflows and the overall ML lifecycle. The associated constructs are realised by a corresponding notation and respective semantic. They incorporate the desired functionality shaped by the identified requirements with regards to the objective for standardized modeling, analysis and communication of machine learning workflows. A secondary artefact proposes a conceptual mapping to convert a modeled diagram based on the new BPMN extension to a deployment diagram that can facilitate serverless deployment orchestration via TOSCA. Combining both helps in realizing technology independent and interoperable modeling and serverless deployment orchestration of ML workflows.

\subsubsection{4) Demonstration}
Refers to demonstration of the artefact's capability to solve instances of the researched problem by means of an appropriate activity. 

Various exemplary snippets of the core artefact are depicted throughout its presentation and explanation. Further, an illustrative use case is taken as form of demonstration. It showcases the possibility of modeling both FaaS-based machine learning aspects as well as offloaded capabilities in an end-to-end process model and associates the overall ML workflow with related services (i.e. in the case of the use case a resource provisioner and external monitoring component). Extensive simulations or experimentation cycles were limited in the scope and time of this work. 

Further, a conceptual mapping scenario is demonstrated that converts a simple BPMN4sML diagram of a credit default prediction machine learning workflow into a corresponding TOSCA template to orchestrate it on a public cloud provider, namely AWS. The serverless nature of the workflow is realized through FaaS functions orchestrated by the AWS StepFunction engine.  

\subsubsection{5) Evaluation}
References observation and measurement regarding effectiveness of the artifact's support for a solution to the studied problem and ties in with the foregoing demonstration. Evaluation of design artefacts can follow different methodologies such as elaborate evaluation frameworks that draw from previously established requirements to assess the artefact. Depending on available capabilities and constraints different methods can be pursued. In place of a sophisticated evaluation set-up, a descriptive method can be applied which however constitutes a limitation and a call for future research. 

In the course and scope of this thesis, a descriptive evaluation method is considered by leveraging the preceding illustrative scenarios. The use case references an existing implementation of an online machine learning solution proposed in peer-reviewed related literature to highlight the artefact's ability of depicting such a system as a BPMN4sML workflow model. It further validates the developed notation to represent existing machine learning tasks, data artefacts and events while accounting for appropriate semantics.

Further, the conceptual mapping scenario validates the possibility of converting a technology independent and interoperable machine learning process model into a corresponding deployment model for serverless orchestration. Taken together, the illustrative use case referenced from related literature and the mapping scenario realizing the conceptual mapping from BPMN4sML to TOSCA are considered to represent a sufficiently convincing prototype that addresses the specified objectives and answers to the identified research problem.
Synthesis of requirements through analysis of literature and industry publications can be understood as its own artefact and contribution, albeit an immeasurable one.

With regards to novelty, the artefacts are new to existing solutions - BPMN4sML incorporates (serverless) machine learning previously unaddressed by the standard; the conceptual mapping and transformation to TOSCA realizes a new mapping from the novel BPMN elements to existing TOSCA counterparts. On their own, the artefacts represent a high level of novelty. As an extension to current solutions, i.e. BPMN and TOSCA, the entirety of the artefacts and solution could however be viewed as limited in its novelty given that they extend current solutions. 

The artefacts contribute to the knowledge base of (serverless) machine learning workflows as well as to general process modeling and deployment modeling as part of model-driven engineering. They address both, practical as well as academic perspectives.


\subsubsection{6) Communication}
Refers to the communication of the research process, its artefacts and results as well as the developed knowledge and overall research contribution to different audiences such as technology-savvy stakeholders as well as managerial ones. 

As this study is conducted outside of an enterprise, stakeholders cannot be considered directly. 
The communication is realized by means of this thesis which explains the processes, generated artefacts and results as well as illustrates and discusses them. A scientific publication would be necessary to fully adhere to communication as it is addressed by DSRM.

\subsubsection{Overall Considerations}
While working towards the objectives and generally throughout the entirety of the DSRM process, research rigor needs to be ensured. For the generation of the artefacts, mathematical knowledge or a similar formalism to establish proofs is not suitable. Instead a well-established theoretical foundation is created while adhering to the research methodology. The theoretical foundation of this work takes in particular from machine learning research and industry practices. Further, serverless computing concepts are formed and organized, especially with regards to Functions-as-a-Service. Business process management practices, workflow patterns and modeling languages are reviewed. 

To form the theoretical knowledge foundation and understand the current state of research, this work conducts a white and grey literature review that adheres (to the best extend possible given the scope of this work) to principles brought forward by Kaiwartya et al.~\cite{Kaiwartya2016}. A systematic multi-vocal literature review as proposed by Petersen et al.~\cite{systematicLiterature} and realized by for instance Cascavilla et al.~\cite{GiuWJ} was infeasible. To circumvent ensuing limitations, existing literature reviews, surveys and reference architectures are leveraged and snow-balled upon.

To create the BPMN extension, next to the DSRM process a proposed methodology from and applied in related literature is followed~\cite{uBPMN,BPMNExclinical} - 1) the target domain for the extension is analysed in-depth, 2) the scope of and requirements for the extension are established, 3) the core structure of BPMN is extended (i.e. meta-models) and ultimately 4) the notation is extended.

%% file: chapters/4.requirement_analyses.tex
In order to properly represent machine learning workflows in a process model diagram, their constituent characteristics such as artefacts generated or accessed throughout the process or activities recurring over various ML workflows need to be identified so as to derive overall modeling elements accounting for the domain. The corresponding requirement analysis builds up on the previous sections w.r.t. business process modeling, serverless paradigms and machine learning in chapter 2. Machine learning specificities are elaborated upon by revisiting the machine learning lifecycle in detail in section~\ref{MLrequirements}. Moreover, given that the present research pursues a serverless deployment approach, certain serverless characteristics come into play that potentially need to be considered when creating the modeling extension elements. Consequently, section~\ref{FAASCDReq} discusses prerequisites associated particularly with the Function-as-a-Service paradigm. Ultimately, BPMN functionality and extensions for related domains need to be taken into account, ensuring that already established concepts and elements are re-used, built upon or referenced. Therefore, this chapter concludes with a synthesis of the identified requirements and an equivalence check to existing BPMN functionality that informs the subsequent work. Note that the provided requirement indices in this chapter are structured according to Table~\ref{tab:requirement_synthesis} for clarity's sake and are thus potentially unsorted throughout the following sections. 
 

\section{Machine Learning}
\label{MLrequirements}
As previously elaborated upon, machine learning as a domain can incorporate various different characteristics depending on methodology and purpose of the ML solution. Thus, to better guide the process of identifying ML activities, decisions, artefacts and events, it can be structured alongside the ML lifecycle and, with regards to this research, particularly alongside ML pipelines as these are of primary interest. Consequently, this section studies each phase of the ML lifecycle more thoroughly, i.e. requirement analysis, data management, model learning, verification, deployment and model monitoring and inference with respect to representative characteristics for conceptual modeling in order to establish a common ground of necessary elements. Note that key events for each phase are extrapolated from literature analysis and industry practices. Focus lies on creating a foundation that can be associated with the respective counterparts of the BPMN language.  

\subsection{Requirement Analysis}
As section~\ref{MLLRA} elaborates, activities during the requirement analysis inform on constraints and possibilities for the prospective machine learning service. To do that, mostly domain experts such as data scientists perform analytical tasks both on available data, i.e. exploratory data analysis, and on surrounding business processes and goals in order to establish a boundary frame in which the developed ML service can be placed. The respective activities can be considered outside the actual machine learning workflow, in the sense of a pipeline that is to be deployed, and thus are not explicitly limited by currently available modeling elements - for instance a user task in BPMN can reflect the activities a data scientist performs. In contrast, potential information in form of artefacts that are generated through these tasks may need to be referenced later on in the ML workflow. An example poses a requirements document which formalizes the constraints for the ML solution through performance thresholds or other explicit standards as elaborated by Ashmore et al.~\cite{AssuringMLLifecycle} (\textbf{R 24}).  Additionally, decisions made with regards to which overall ML methodology is applicable to the respective scenario may inform the subsequent ML workflow~\cite{MicrosoftMLWorkflow}.

A relevant event occurring throughout this phase can be the establishment of new requirements, e.g. a higher threshold for ML model performance, or the identification of new relevant features that shall be added to the dataset used for model training. If new requirements present themselves, the entire ML pipeline or selected activities such as model learning or model verification may need to be initialized (\textbf{R 7}). Furthermore, a decision for actively triggering the subsequent ML workflow phases and activities, i.e. data preparation, ML learning and verification and potential deployment, can be taken (\textbf{R 6}). \\

\subsection{Data Management}
Section~\ref{MLLDM} introduces data management practices in the context of the ML lifecycle. Notably, a large range of activities, artefacts and events can be created throughout this phase that require representation in a corresponding process model.

As Idowu et al.~\cite{AssetML} postulate, quality of datasets is crucial for ML model development which therefore justifies representation of corresponding aspects and activities. Conventional data sourcing operations are fundamental, e.g. an extract-transform-load (ETL) operation as an activity or process representing extraction, potential first transformation and subsequent loading steps of raw data into a data storage such as a relational database as part of a data warehouse or a more generic data lake~\cite{ETLPatterns}. Additionally, also data fusion operations may be required if for instance, data from several sources reflect different aspects and perspectives of the same phenomenon, e.g. capturing a person walking across a street through multiple cameras. Fusing these data sources for better overall data quality requires techniques to correctly associate separate data points with one another to form a 'combined' observation~\cite{IotMLLifecycle}. A data sourcing activity can theoretically be modeled in a granular manner, accounting for detailed operations and step-by-step logic, for instance as illustrated by Awiti et al.~\cite{BPMNExtensionETL}. Similarly, a data fusion activity may itself be a sophisticated process~\cite{DataFusionExample}. In the case of machine learning workflow modeling, modeling these intricacies is not the main focus. Instead, respective overarching representations under a collective name are appropriate - such as a fusion or sourcing operation (\textbf{R 25, 26})~\cite{GoogleMLOPsDoc,AWSmlconcepts}. If necessary, these operations can then be associated with other process modeling diagrams that hold detailed descriptions. 
In contrast to regular data sourcing operations, a fusion operation may not only be directed at raw data but can also occur in later stages of the ML workflow to fuse for instance features originating from different sources for the same phenomenon or to 'fuse' already processed results based on data of different data sources into one final data fusion result~\cite{DataFusionSensing}. Hence, this presents further requirements to the modeling language.

Apart from data sourcing, data management can also account for validation activities to prevent training a ML model on corrupted data as described by Google's MLOps white paper (\textbf{R 27})~\cite{googleML,GoogleMLOPsDoc}.
Artefacts that can be generated as results of these operations range from raw data that is stored (if required) in corresponding raw data stores to feature sets which can be stored in a corresponding feature repository. A feature repository represents a centralized storage that holds standardized feature sets which can be re-used for different ML workflows (in this case a feature refers to a mostly cleansed and prepared attribute of some entity). Notably, following Google Cloud's definition~\cite{googleML}, a feature repository does not directly require labeled data instances as part of the feature sets (they may instead need to be combined). Thus, in case of supervised learning, another storage for the final dataset used for the ML learning phase is can be necessary, i.e. a dataset repository, (\textbf{R 28, 29})~\cite{AssetML}. Note that this is not an explicit constraint but more of a guideline as feature sets may indeed also hold a target variable.

Different from sourcing and fusion operations, data management and preparation in the context of machine learning requires separated representation. As indicated in section~\ref{MLLDM} a large number of data pre-processing steps can be conducted to shape the stored data or features into the ideal format for a ML task. While not each potential technique needs to be differentiated, on a modeling level it is still beneficial to account for heterogeneous outputs of these operations. For instance, a conventional pre-processing activity such as data cleaning (removing missing values etc.) takes raw data or a feature set and outputs a processed or 'cleaned' dataset or a further processed feature set (\textbf{R 30})~\cite{AssuringMLLifecycle, MicrosoftMLWorkflow,AWSmltaming}. In contrast, a feature engineering activity may provide a dataset or a feature set which is differently structured from the one used as input, i.e. only a selected number of new, changed features is returned - for instance by combining existing features through some numerical computations or by selecting a set of features through some technique (\textbf{R 31})~\cite{AssuringMLLifecycle,MicrosoftMLWorkflow,IotMLLifecycle}. Moreover, the processed dataset or features can be enriched, e.g. by combining disjoint features on different aspects of the phenomenon under study into one larger feature or dataset in order to create a stronger, enriched signal for the ML algorithm to capture (\textbf{R 32}). 
In the case of supervised learning, an important operation during the data management phase is splitting the processed datasets into subsets used for 1) training, 2) potential validation and 3) testing (\textbf{R 29, 33})~\cite{googleML,ManagementModelingLifecycleinDL,AssuringMLLifecycle}.

Notably, not all data preparation activities can be conducted on the dataset prior to splitting it. In cases such as standardization, dimensionality reduction or imputation, data leakage must be avoided and thus observations considered for such techniques can only do so in their bounded context for either training, validation or testing. In literature, these operations still count as a pre-processing, feature engineering or enrichment activity and their correct application lies in the hands of ML engineers or data scientists~\cite{preprocessing}. Overall the discussed activities (i.e. general overall data preparation) can be understood as a sequence of immutable operations (or one overall operation) solely aimed at the provision of the right datasets as input for the ML task. \\
Further, pre-processing techniques such as data balancing can lead to a new, larger dataset carrying synthetic data either stored in a dataset repository or directly fed to the ML algorithm. Hence, these activities may also be associated with an embedded operation as part of the actual model learning~\cite{AssuringMLLifecycle,IotMLLifecycle}. Thus, next to defining such pre-processing steps as activities, for instance in the case of job offloading (see section~\ref{FAASCDReq}) the actual processing can be handed over as a recipe, i.e. a source code file or string containing the code, to the respective training or tuning operation (\textbf{34}). Besides proper representation of such code recipes sent to offloading jobs, storing them for future access can also be necessary, e.g. through a metadata repository (\textbf{R 35})~\cite{AssetML}.

Relevant events occurring throughout this phase can be deduced. The presence of latest available raw data or an update to raw data, feature sets or datasets in the respective repository can trigger subsequent operations (\textbf{R 8})~\cite{googleML}. In the case of infrequent data updates, scheduled events for data sourcing or preparation may be put into place to trigger a new run of activities (\textbf{R 9})~\cite{googleML}. Moreover, if a data scientist or engineer decides to update the data preparation sequence, they may require to manually (or automatically) trigger a new run - this action-based trigger is of the same concept as the previous one in \textbf{R 6}.  Further, enabling the modeling of events triggering not just the beginning of a phase (or the entire ML pipeline) but also the actual operation should be supported - in case of large volumes of data, only the very necessary set of steps should be instantiated to avoid needless computations (\textbf{R 22}). Due to being mostly event-driven, FaaS-based architectures support this requirement as well, see section~\ref{FAASCDReq}.

\subsection{Model Learning}
Model learning, as described in section~\ref{MLLMLV}, involves all directly involved activities for producing a ML model, i.e. set-up of the algorithm or learning strategy, actual training (in case of unsupervised learning solely the application of the algorithm on the dataset), scoring the trained model or evaluating a prospective ML algorithm and respective tuning of hyperparameters where applicable. Section~\ref{MLMethods} gave an overview of a number of existing ML methods with varying operations involved. Outlining in-depth requirements for all methods and techniques is out of scope for this work, thus identifying a minimal sufficient set of recurring operations is pursued.

Base operations remain the training of a single algorithm with subsequent scoring (\textbf{R 36, 37})~\cite{AssuringMLLifecycle,IotMLLifecycle,MicrosoftMLWorkflow,AWSmlconcepts,AzureMLOpsref}. This requires the output of the previous phase, i.e. training and potential validation datasets. Optionally, also other code artefacts to guide final pre-processing steps may be ingested. Further, the design of the algorithm, i.e. set of hyperparameters, may either be hard-coded as part of the training operation or handed to it beforehand as configuration (\textbf{R 43})~\cite{AssetML,IotMLLifecycle,googleML}. The main artefact produced by this operation is a ML model next to potential metadata and logging information such as where the model is stored or how long the training task took as presented by Schelter et al.~\cite{MLOpsMetadata}, Gharibi et al.~\cite{ManagementModelingLifecycleinDL} or industrial tools~\cite{GoogleMLOPsDoc,AWSmlarchitecting} (\textbf{R 39, 40, 41}). Logs can be stored in respective repositories~\cite{googleML} (\textbf{R 57}).
According to Idowu et al.~\cite{AssetML}, datasets and models are the most critical artefacts of ML workflows. Similar to Schelter et al. \cite{MLOpsMetadata} they further differentiate between other assets such as training configuration artefacts and associated metadata thereby justifying the respective representation on a modeling level. An important aspect in any machine learning workflow is its reproducibility, i.e. the capability to replicate previous pipeline instances (e.g. data preparation, training, tuning operations and resulting ML models). Next to metadata that informs about the operations and artefacts, configuration files such as training or tuning configuration as well as data processing recipes require a respective entity to be stored at. Two options present themselves - 1) store the configuration files alongside the metadata in a metadata repository or 2) store the configuration files in a new entity to clarify their importance and differentiate between conventional metadata. Related work on MLOps such as Salama et al.~\cite{googleML} understand configuration files as ML artefacts which can be stored in a combined metadata \& artefact repository whereas works such as Giray~\cite{SEperspectiveML} tend to favor the latter option, not least as a configuration storage allows to also store external infrastructure configuration files to specify resources or execution environments of the entire ML solution. As such configuration is external to the modeling of ML workflows, the former option (i.e. storing ML configuration files within a metadata repository) is chosen for this work.
When scoring the predictive performance of a model, the requirements and constraints defined in the requirement analysis may impose which scoring metrics to run. Typically, a scoring operation takes a validation dataset next to the computed ML model (and potentially a pre-defined scoring function) and outputs a performance score~\cite{googleML,SEperspectiveML}. In simple cases this manifests for instance through metrics such as accuracy or ROC AUC (area under the curve, AUC, of the receiver operator characteristic, ROC). In complex cases, a pre-defined scoring function such as a cost function based on business goals can be created and ingested. For instance, a type-1 error may be more expensive than a type-2 error (or vice versa) when identifying and triggering operational errors and respective maintenance actions for heavy machinery.

If the fit of a prospective ML algorithm is to be estimated for a given machine learning problem, an evaluation can be conducted to train the algorithm with the same parameters repeatedly on subsets of the training data by means of a resampling strategy and directly score each produced model. The scores are averaged producing a more robust estimation on how the algorithm will perform on new data (\textbf{R 38}). 

Similar to regular training, more sophisticated hyperparameter tuning operations can be structured in various ways. The precise formulation of these operations depends on their implementation - offloading the activity on a specific ML service a cloud provider offers such as AWS SageMaker can simplify its design with only a tuning configuration required (\textbf{R 42, 43})~\cite{HyperparamTuning,DMLSurvey1,IotMLLifecycle}. 
In complex tuning scenarios, model evaluation is an integrated and automated aspect of the operation, i.e. many models are trained and scored on different hyperparameter sets. In this case, the tuning operation incorporates required splits for validation sets - as part of pre-defined resampling strategies that are part of the overall tuning configuration such as nested resampling~\cite{NestedResampling}. Consequently, any tuning operation requires a training dataset and potentially a tuning configuration recipe~\cite{googleML}. Similar to regular training, the tuning configuration may incorporate an identifier to locate the training dataset instead. Produced artefacts can be the sets of hyperparameters alongside the aggregated performance scores of the models, i.e. the tuning results, (\textbf{R 24})~\cite{AssetML,AssuringMLLifecycle}. Further, also the trained models can be kept and stored for future access. In case a large number of models is generated, it is more efficient to only keep the list of hyperparameters and performance scores. 

Subsequently, a final regular training job can be run on the full dataset (i.e. training + validation) that takes the best performing hyperparameter configuration and returns the possibly 'best' model in respect of the context. Ultimately, the ML solution is stored in a model repository, triggering the verification phase given that it meets the requirement thresholds(\textbf{47})~\cite{googleML,SEperspectiveML}.  If evaluation requirements still cannot be met, a performance deficit report may be generated (\textbf{24})~\cite{AssuringMLLifecycle}.

Apart from model training, potential operations preceding or subsequent to it can occur, referenced as further pre-processing or post-processing. These range from miscellaneous but necessary activities such as identifying and extracting the best performing set of hyperparameters and models after a tuning strategy (\textbf{R 44})~\cite{googleML} to more complex ones depending on the ML method in place. As such, transfer learning requires loading an existing ML model as part of the training job to learn the model on the new dataset (\textbf{R 45})~\cite{Transferlearning2}. Next to this, in the case of ensemble learning several ML algorithms are run, each trained and scored or tuned directly. Ultimately their inferences are merged through a voting or consensus operation that may aggregate the predictions in some manner or even eliminate the propositions of some of the models~\cite{EnsembleLearningSurvey1}. Principally, ensemble learning can be applied as part of a single algorithm, e.g. Random Forest or AdaBoost, or across algorithms. In the case of modeling, it is of interest to represent the cross-algorithm ensemble functionality whereas the former one is already covered by a single learning operation (\textbf{R 46}).

Relevant events occurring throughout the model learning phase are manifold. Re-training based on an existing configuration can be necessary in response to several ML lifecycle scenarios (\textbf{R 10}). Likewise re-tuning the entire solution may be required in case the configuration of hyperparameter values is no longer deemed applicable (\textbf{R 11})~\cite{googleML}. Reasons include changes in raw data, feature sets or full datasets. At large, a degradation of a deployed model's performance~\cite{TowardsMLOps} caused by events related to concept and data drift requires proper representation on a modeling level as such events can trigger the model learning phase as well as possible data preparation activities (\textbf{R 12, 13}). Next to restarts in response to ML lifecycle situations, also scheduled re-training and tuning is common practice (see \textbf{R 9})~\cite{TowardsMLOps,googleML,uber}. Just as for changes in the implementation of data preparation activities, a manual trigger by action of a data scientist or engineer should allow restarting the model learning phase (or selected activities) from the beginning as well~(\textbf{R 6}). Having fully trained, tuned and evaluated a model, two types of events can occur - 1) a model learning deficit event holding the generated performance deficit report to request intervention of an expert or 2) a request for model verification (possibly triggered by writing the model artefacts to a corresponding directory) (\textbf{R 15, 16})~\cite{AssuringMLLifecycle,MicrosoftMLWorkflow,googleML}.

Next to triggers for initialisation of operations, in machine learning workflows, exceptions and execution errors play an ever-present role. Consequently, accounting for mitigating actions within a process model should be supported (\textbf{R 48})~\cite{DMLSurvey1,googleML}. While understanding all possible manifestations of errors is not focused by this work, a selected sample for illustration purposes can be considered - this includes: 1) a training or tuning job failure due to convergence problems; 2) tuning or evaluation error e.g. due to new unseen classes of categorical features that the trained model is unable to process; 3) computational failures (or exceptions) outside from mathematical problems, such as a training task running out of allocated memory or crossing a threshold of allocated time (see section~\ref{FAASCDReq}). Considering such training, tuning or evaluation errors, forwarding them to data scientists for inspection or automatically handling them may be necessary on a modeling level.


\subsection{Model Verification}
Once a (set of) model(s) passes the model learning phase, they are forwarded to the verification phase for rigorous testing before officially accepting the new solution and integrating it into the infrastructure for provisioning, see section~\ref{MLLMLV}. The verification phase fundamentally consists of a verification operation which uses trained models and (in case of supervised learning) the verification dataset(s) (\textbf{R 49})~\cite{AssuringMLLifecycle,MicrosoftMLWorkflow}. Moreover, similar to evaluation operations, the constraints and thresholds that are to be tested can be specified by requirement documents (\textbf{R 22})~\cite{MLOpsDataQuality}. Additionally, in operational systems, also performance statistics of currently deployed, i.e. active, models can be sourced to further strengthen the gathering of evidence for decision-making - to deploy the model or to request a change in the data preparation or model learning routine (\textbf{R 24})~\cite{googleML}. Depending on complexity of the ML use case, verification can be defined through more sophisticated logic to ensure consistency of the ML solution's performance on different aspects of test datasets that incorporate varying characteristics of the actual population statistics~\cite{googleML}. In security critical environments instead of an automated verification activity, an expert may have to manually confirm the model prior to deployment (\textbf{R 50})~\cite{MicrosoftMLWorkflow,AssuringMLLifecycle}.

Verifying a model (ML solution) can create artefacts such as information on the overall result of the experiment, i.e. combination of data preparation, model learning and tuning activities, and other auditable documentation, i.e. verification results, which can be written to a metadata storage (\textbf{R 24, 40})~\cite{MLOpsMetadata,AssuringMLLifecycle,uber}. Such information guides future activities and decisions by data scientists and engineers on how to improve the ML solution.

Once a model passes all verification obstacles, if necessary post-processing operations may take care of writing the model as a verified ML solution to a model registry for future access and deployment (\textbf{R 44}). Next to accessing the model and potentially changing its location (e.g. in terms of storage directories) such operations may propagate information such as the new path to find the verified model at. Once written to model registry as a verified model, subsequent deployment follows~\cite{AssuringMLLifecycle,MLOpsDataQuality}.

Relevant events occurring throughout the model verification phase can be - 1) a verification failure to initiate intervention by a human expert and to forward information on the current ML workflow~\cite{AssuringMLLifecycle} and 2) a deployment trigger to initiate subsequent activities for ML solution integration (\textbf{R 17, 18})~\cite{uber,googleML,AWSmltaming}.

\subsection{Deployment}
To provision a ML model and make it accessible to other components of the broader system infrastructure of an organization (e.g. virtually as a service or physically on an edge device), different storage and provisioning solutions are available. Similar to operations and artefacts described previously, the deployment operation warrants a representation as a modeling element \textbf{R 51}~\cite{AssuringMLLifecycle,uber}. Depending on the scenario, the actual model deployment or integration activity may be constrained by available computing resources or e.g. by specific demands for short-time inference responses. Further, ML model deployment may consist of more than a singular operation - in case of edge-ML, deploying the model can involve elaborate pre-processing and optimization techniques w.r.t. the ML model to specify and optimize the deployment plan~\cite{IotMLLifecycle}. Adhering to these constraints can however be understood as a matter of implementing the modeled deployment operation correctly prior to running it and does therefore not necessitate an explicitly new modeling statement to arrive at a minimal set of requirements for conceptual modeling of ML workflows. The deployment operation can either be informed by a model identifier (e.g. a path to locate the verified ML model) or by actually receiving the model as a file, for instance in case of edge-based scenarios. It then correctly integrates the model in the environment and returns identifiable information w.r.t. the model's access points~\cite{googleML}.\\
In security-critical environments, prior to the actual model deployment, the ML solution may instead be integrated into a staging environment resembling the production environment (e.g. in form of shadowing). The purpose is to fireproof the model's adherence to requirements as well as to fully validate its operational fit~\cite{TowardsMLOps,googleML}. A staging operation can be synonymous to a deployment operation, the only difference being the value of a variable specifying the deployment environment.

In case a change or update of an already deployed ML model is required, after deployment of the latest ML solution, the previous one(s) are to be deprecated (\textbf{R 52})~\cite{uber}. This involves removing the access to the now retired model(s), documenting the process and potentially saving the retired models in a different location. Thus, also new metadata is created such as a new identifier or path and ML model status (e.g. \textit{deprecated}).

Relevant events occurring throughout the deployment phase are mainly 1) informing connected services of the now active ML solution depending on the mode of deployment - for example initiating a monitoring service (\textbf{R 20}) - and 2) potentially triggering deprecation of an existing model (\textbf{R 19}). 

\subsection{Monitoring and Inference}
Once deployed, a ML solution is ready to operate its inference services. Thus, next to making the actual inference (\textbf{R 54}), in certain scenarios incoming data first requires pre-processing to format it into a shape which the ML model accepts. This may manifest in enriching the request (\textbf{29}) with data from the feature repository if only a subset of information is given~\cite{googleML}. Furthermore, modeling the actual inference result may be necessary (\textbf{24}). In case of complex ML solutions, several machine learning models can be available and selected based on best prospective performance with regards to the current inference request - Alipour et al.~\cite{UseCaseOnlineML} and Karn et al.~\cite{UseCaseDynamicAutoselection} propose examples to such model selection scenarios in online systems. Consequently, such a model selection operation warrants its own representation on a modeling level (\textbf{R 53}).

Next to the actual inference, certain scenarios require the ML solution to provide explanations on its inference results - also known as explainable artificial intelligence. Respective modeling requires representation of an explanation activity, e.g. computing SHAP values, that may access the information of the inference request alongside the applied ML solution (\textbf{R 55}). Explaining the model's inference creates an explanation result (\textbf{R 24}). The field of explainable artificial intelligence involves more than a single explanation task - documenting the entire machine learning process, making it transparent and auditable as well as applying other best practices is equally important~\cite{XAISurvey}. This however goes beyond the scope of modeling a machine learning workflow via BPMN - in fact, creating transparency of the workflow by applying standard-based modeling contributes to achieving explainable AI. 

While the ML solution is operational, continuous monitoring can take place on various levels (\textbf{R 56})~\cite{AssuringMLLifecycle}:
\begin{itemize}
\item monitoring data distributions of incoming data streams or batch updates
\item monitoring feature and target relationships, e.g. training data correlations 
\item monitoring the model's environment, i.e. system.
\end{itemize}
Monitoring may capture data and concept drifts or production skews - changes in distributions of raw data or features, their respective importance, statistical properties of the target or similar that may violate the assumptions made during model training ~\cite{googleML,uber}. Such monitoring activities can be fueled by documents (e.g. data and performance statistics) or general logging information that any operation throughout the ML workflow produces (\textbf{R 24, 41}).
Moreover, overall model performance variations on a system level (e.g. response time) can decrease to such an extent that intervention becomes necessary.
Modeling monitoring as an atomic operation does not fully capture the component's actual purpose as it goes hand in hand with overall system monitoring capabilities which may involve operations by components outside of the actual ML workflow. The activities can however be proxied, for instance by modeling a ML specific monitoring operation which is specifically directed towards the ML workflow. This way, certain monitoring situations that result in events relevant to the machine learning lifecycle can be highlighted included in the modeling diagram. 

Relevant events occurring throughout the monitoring and inference phase are related to concept or data drifts which in turn can trigger previous steps of the ML pipeline. Thus, a data drift may initialize data preparation and subsequent model training activities. Similarly, registering concept drift indicating upcoming model performance degradation can require a data scientist to step in or to automatically perform re-training or re-tuning (see \textbf{R 12, 13}). 
Further, in case the model's operational performance, e.g. response time, degrades due to shortcomings in the overall system infrastructure an event requesting an expert's intervention to the ML solution can be necessary (\textbf{R 14}).
An inference activity may be automatically initialized in some scenarios and thus justified a representative event (\textbf{21}).
A violation of data structure or format of the incoming inference request may result in an inference error (\textbf{R 48}).

\subsection{An Outlook on Federated Learning}
In section~\ref{FL} we introduced an upcoming machine learning method, federated learning, and used it to showcase how different ML methods can vary from the established machine learning lifecycle and workflow. Inclusion of such methods to our requirement analysis is not necessarily binding to arrive at a collection of generally valid ML concepts for the conceptual modeling of ML workflows. In this sub-section, we outline how the various characteristics of FL relate to and could theoretically build on top of the previously identified ML concepts for inclusion. Nonetheless, due to the scope limitations of this thesis, integration of these requirements will not be further pursued and is deferred to future research.

As indicated, the two main parties involved in the federated learning workflow are \textit{clients} and \textit{central server} which consequently requires separate modeling. Clients can be referred to as local entities whereas the central server represents the global entity~\cite{FLArch}. The key differentiator among activities, data artefacts or repositories is therefore whether it takes place locally on a client or globally on the server side. Most of the identified ML activities can be run on both sides. Exceptions are in certain cases data management operations if the federated learning scenario is constructed in a way that entirely prohibits data processing on a centralized node, for instance if data is highly sensitive such as medical records.

Next to the already identified ML activities, federated learning necessitates some more specific operations. An \textit{authentication activity} ensures that a client is registered and authenticated prior to participation in a shared model learning round in order to mitigate data or model corruption~\cite{FedLess}. Further, a model upload can represent the operation of returning a trained model (or the parameters of a trained model) back to the central server who then aggregates the received local model instances~\cite{FLRA,FedLearnSurv1}. In large federated learning systems with many clients, a client selection can help in identifying 'high-quality' devices that are chosen for the next training round. Similarly, a deployment selection operation can support propagation of the global model to only a subset of clients based on some criteria - for instance if a set of clients indicates a bad model performance and requires the latest (or an updated) version of the global ML model.

In that context, the previously identified monitoring activity can be adapted to a local model monitoring operation to recognize the need for local model replacement. Such situations can occur in case of data drift, concept drift or general model degradation.

Next, to realize a federated learning process additional or updated repositories and data artefacts are required. A ML model can be global or local similar to datasets, data repositories and model registries. Further, a global client registry is necessary to keep track of devices that are involved in the federated learning workflow~\cite{FLRA}. In turn, the job creator operation (i.e. job configuration or client selection activity) can access the client registry to source client identifiers for the next model learning round.

Much more sophisticated federated learning workflows are actively researched, for instance to better account for data privacy by including mechanisms to encrypt the local model parameters prior to sending them to the central server for aggregation~\cite{FedLearnSurv1}. Also combinations of federated learning with other ML methodologies such as transfer learning are possible or entirely new ways of incorporating model and data partitioning~\cite{Yu2021GillisSL}. Such scenarios can be considered sub-classes or rather instances of the identified general federated learning operations and artefacts. Hence, in case the explained generic federated learning elements do not capture the scenarios sufficiently, additional extensions can be created building on top of the existing one in future work.

\section{FaaS Function Composition and Deployment }
 \label{FAASCDReq}
With Function-as-a-Service as method of choice for the deployment of ML workflows, inherent serverless characteristics require consideration when extending the modeling language. Therefore this section outlines general BPMN-relevant FaaS aspects as well as aspects directly related to either function composition or function orchestration. 

As elaborated upon in section~\ref{FaaS}, serverless functions come with computational constraints, mainly - runtime and memory limits as well as statelessness. This consequently imposes that potential BPMN activities carried out via serverless functions are constricted to that end. Thus, a need arises to account for such activities also on a process modeling level given that available task types do not represent these characteristics sufficiently (\textbf{R 1}). 

With respect to the ephemeral nature of FaaS, any data artefact or result produced by a FaaS operation throughout the machine learning lifecycle needs to be externalized to a respective storage if it is meant to be persistent beyond that single operation. A BPMN extension therefore needs to be able to explicitly represent and differentiate such data artefacts (\textbf{R 3}). In the same vein, the concept of BPMN data objects existing for the lifetime of a single process instance is amplified in the case of FaaS-based systems, theoretically making them (if produced by a serverless function) inaccessible as soon as the activity, i.e. function, finishes. As a result, to enable access of data objects across activities respective, domain-specific data stores are required (\textbf{R 2}). Similarly, modeling the hand-over of identifiers to allow retrieval of such artefacts in subsequent activities when traversing the ML workflow needs to be accounted for - as mentioned in previous ML lifecycle sections (e.g. via metadata or config files) (\textbf{R 44}). In the case of FaaS, such identifiers are typically contained as part of the context information a produced event carries.

In addition, a serverless function's memory restriction can be in opposition to the usually high requirements that machine learning operations hold. As Carreira et al.~\cite{ServerlessCirrus} describe, even hosting current ML libraries such as TensorFlow in a serverless function can be challenging if at all possible. 
For any extensive tuning set-up, training of a complex ML model or data preparation activity, sophisticated distributed approaches may need to be taken to leverage FaaS functions in such ML settings. While an interesting and promising avenue of upcoming research, it is not the goal of the BPMN extension to fully model such distributed strategies, albeit options for future incorporation should be possible. Many of the proposed approaches focus on specific sub-domains of machine learning and are thus out of the scope of a general ML workflow modeling extension (within in the thesis boundary) - see for instance propositions for serverless edge-ML or specific serverless deep learning architectures~\cite{ServerlessEdge,ServerlessDL}. 

To still mitigate said challenges around economics of deployment, modeling offloading operations, i.e. job offloading, from a serverless function to a more applicable ML service or comparable execution platform is necessary (\textbf{R 1, 23})~\cite[p.~81]{ServerlessComputingReport}. In certain offloading scenarios this comes with a preceeding job creation task and a concomitant code file describing which operations to execute during the offloaded job (\textbf{R 5, 34}). Considering such functionality on a modeling level further helps in accounting for critiques. For instance, drawbacks in response-time of purely FaaS-based systems compared to ML optimized deployment technologies can be mitigated via offloaded jobs~\cite{ServerlessBackwards}.
Aside from the mentioned constraints, with regards to section~\ref{FaaS} serverless operations facilitate ML lifecycle activities such as 1) data partitioning, 2) model partitioning such as federated learning or 3) aggregations allowing to merge the results of a set of incoming functions and processing it further~\cite[p.81]{ServerlessComputingReport}. Likewise, parallelization can be achieved through automated initialization of multiple instances of a function. Consequently, to leverage such concepts a modeling extension should enable their representation, i.e. enable modeling FaaS-supported workflow patterns (\textbf{R 4})~\cite{FaaSBPMNOrchestration,FaaSPatterns}.
 

To transition between activities and phases in a FaaS-based ML workflow, explicitly modeling the respective events that inform and trigger subsequent components can be necessary. As elaborated in sections~\ref{FaaSChoreography} and~\ref{FaaSOrchestration}, in FaaS environments this flow of communication and information is achieved by 1) event providers, brokers and consumers in event-driven contexts or 2) orchestration engines in function orchestration contexts. A BPMN diagram of ML workflows on FaaS should consequently allow modeling such transitions (\textbf{R 4}). 

In the case of function orchestrations, this can be achieved to a certain extend by analysing the control flow of the BPMN model and directly mapping it as an orchestration of functions with for instance each task representing a single function as presented by Yussopov et al.~\cite{FaaSBPMNOrchestration} (\textbf{R 1, 4}). Notably, this only incorporates a single function orchestration whereas a typical ML workflow diagram may consist of several such orchestrations. As explained by  Beswick~\cite{awsFOBlog3} an orchestration is most-applicable to a workflow within a bounded context - in the case of ML workflows this can be a single phase such as model learning (depending on ML workflow complexity - in smaller scenarios, the entire workflow may be understood as one bounded context). Coordinating state changes across bounded contexts can instead be achieved through loose-coupling via choreography which in turn may be indicated by explicitly modeling the responsible events (\textbf{R 7 - 23}). 

\section{Requirement Synthesis}
\label{reqSynthesis}
Despite the multitude of BPMN extensions available in literature, while some can be considered for and related to the context of ML pipelines, ML particularities still require and thereby justify an own representation in form of an extension artefact. We refer back to section~\ref{relatedWork} for an overview of established and potentially relevant BPMN extensions.
With an established need for a modeling extension to properly account for machine learning characteristics, the identified requirements can be synthesised to guide the extension process.

Overall the workflow modeling language needs to adhere to certain foundational requirements so that a serverless machine learning workflow can be fully represented. Similar to the argumentation by Weder et al.~\cite{BPMNExQuant}, the workflow language that is extended must support:
\begin{itemize}
\item definition of a task (function) / an execution step in the process
\item definition of properties/attributes that shape and configure the activities
\item establishment of control flow to define order between activities
\item modeling of data artefacts and data transfer between activities and within the process overall
\item exception handling in case the conducted activity fails
\end{itemize}
The Business Process Model and Notation standard is compliant (or can be compliant) with respect to all listed elements. Moreover, similar to reasoning in related works~\cite{FaaSBPMNOrchestration,BPMNExclinical}, BPMN 1) is a well-established technology independent and interoperable process modeling standard, thereby lending itself well for machine learning workflow modeling, 2) supports a machine-readable format of the modeled workflow incorporating modeling constructs and workflow artefacts and 3) helps in explaining the workflows to both non-technical and technical stakeholders through a graphical notation.

To conclude the requirements analysis, a summary next to an equivalence check with BPMN on the identified machine learning specific requirements is provided in listing~\ref{tab:requirement_synthesis}.
\newpage
\begingroup
\footnotesize  
\setlength\tabcolsep{3pt}
\setcellgapes{2pt}
\makegapedcells
\setlength\LTcapwidth{\dimexpr\linewidth-3pt}
    \begin{xltabular}{\linewidth}{>{\bfseries}c p{2cm} L L c }
\caption{Aggregated analysis and BPMN equivalence check of generic FaaS and workflow concepts and ML workflow and lifecycle concepts for derivation of extension element requirements}
\label{tab:requirement_synthesis} \\
    \Xhline{1.2pt}

\thead{Req.}  & \thead{Concept}& \thead{Description}  & \thead{Equivalence Check}
            & \thead{BPMN - Ext} \\
\endfirsthead
\Xhline{1.2pt}
\thead{Req.}  & \thead{Concept}& \thead{Description}  & \thead{Equivalence Check}
            & \thead{BPMN - Ext} \\ \hline 
\endhead
            
    \Xhline{0.8pt} 
1	&	Serverless, e.g. FaaS, operations	&	Modeling of serverless (FaaS) or offloaded activities and differentiation to conventional shall be possible	&	 Limited equivalence - \textit{service task} represents basic concept of services such as as a web-service but is still unable to account for FaaS / serverless characteristics~\cite[p.156ff.]{OMGBPMN} 	&	Extension required	\\	\hline
2	&	FaaS persistent storage	&	Modeling persistent storage shall be possible	&	Equivalence - \textit{data stores} represent persistent storage across process instances~\cite[p.207ff.]{OMGBPMN}	&	BPMN concept	\\	\hline
3	&	FaaS data artefacts	&	Modeling any artefact that is written to a persistent storage or metadata that informs about the artefact shall be possible	&	Limited equivalence - \textit{data objects} can abstractly represent artefacts. However, they would lead to a high level of abstraction. The use of \textit{properties} is not possible as only \textit{processes}, \textit{activities} and \textit{events} may be associated with them~\cite[p.208]{OMGBPMN},\cite{BPMNExclinical}	&	Extension required	\\	\hline
4	&	FaaS \& event-driven workflow patterns	&	Modeling FaaS and event-driven workflow patterns shall be possible	&	Equivalence - BPMN allows for workflow pattern modeling~\cite[p.426ff.]{OMGBPMN}	&	BPMN concept	\\	\hline
5	&	Job preparation	&	Modeling the activity of configuring and preparing offloaded jobs prior to submitting them shall be possible	&	No equivalence - no appropriate marker	&	Extension required	\\	\hline
6	&	Manual trigger	&	Modeling a situation in which a domain expert is able to manually or automatically trigger an operation (e.g. after updating a training configuration) shall be possible e.g. through an event.	&	Conditional equivalence - a \textit{message event}, such as HTTP request to invoke a service directly, can be modeled.	&	BPMN Concept 	\\	\hline
7	&	Requirement change trigger	&	Modeling an event triggering subsequent operations such as model verification after a change in a requirement document shall be possible.	&	No equivalence - no respective event for this particular situation	&	Extension required	\\	\hline
8	&	Data update trigger	&	Modeling events to trigger operations based on the presence of new raw data, feature sets or datasets in a repository shall be possible	&	No equivalence - uBPMN~\cite{uBPMN} \textit{collection event} resembles functionality to a certain extend but its focus lies on singular observations derived through ubiquitous technology	&	Extension required	\\	\hline
9	&	Scheduled trigger	&	Modeling an event to trigger operations based on a pre-defined schedule shall be possible	&	Equivalence - \textit{timer event}	&	BPMN Concept	\\	\hline
10	&	Training trigger	&	Modeling an event to initialize the training operation or ML learning phase shall be possible	&	Conditional equivalence - a training operation can be triggered either according to previous events (e.g. new dataset available) or upon request (HTTPS) visualized by a BPMN \textit{message event} that carries the training configuration	&	BPMN Concept	\\	\hline
11	&	Tuning trigger	&	Modeling an event to initialize the tuning operation shall be possible	&	Conditional equivalence - a tuning operation can be triggered either according to previous events (e.g. concept drift detected) or upon request (HTTPS) visualized by a BPMN \textit{message event} that carries the tuning configuration	&	BPMN Concept	\\	\hline
12	&	Concept drift trigger	&	Modeling an event to showcase that a concept drift is happening shall be possible	&	No equivalence - no appropriate event, \textit{signal event} can be used but abstracts away the domain-specific context	&	Extension required	\\	\hline
13	&	Data drift trigger	&	Modeling an event to showcase that a data drift is happening shall be possible	&	No equivalence - no appropriate event, \textit{signal event} can be used but abstracts away the domain-specific context	&	Extension required	\\	\hline
14	&	Operational performance degradation	&	Modeling an event triggered by degradation of a deployed model's operational performance shall be possible	&	No equivalence - no appropriate event	&	Extension required	\\	\hline
15	&	Performance deficit event	&	Modeling an event triggered by the performance deficit report and requesting expert intervention shall be possible	&	No equivalence - no appropriate event	&	Extension required	\\	\hline
16	&	Verification	&	Modeling an event to trigger automatic or manual verification based on the presence of a newly trained / tuned model shall be possible	&	No equivalence - no appropriate event	&	Extension required	\\	\hline
17	&	Verification failure event	&	Modeling an event triggered by a verification (failure) report and requesting expert intervention shall be possible	&	No equivalence - no appropriate event	&	Extension required	\\	\hline
18	&	Deployment trigger	&	Modeling an event that takes a model identifier for a verified model and triggers deployment shall be possible	&	No equivalence - no appropriate event	&	Extension required	\\	\hline
19	&	Deprecation trigger	&	Modeling an event that takes a model identifier of a deployed model and triggers deprecation of the model shall be possible	&	No equivalence - no appropriate event	&	Extension required	\\	\hline
20	&	Broadcasting event	&	Modeling an event that informs any other services about a deployed model shall be possible	&	Equivalence - \textit{signal event}	&	BPMN Concept	\\	\hline
21	&	Inference trigger	&	Modeling an event triggered by an inference request shall be possible	&	No equivalence - no appropriate event	&	Extension required	\\	\hline
22	&	Data preparation trigger	&	Modeling an event to initialize data preparation shall be possible	&	Conditional equivalence - a data preparation operation can be triggered either according to previous events (e.g. raw data or feature set update) or upon request (HTTPS) through a \textit{message event}  that carries the tuning configuration	&	BPMN Concept	\\	\hline
23	&	Job offloading	&	Modeling an event triggered by offloaded jobs (e.g. job completed) shall be possible	&	No equivalence - no appropriate event	&	Extension required	\\	\hline
24	&	Document Object	&	Modeling documents created throughout the ML lifecycle (i.e. requirement document, tuning results, evaluation results, deficit report, verification result, inference result, model performance \& data statistic reports) shall be possible	&	Limited equivalence - \textit{data objects} can abstractly represent artefacts. However, they would lead to a high level of abstraction. The use of \textit{properties} is not possible as only \textit{processes}, \textit{activities} and \textit{events} may be associated with them~\cite[p.208]{OMGBPMN},\cite{BPMNExclinical}	&	Extension required	\\	\hline
25	&	Data sourcing operation	&	Modeling data sourcing from raw data providers into a raw data repository or a feature repository shall be possible	&	Limited equivalence - Yousfi et al.~\cite{uBPMN} introduce extensions for specific data sourcing tasks, however solely for ubiquitous computing and limited to one observation at a time. Akkaoui et al.~\cite{BPMNETL} introduce a potential process for BPMN based ETL operations, however a BPMN extension accounting for such tasks is lacking.	&	Extension required	\\	\hline
26	&	Data fusion	&	Modeling fusion of (processed) data from different sources shall be possible	&	No equivalence - no appropriate marker	&	Extension required	\\	\hline
27	&	Data validation	&	Modeling data validation activities for raw data shall be possible	&	No equivalence - no appropriate marker	&	Extension required	\\	\hline
28	&	Data repositories	&	Modeling data repositories, i.e. raw data repository, feature repository, dataset repository, shall be possible	&	No equivalence - no appropriate representation (data store cannot reflect characteristics)	&	Extension required	\\	\hline
29	&	Differentiation of data	&	Modeling data at various stages, i.e. raw data, feature sets, complete datasets (train, validate, verify), inference request datasets shall be possible	&	Limited equivalence - \textit{data objects} can abstractly represent artefacts. However, they would lead to a high level of abstraction whereas reflecting the domain-specific nature of raw data, feature sets, and (train, validation, verification) datasets is necessary. The use of \textit{properties} is not possible as only \textit{processes}, \textit{activities} and \textit{events} may be associated with them~\cite[p.208]{OMGBPMN},\cite{BPMNExclinical}	&	Extension required	\\	\hline
30	&	Pre-processing	&	Modeling data processing from raw data or feature sets into more processed feature sets or complete datasets shall be possible	&	No equivalence - no appropriate marker	&	Extension required	\\	\hline
31	&	Feature engineering	&	Modeling engineering operations for structural changes of a feature set (or full dataset) shall be possible	&	No equivalence - no appropriate marker	&	Extension required	\\	\hline
32	&	Data enrichment	&	Modeling feature or dataset enrichment to combine or add several feature or datasets into one overall dataset to improve its signal shall be possible	&	No equivalence - no appropriate marker	&	Extension required	\\	\hline
33	&	Data splitting	&	Modeling creation of training, validation and verification data subsets shall be possible	&	No equivalence - no appropriate marker	&	Extension required	\\	\hline
34	&	Code objects	&	Modeling code files representing e.g. data processing operations or entire scripts to execute when offloading a job shall be possible	&	Limited equivalence - \textit{data objects} can abstractly represent artefacts. However, they would lead to a high level of abstraction. The use of \textit{properties} is not possible. Similar reasoning to \textbf{R 25}.	&	Extension required	\\	\hline
35	&	Metadata repository	&	Modeling a repository to store metadata shall be possible	&	No equivalence - no appropriate representation (data store requires lower level of abstraction)	&	Extension required	\\	\hline
36	&	Model training	&	Modeling algorithm training based on training (validation) datasets and potentially a configuration file shall be possible	&	No equivalence - no appropriate marker	&	Extension required	\\	\hline
37	&	Model scoring	&	Modeling scoring of trained ML models based on some metric either as input file or coded as part of the task shall be possible	&	No equivalence - no appropriate marker	&	Extension required	\\	\hline
38	&	Model evaluation	&	Modeling evaluation operations to estimate the fit of ML algorithms for a ML problem based on some resampling strategy shall be possible	&	No equivalence - no appropriate marker	&	Extension required	\\	\hline
39	&	ML Model	&	Modeling a machine learning model object shall be possible	&	No equivalence	&	Extension required	\\	\hline
40	&	Metadata object	&	Modeling metadata on different ML artefacts (e.g. data schemas, model identifiers etc.) shall be possible	&	Limited equivalence - \textit{data objects} can abstractly represent metadata. However, they would lead to a high level of abstraction. The use of \textit{properties} is not possible. Similar reasoning to \textbf{R 25}.	&	Extension required	\\	\hline
41	&	Log object	&	Modeling logs created throughout the ML lifecycle shall be possible	&	Limited equivalence - \textit{data objects} can abstractly represent log objects. However, they would lead to a high level of abstraction. The use of \textit{properties} is not possible. Similar reasoning to \textbf{R 25}.	&	Extension required	\\	\hline
42	&	Hyperparameter tuning	&	Modeling a tuning operation shall be possible	&	No equivalence - no appropriate marker	&	Extension required	\\	\hline
43	&	Training, Evaluation or Tuning configuration	&	Modeling an evaluation strategy in an evaluation configuration; a set of hyperparameter values, search strategies, resampling strategies, scoring metrics, dataset IDs and related necessary configuration information in a tuning configuration or modeling hyperparameters, constraints, dataset IDs in a training configuration shall be possible	&	Limited equivalence - can be represented by a data object but requires lower level of abstraction to represent characteristic attributes. Similar reasoning to \textbf{R 25}.	&	Extension required	\\	\hline
44	&	Post-processing	&	Modeling post-processing operations (such as retrieval of best hyperparameter sets or model location or writing a model to a different location) shall be possible	&	No equivalence - no appropriate marker available; can be realised by a regular serverless function (i.e. no ML specific task)	&	Extension required	\\	\hline
45	&	Transfer learning	&	Modeling an operation to ingest and further train a transfer model shall be possible	&	No equivalence - no appropriate marker	&	Extension required	\\	\hline
46	&	Voting or Consensus operation	&	Modeling a cross-algorithm voting operation shall be possible	&	No equivalence - no appropriate marker	&	Extension required	\\	\hline
47	&	Model registry	&	Modeling a model registry shall be possible	&	No equivalence - no appropriate representation (data store requires lower level of abstraction)	&	Extension required	\\	\hline
48	&	ML lifecycle \& FaaS exceptions	&	Handling common exceptions and errors occurring throughout the ML lifecycle and FaaS-based deployment such as training, tuning, validation or inference errors or exceptions (e.g. violation of set constraints) shall be possible	&	Equivalence - \textit{cancel events} (e.g. cancellation of a tuning operation after crossing a time constraint), \textit{error events} (e.g. new class in validation set), \textit{escalation events} (informing data scientist), \textit{compensation events} (automatic mitigation of exception or error - e.g. by re-running the FaaS function in case of one-time unavailability)~\cite[p.254ff.]{OMGBPMN}	&	BPMN Concept	\\	\hline
49	&	Automated verification	&	Modeling the automated verification operation of trained ML models against verification criteria and a verification set shall be possible	&	No equivalence - no appropriate marker	&	Extension required	\\	\hline
50	&	Expert verification	&	Modeling the verification confirmation activity of trained ML models by a domain expert shall be possible 	&	Equivalence - a \textit{UserTask} allows to model the operation	&	BPMN Concept	\\	\hline
51	&	Model deployment	&	Modeling the deployment operation of a verified ML model shall be possible	&	No equivalence - no appropriate marker	&	Extension required	\\	\hline
52	&	Model deprecation	&	Modeling the deprecation of a deployed model shall be possible	&	No equivalence - no appropriate marker	&	Extension required	\\	\hline
53	&	Model selection	&	Modeling the selection of an ML model out of several deployed ones shall be possible	&	No equivalence - no appropriate marker	&	Extension required	\\	\hline
54	&	Model inference	&	Modeling the inference operation of a deployed model shall be possible	&	No equivalence - no appropriate marker	&	Extension required	\\	\hline
55	&	Model explanation activity	&	Modeling an activity to explain a model and its inference shall be possible	&	No equivalence - no appropriate marker	&	Extension required	\\	\hline
56	&	Model monitoring	&	Modeling an operation to monitor or check on changes in data, model and environment shall be possible	&	No equivalence - no appropriate marker	&	Extension required	\\	\hline
57	&	Log Storage	&	Modeling a repository that holds logging information produced by any activity throughout the ML workflow shall be possible	&	No equivalence - no appropriate representation (data store requires lower level of abstraction to allow differentiation to other types of repositories)	&	Extension required	\\	\Xhline{1.2pt}
    \end{xltabular}
\endgroup

%% file: chapters/5.results.tex
This chapter consolidates the preceding requirement analysis and introduces the BPMN extension for modeling FaaS-based (serverless) machine learning workflows (from now on abbreviated as BPMN4sML). The addition of sML specific BPMN elements targets improved modeling, analysis, visualization and communication of ML workflows in a standardized manner, further increasing transparency of activities involved in a ML solution. Moreover, the extension elements shall help in mapping and translating ML workflows to equivalent artefacts of deployment models such that a deployment model can eventually be derived from its preceding BPMN process model. In line with the requirement analysis, the extension elements build on top of concepts from the ML lifecycle, incorporating phases within and outside of ML pipelines, and on top of underlying concepts of the Function-as-a-Service paradigm. Distinctive ML lifecycle and workflow elements can be identified with respect to activities (i.e. BPMN \textit{tasks}) alongside their inputs and outputs or produced artefacts (i.e. \textit{data objects}) as well as artefact repositories (i.e. \textit{data stores}). Further, certain key occurrences (i.e. \textit{events}) can be generalized across ML workflow instances. In line with Yousfi et al.~\cite{uBPMN} the added elements are directed towards a conservative extension, that is to say an extension which does not alter nor contradict the BPMN semantics of the OMG standard. 

In the following the extended BPMN metamodel is presented in section~\ref{metamodelExt} with a subsequent presentation of a corresponding notation, i.e. graphical elements that allow modeling the added metamodel components, in section~\ref{NotationExt}. The notation goes along with exemplary modeling snippets to showcase usage and further validate applicability of the introduced extension for specific ML workflow scenarios.

\section{BPMN Metamodel Extension}
\label{metamodelExt}
The metamodel extension is based on existing elements of the established BPMN metamodel. Notably, only aggregated segments of the BPMN specification, i.e. the most important metamodel elements of the standard, are displayed due to space limitations. They represent the bindings for the extension elements. The introduced BPMN4sML metamodel extension is split into three views:
\begin{itemize}
\item Figure~\ref{fig:BPMN4sMLTasks} introduces new \textit{tasks} as extension to \textit{Activity} belonging to the \textit{Flow Object} category  
\item Figure~\ref{fig:BPMN4sMLEvents} extends the BPMN \textit{Event} from the category \textit{Flow Object}
\item Figure~\ref{fig:BPMN4sMLDataArtefacts} introduces the extension of \textit{DataObject} and \textit{DataStore} from the category \textit{Data}.
\end{itemize}
All extended elements are fundamental to represent a ML workflow with varying flavors - for example a fully FaaS-based pipeline, a hybrid pipeline that leverages job offloading or a service that only focuses on how a deployed model is integrated with another business process. Grey coloured elements highlight the extension whereas white coloured elements are part of the BPMN standard metamodel. 

\subsection{Metamodel Task Extension}

\begin{figure}
    \centering
    \includegraphics[width=1\linewidth]{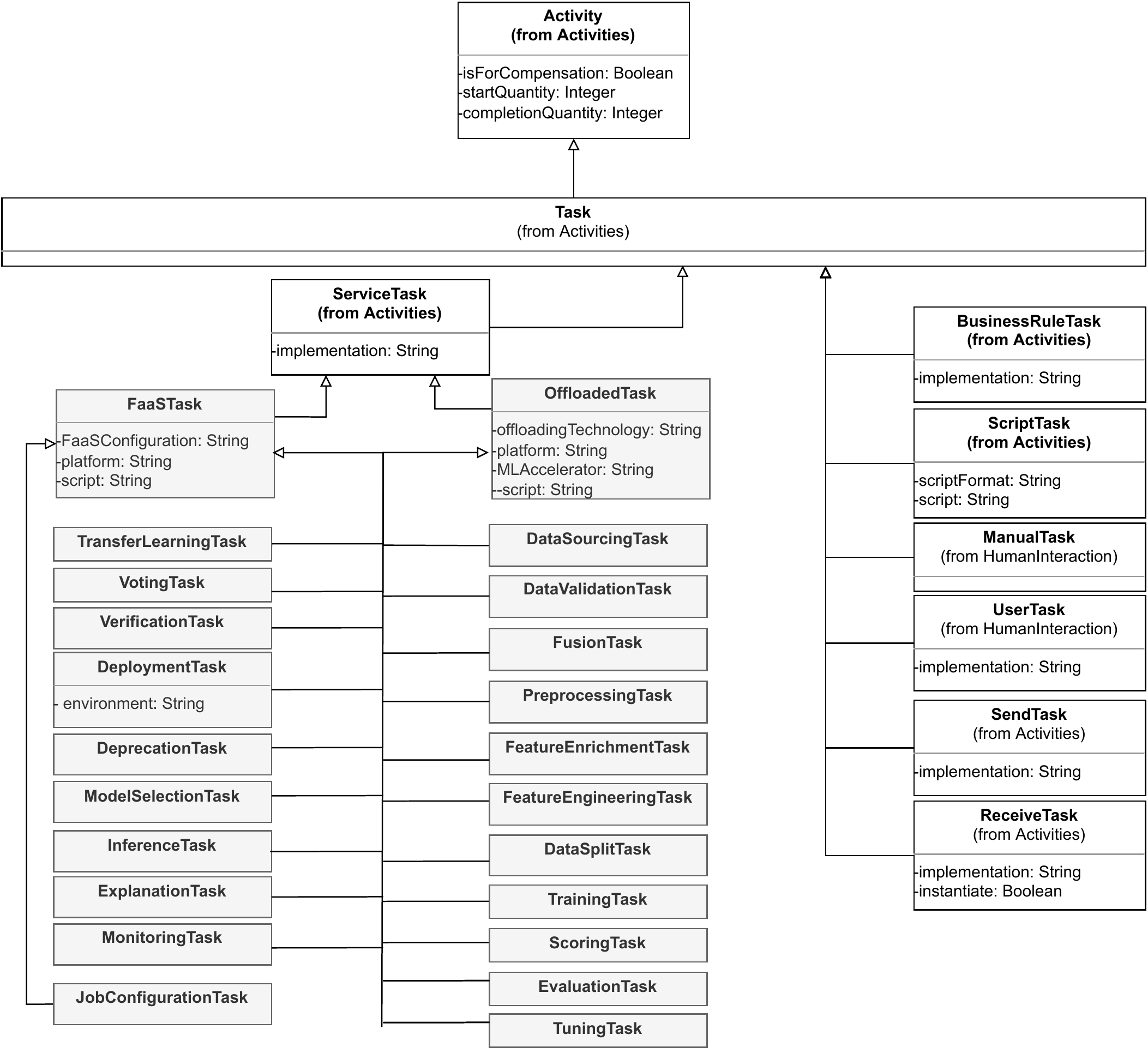}
    \caption[BPMN4sML metamodel task extension]%
    {BPMN4sML metamodel task extension \par \small Extended task elements covering common steps occurring throughout the machine learning lifecycle, particularly throughout ML pipelines. The base BPMN \textit{ServiceTask} is extended to better represent serverless environments covering both offloaded jobs and serverless functions as new tasks types which the remaining ML tasks inherit from.}
    \label{fig:BPMN4sMLTasks}
\end{figure}

With regards to Figure~\ref{fig:BPMN4sMLTasks} two overarching \textit{Task} structures are added to account for the previously identified characteristics of tasks executed by severless functions or realized as an offloaded job. The corresponding meta-classes \textit{FaaSTask} and \textit{OffloadedTask} inherit the attributes and model associations of the BPMN \textit{Activity}. Further, they are sub-classes of the BPMN \textit{ServiceTask} and therefore inherit the attributes and constraints that come with it. A BPMN \textit{ServiceTask} is mostly referenced in the case of web-services or automated applications and aligns itself well with the concept of a serverless function. It defines an \textit{implementation} of type \textit{String} to specify the implementation technology used~\cite[p.156ff.]{OMGBPMN}.

A \textit{FaaSTask} further defines two other required \textit{attributes} that constrain the \textit{implementation} attribute of the \textit{ServiceTask} as well as an optional one: 1) \textit{platform} of type \textit{String}, 2) \textit{FaaSConfiguration} of type \textit{String} and 3) \textit{script} of type \textit{String}. The \textit{platform} property is specified to account for differences in function configuration between FaaS-providers. As such if a \textit{FaaSTask} is mapped to an element of a deployment model, different verification checks for its configuration can be run depending on the specified platform on which the serverless function is deployed. The \textit{FaaSConfiguration} refers to implementation details that define the serverless function (for instance an identifier such as an AWS ARN number or allocated memory). Finally, an optional \textit{script} may be included (similar to the \textit{ScriptTask}) which can be referenced if the serverless function needs to be freshly deployed and implemented. The \textit{script} can be used to define the business logic of the \textit{FaaSTask}. A \textit{script} may not be necessary if the serverless function already is deployed and its logic is implemented. 

An \textit{OffloadedTask} comes with three \textit{attributes}: 1) \textit{offloadingTechnology} of type \textit{String}, 2) \textit{MLPlatform} of type \textit{String} and similar to the \textit{FaaSTask} a \textit{script} of type \textit{String}. The \textit{offloadingTechnology} takes from the \textit{implementation} of the \textit{ServiceTask} and is used to differentiate on an abstract level between technology types, e.g. \textit{cloud} or \textit{edge}. In case of \textit{cloud} the remaining attributes \textit{MLPlatform} and \textit{script} may be defined. They allow to specify the 1) machine learning platform such as Azure Machine Learning or Amazon SageMaker which can be directly included in a machine learning workflow instead of a serverless function and 2) a potential script which can hold configuration instructions of the offloaded job. 

Next to the two overarching Tasks, different task types typical to the machine learning domain are proposed as an extension to the standard. The identified tasks relate to the requirement synthesis and its preceding analysis in section~\ref{reqSynthesis}.
The new task types can inherit from both the \textit{FaasTask} or the \textit{OffloadedTask}, this is to show that the actual activity can be realized as a serverless function or as an offloaded job. Thus, depending on the modeled scenario different \textit{attributes} may need to be specified. An exception is the \textit{JobConfigurationTask} which only inherits from the \textit{FaaSTask} and can precede an offloaded task to generate any artefacts that are required to fulfill the job. Note that the \textit{DeploymentTask} defines \textit{environment} as an optional \textit{attribute} of type \textit{String}. The attribute allows to specify the target environment of the deployment activity and can take the form of a staging or a production environment in which the verified machine learning model should be placed.  

\subsection{Metamodel Event Extension}
\begin{figure}
    \centering
    \includegraphics[width=1\linewidth]{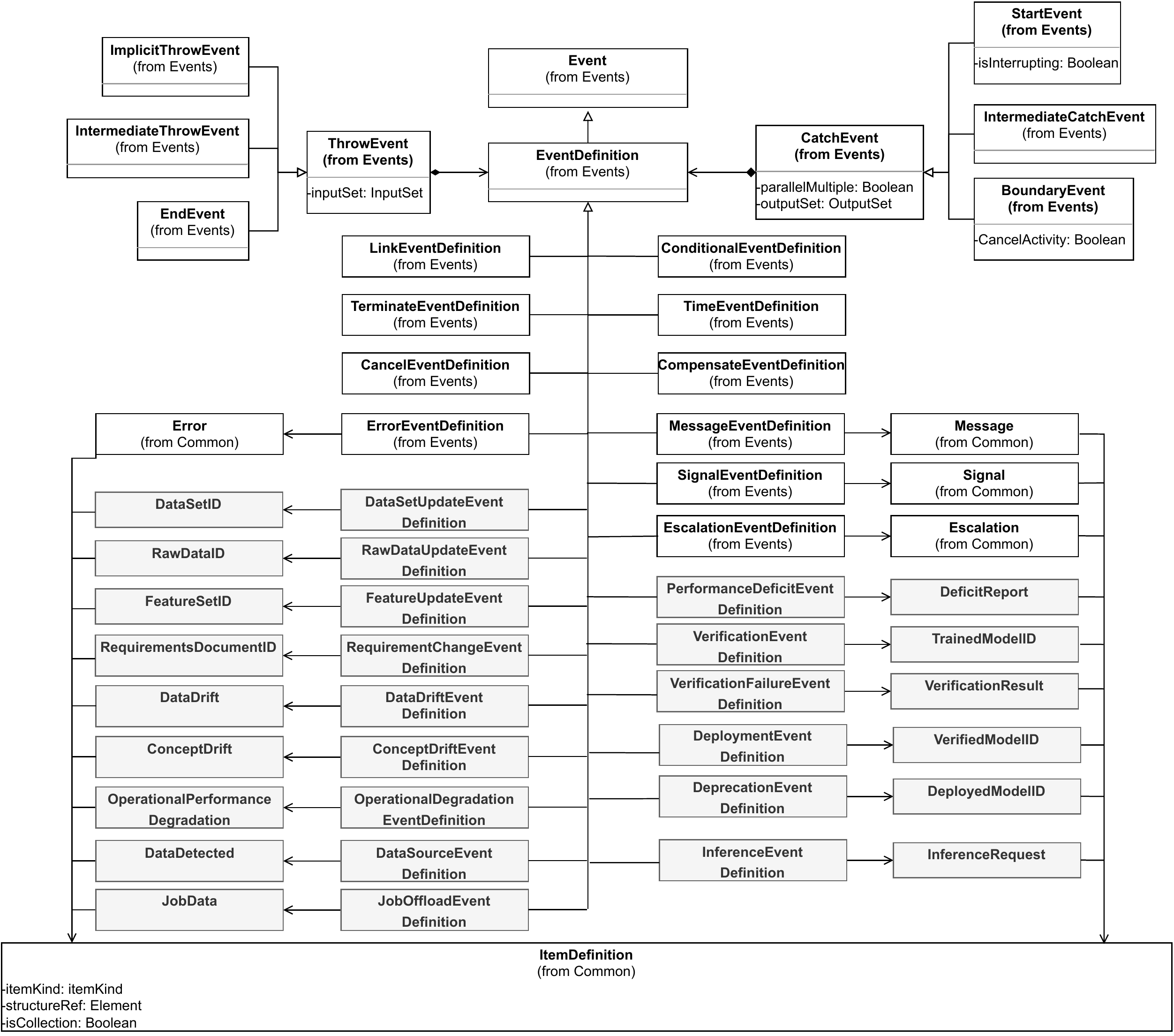}
    \caption[BPMN4sML metamodel event extension]%
    {BPMN4sML metamodel event extension \par \small Extended BPMN events relating to situations occurring throughout the machine learning lifecycle, particularly throughout ML pipelines, as well as to serverless deployments. The new events inherit from the base BPMN \textit{EventDefinition}. They carry data and thus depend on the BPMN  \textit{ItemDefinition}.}
    \label{fig:BPMN4sMLEvents}
\end{figure}

Next to new \textit{Task} types, Figure~\ref{fig:BPMN4sMLEvents} proposes several new events to better incorporate machine learning or serverless specific situations that can trigger (or be triggered by) activities throughout the ML workflow. A total of fifteen new \textit{Event} classes is defined which inherit model associations and respective attributes of the BPMN \textit{BaseElement} given through the \textit{EventDefinition}. The new \textit{Event} structures are an answer to the needs identified in the requirement synthesis and come as new \textit{EventDefinition} sub-classes. Note that the two verification-associated \textit{EventDefinitions} specify different situations occurring throughout the model verification phase which creates the need of separate \textit{EventDefinitions}. Moreoever, all the \textit{Event} extensions carry data, similar to the BPMN \textit{Message}, \textit{Escalation}, \textit{Error}, \textit{Signal} and \textit{Multiple	Event}~\cite[p.234]{OMGBPMN}. This is represented by the dependency of  the elements linking each new \textit{EventDefinition} to the BPMN \textit{ItemDefinition} - see for instance the \textit{RawDataID} or the \textit{DataSetID} meta-class. Note that these new meta-classes define a \textit{name} attribute of type \textit{String} similar to the BPMN events that carry data such as \textit{Error}~\cite[p.80ff.]{OMGBPMN}. In order to save space these attributes are not displayed on Figure~\ref{fig:BPMN4sMLEvents}.

Events that have been proposed by related extensions such as uBPMN can be incorporated in the context of the current metamodel. For instance, the uBPMN \textit{SensorEventDefinition} or \textit{CollectorEventDefinition} can function as limited alternatives to the proposed \textit{DataSourceEventDefinition} of this work. They may be used to represent specific scenarios in which an observation is sourced from a ubiquitous smart technology. The \textit{DataSourceEventDefinition} aggregates these specific scenarios to allow for modeling of an event triggered by the presence of new data of any data source.

\subsection{Metamodel Data Artefact Extension}
\begin{figure}
    \centering
    \includegraphics[width=1\linewidth]{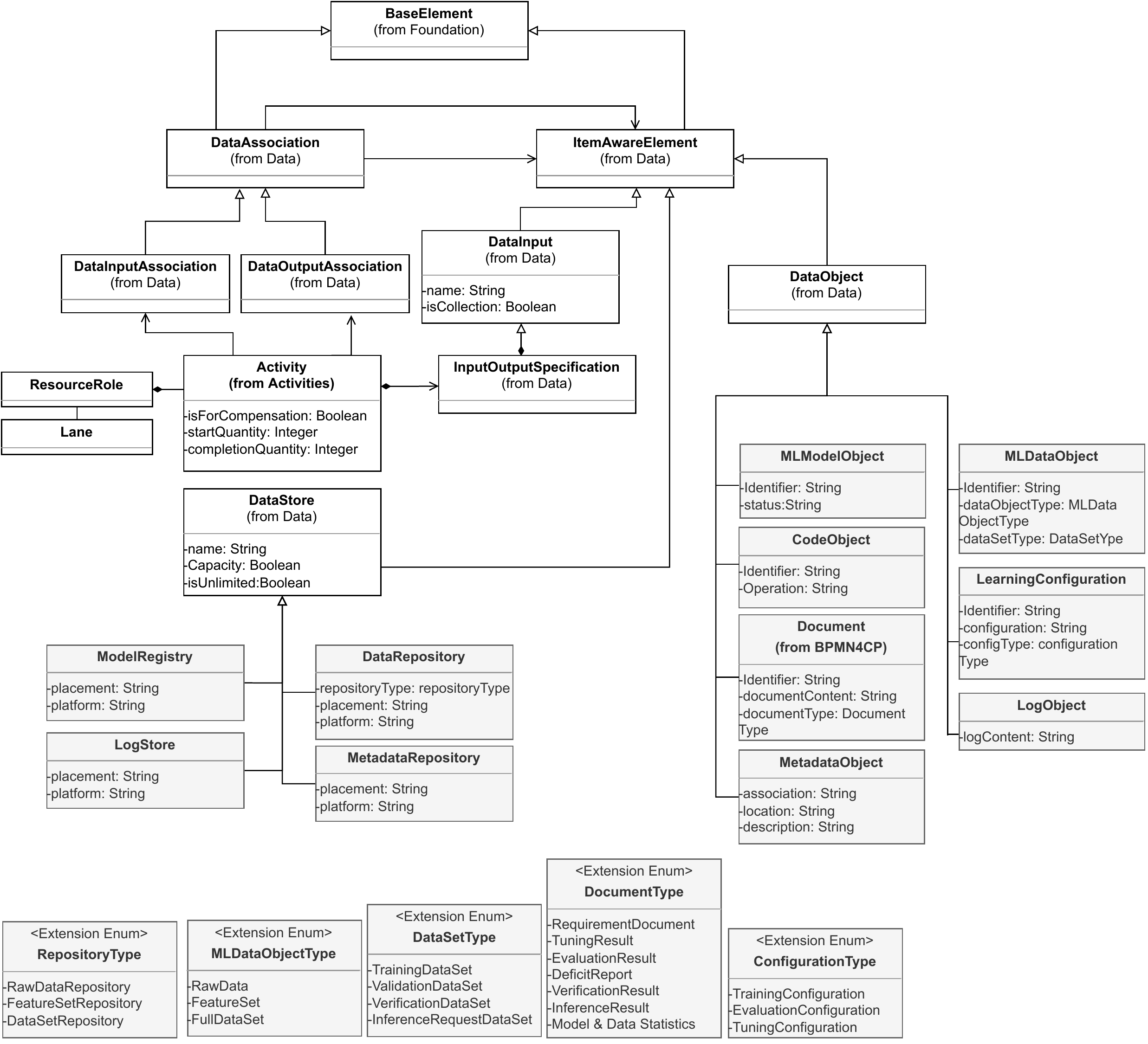}
    \caption[BPMN4sML metamodel data object \& data store extension]%
    {BPMN4sML metamodel data object \& data store extension \par \small Extended BPMN data artefacts relating to artefacts produced throughout the machine learning lifecycle, particularly throughout ML pipelines, as well as to artefact repositories relevant to machine learning in operation. The new artefacts inherit from the base BPMN \textit{DataObject} or from \textit{DataStore}. They define properties thereby lowering the level of abstraction to accommodate ML and FaaS characteristics.}
    \label{fig:BPMN4sMLDataArtefacts}
\end{figure}
The BPMN4sML extension introduces several new \textit{Data} structures based on typical ML artefacts as well as artefacts common to MLOps practices which are moreover necessary to realize a machine learning workflow on a FaaS-based infrastructure, see Figure~\ref{fig:BPMN4sMLDataArtefacts}. The new data artefact meta-classes either inherit from the model associations and attributes of \textit{DataObject} or from the ones of \textit{DataStore}. By defining new data artefacts as sub-classes of \textit{DataObject} they can be referenced both as \textit{DataInputObject} or \textit{DataOuputObject} of a modeled task, allowing to propagate the generated artefacts within a process instance. The artefacts inheriting from \textit{DataStore} realize the required persistence of the former artefacts beyond a process (or function) instance. 

\subsubsection{DataObject Extension}
In total, seven new elements inheriting from \textit{DataObject} are proposed. Extensions to the BPMN \textit{DataObject} are necessary as a \textit{DataObject} would otherwise lead to a high level of abstraction. Further, the use of a \textit{Property} as an alternative \textit{DataAwareElement} is not possible as only \textit{processes}, \textit{activities} and \textit{events}, i.e. \textit{FlowElements}, may be associated with it~\cite[p.208]{OMGBPMN},\cite{BPMNExclinical}. Consequently,  lowering the level of abstraction can be achieved by creation of sub-classes of \textit{DataObject} and defining the necessary \textit{attributes}.

A \textit{MLModelObject} represents the machine learning models created throughout the ML workflow. It further specifies two attributes, an \textit{identifier} of type \textit{String} for explicit identification of the model artefact and an optional \textit{status} of type \textit{String} to allow differentiating between a trained, verified, deployed or deprecated ML model. 

Apart from the ML model, other core ML artefacts are accounted for by means of \textit{MLDataObject}. The extension element provides an \textit{identifier} attribute of type \textit{String} to explicitly describe it. Further, a \textit{MLDataObject} defines two other attributes, a \textit{dataObjectType} of type \textit{RawData}, \textit{FeatureSet} or \textit{FullDataSet} and an optional \textit{dataSetType} of type   \textit{TrainingDataSet}, \textit{ValidationDataSet}, \textit{VerificationDataSet} or \textit{InferenceRequestDataSet}. The \textit{dataSetType} may be specified in case the \textit{dataObjectType} is set as \textit{FullDataSet}. A choice has been made to account for the different types by means of an attribute in order to avoid cluttering the metamodel.

Further, a \textit{CodeObject} is defined with two attributes, 1) an \textit{identifier} of type \textit{String} and 2) an \textit{operation} of type \textit{String}. The \textit{operation} attribute allows to specify the logic or set of programming commands that the artefacts holds. \\
Next, BPMN4sML proposes a \textit{LearningConfiguration} with three attributes, 1) an \textit{identifier} of type \textit{String}, 2) a \textit{configuration} of type \textit{String} and 3) a \textit{configType} of type \textit{configurationType} which can take the values \textit{TrainingConfiguration}, \textit{EvaluationConfiguration} or \textit{TuningConfiguration}. The \textit{configuration} describes the actual training, evaluation or tuning configuration. 

To further address the identified requirements a \textit{LogObject} is introduced with a \textit{LogContent} attribute of type \textit{String} which holds the information and content of that log file. In a similar vein, a \textit{MetadataObject} is proposed alongside three attributes. An \textit{association} of type \textit{String} relates the metadata artefact to the corresponding artefact that the metadata pertain to. A \textit{location} of type \textit{String} holds the location of the related artefact. Optionally, further information can be described via the \textit{description} attribute of type \textit{String} to provide enough flexibility to account for the various information a metadata object may hold, as elaborated by for instance Schelter et al.~\cite{MLOpsMetadata}.  

Finally, a \textit{Document} artefact is proposed, inspired by the extension by Braun et al.~\cite{BPMNExclinical}. Similar to the preceding extension elements, the \textit{Document} is an extension of \textit{DataObject}. It holds three attributes. An \textit{identifier} of type \textit{String} as well as a \textit{documentContent} of type \textit{String} which holds the information of the document. Further, a \textit{documentType} allows referencing a specific type. The different types relate to the various documents that are relevant to the machine learning lifecycle or are created by ML tasks - they can be realised as \textit{RequirementDocument}, \textit{TuningResult}, \textit{EvaluationResult}, \textit{DeficitReport}, \textit{VerificationResult}, \textit{InferenceResult} or \textit{Model \& Data Statistics}.  \\

\subsubsection{DataStore Extension}
Next to the \textit{DataObject} extensions, \textit{DataStore} extensions provide possibilities for modeling artefact-specific storage. A \textit{ModelRegistry} defines a repository to store \textit{MLModelObjects} at. It specifies two attributes of type \textit{String}, a \textit{placement} and a \textit{platform}. A \textit{placement} allows differentiation between local, i.e. internally hosted, or external cloud-based model registries. The \textit{platform} further allows to describe the cloud provider in case of an external cloud-based storage solution. Two other types of persistent storage are defined with the same attribute semantic. A \textit{LogStore} allows modeling a repository to (or from) which \textit{LogObjects} can be written (read). A \textit{MetadataRepository} allows to formally model a storage that can hold artefacts pertaining to the ML workflow such as \textit{MetadataObjects}, \textit{CodeObjects}, \textit{LearningConfigurations}, or other \textit{Documents}. Furthermore, a \textit{DataRepository} extends the abstract \textit{DataStore}. Next to a \textit{placement} and \textit{platform} attribute, it allows specifying a \textit{repositoryType} which can take the values \textit{RawDataRepository}, \textit{FeatureSetRepository} or \textit{DataSetRepository}. While the metamodel extension does not specify an individual element for each data repository, the notation extension supports differentiation by means of different visual elements.




\section{BPMN Notation Extension}
\label{NotationExt}

As elaborated in section~\ref{BPMN} BPMN defines a set of graphical modeling elements, i.e. its notation, to visually describe a process in a standardized manner for a heterogeneous audience. So far, the requirements to represent machine learning workflows have been derived and formalized in respective metamodels. To now leverage the derived definitions and make them accessible through provision of appropriate modeling concepts for process modelers, analysts or data scientists and ML engineers, it is necessary to introduce respective notation elements that follow the structure of the metamodels. A BPMN notation extension allows to integrate ML specific behaviour in business processes as well as to fully design a machine learning workflow through BPMN4sML.

Similar to the preceding metamodels, the notation extension covers \textit{Activity} as well as \textit{Event} of the BPMN \textit{FlowObject} category and \textit{DataObject} and \textit{DataStore} of the BPMN \textit{Data} category. In line with the described BPMN extension mechanism in section~\ref{BPMNExtensionMechanism}, the notation extension abides by the standard BPMN notation style and BPMN extension practices. Further, upon elaborating on each added notation element the phrasing pattern introduced by the OMG standard and referenced by several related works is applied~\cite{OMGBPMN,uBPMN,BPMNExclinical}.

\subsection{Data Artefacts}
The OMG standard defines four types to graphically depict data elements in a process model~\cite[p.202ff]{OMGBPMN},~\cite{uBPMN}: \textit{DataObject}, \textit{DataInput} and \textit{DataOutput} as well as \textit{DataStore}. The BPMN4sML notation extension offers several new elements in line with the described metamodel extension, namely \textit{MLModelObject}, \textit{MLDataObject}, \textit{CodeObject}, \textit{DataSetObject}, \textit{Document}, \textit{LearningConfiguration}, \textit{MetadataObject}, \textit{LogObject}, \textit{DataRepository}, \textit{MetadataRepository}, \textit{ModelRegistry} and \textit{LogStore}. 

\subsubsection{MLModelObject and ModelRegistry}
Subsequently, the notation and usage of the \textit{MLModelObject} and the \textit{MLRegistry} are described. The notations of the two metamodel elements are shown in Figure~\ref{fig:Model_and_Meta}. 
\begin{itemize}
\item A \textbf{MLModelObject} is an extended \textit{DataObject} that represents the actual machine learning model that is created during the ML learning phase and referenced in the subsequent ML lifecycle phases. It holds one machine learning algorithm and is the core data artefact of machine learning workflows. It has the same notation as the BPMN 2.0.2 \textit{DataObject}. As addition, a machine learning model icon is placed in the upper left corner of the graphical element, indicating that it is a \textit{MLModelObject}. 
\item A \textbf{ModelRegistry} is an extended \textit{DataStore} that represents a storage for \textit{MLModelObjects}. Its specific purpose is the provision of a universal access point to read and write ML models to. In the context of BPMN4sML it is assumed that a \textit{ModelRegistry} is unique to the modeled ML service such that no conflict can arise with MLModels from other unrelated processes - in practice this can be realised by specifying a directory unique to the ML service. The \textit{ModelRegistry} has the same notation as the BPMN 2.0.2 \textit{DataStore}. As addition, two machine learning model icons are placed in the lower left corner of the graphical element, indicating that it is a \textit{ModelRegistry}. 
\end{itemize}

\subsubsection{MetadataRepository and related Artefacts}
\begin{figure}
    \centering
    \includegraphics[width=0.6\linewidth]{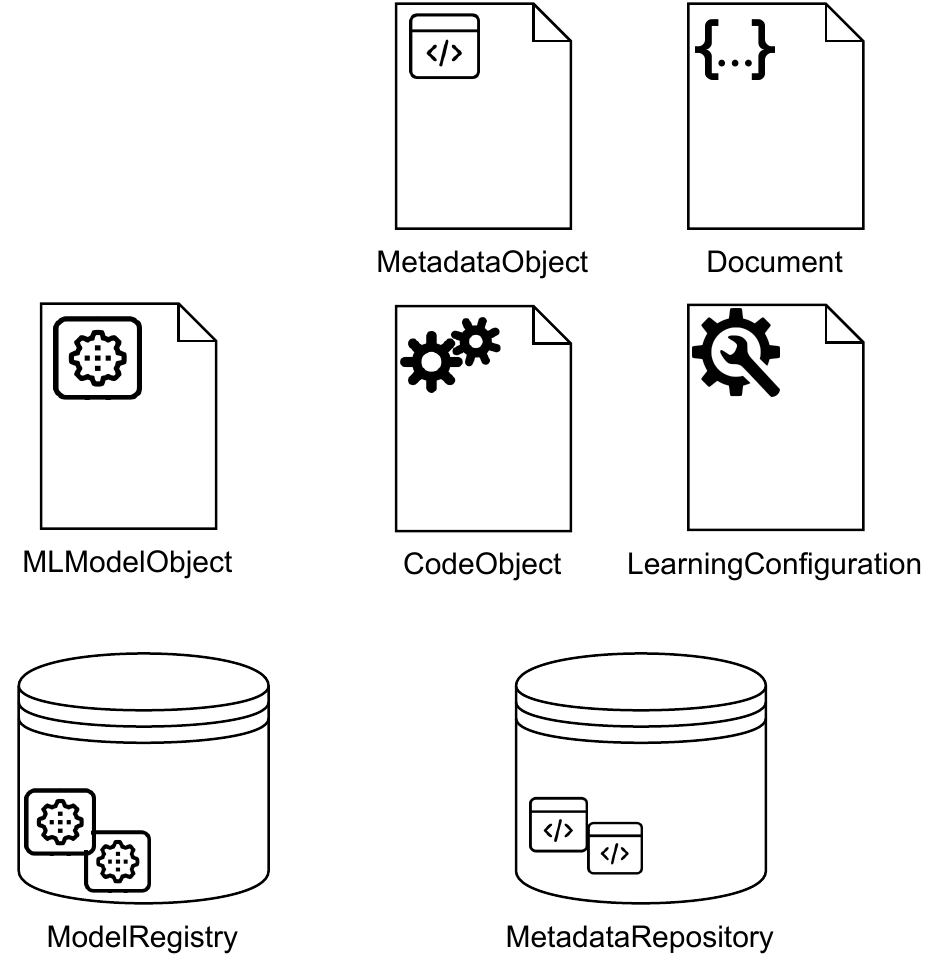}
    \caption[BPMN4sML notation of ModelRegistry, MLModelObject, MetadataRepository \& related Artefacts]%
    {BPMN4sML notation of ModelRegistry, MLModelObject, MetadataRepository \& related Artefacts \par \small }
    \label{fig:Model_and_Meta}
\end{figure}

Subsequently, the notation and usage of the \textit{MetadataRepository}, \textit{MetadataObject}, \textit{Document}, \textit{CodeObject}   and \textit{LearningConfigurationObject} are described. The notations of the two metamodel elements are shown in Figure~\ref{fig:Model_and_Meta}. 
\begin{itemize}
\item A \textbf{MetadataObject} is an extended \textit{DataObject} that represents metadata of the various artefacts such as data schemata of raw data. It can further be produced by the BPMN4sML Tasks throughout the ML lifecycle to e.g. reference structural static information of a learning job. It has the same notation as the BPMN 2.0.2 \textit{DataObject}. As addition, a generic metadata icon is placed in the upper left corner of the graphical element, indicating that it is a \textit{MetadataObject}. 
\item A \textbf{Document} is an extended \textit{DataObject} that represents various types of documentation and result files. It is derived from BPMN4CP~\cite{BPMNExclinical} where it is referenced to specify domain-specific documents for medical purposes. As part of BPMN4sML it is customized in order to model ML workflow specific documents (i.e. requirements, tuning, evaluation, verification and inference results, deficit reports and ML model and data statistics). It is assumed that the document content is written as JSON. Consequently, the element has the same notation as the BPMN 2.0.2 \textit{DataObject}. As addition, in context of BPMN4sML a respective JSON icon is placed in the upper left corner of the graphical element, indicating that it is a \textit{Document}. 
\item A \textbf{CodeObject} is an extended \textit{DataObject} that represents code logic, e.g. to specify programming logic for a data processing operation. Its specific purpose is to enable the provision of such coded logic to tasks as a data artefact - for instance in case of job offloading or data processing embedded in a ML tuning task. The \textit{CodeObject} has the same notation as the BPMN 2.0.2 \textit{DataObject}. As addition, a processing icon is placed in the upper left corner of the graphical element, indicating that it is a \textit{CodeObject}. 
\item A \textbf{LearningConfiguration} is an extended \textit{DataObject} that represents a configuration file to specify a training, evaluation or tuning task via parameters - for instance it can hold a set of different hyperparameter values, search strategies, resampling strategies, evaluation strategies, dataset IDs and related necessary configuration information for a tuning operation; a resampling strategy, algorithm parameter and scoring metric for an evaluation task or a specified set of hyperparameters, constraints and a dataset ID for a training configuration. The \textit{LearningConfiguration} has the same notation as the BPMN 2.0.2 \textit{DataObject}. As addition, a configuration icon is placed in the upper left corner of the graphical element, indicating that it is a \textit{LearningConfiguration}. 
\item A \textbf{MetadataRepository} is an extended {DataStore} that represents a storage for {MetadataObjects}, \textit{Documents}, \textit{CodeObjects} and \textit{LearningConfigurations}.  Its specific purpose is the provision of a universal access point to read and write the specified data artefacts to. It has the same notation as the BPMN 2.0.2 \textit{DataStore}. As addition, a generic metadata icon is placed in the lower left corner of the graphical element, indicating that it is a \textit{MetadataRepository}. 
\end{itemize}

\subsubsection{MLDataObject and DataRepository}
Subsequently, the notation and usage of the \textit{MLDataObjects} and the \textit{DataRepository} are described. The notations of the two metamodel elements are shown in Figure~\ref{fig:Data_and_Log}. 
\begin{itemize}
\item A \textbf{MLDataObject} is an extended \textit{DataObject} that represents the actual types of data which are referenced and processed throughout the ML lifecycle. The \textit{MLDataObject} notation comes in several flavors, as \textit{RawDataObject}, \textit{FeatureSetObject}, \textit{DatasetObject}. \textit{RawDataObjects} represent the mostly unprocessed raw data that can be sourced from various data providers. They act as the one source of truth throughout the ML lifecycle. A \textit{FeatureSetObject} can be derived from a \textit{RawDataObject} or from other \textit{FeatureSetObjects} and represents features of some entity. A \textit{DatasetObject} can be derived from \textit{RawDataObjects}, \textit{FeatureSetObjects} or other \textit{DatasetObjects}. It represents the datasets that are eventually referenced in the learning phase. Through the notation a \textit{DatasetObject} can be further classified into \textit{TrainDatasetObject}, \textit{ValidationDatasetObject}, \textit{VerificationDatasetObject} and an \textit{InferenceRequestDatasetObject}. All seven objects have the same notation as the BPMN 2.0.2 \textit{DataObject}. As addition, for \textit{RawDataObjects} a data icon -, for \textit{FeatureSetObjects} overlapping rectangles (i.e. a feature icon) and for \textit{DatasetObjects} a data frame icon is placed in the upper left corner of the graphical element. To differentiate the \textit{DatasetObject} further specific icons are placed on top of the data frame icon. Consequently, the notation indicates that it is a \textit{MLDataObject} of the corresponding \textit{MLDataObjectType} and \textit{DataSetType}.
\item A \textbf{DataRepository} is an extended \textit{DataStore} that represents a storage for \textit{MLDataObjects}. Its specific purpose is the provision of a universal access point to read and write different \textit{MLDataObjects} to. Similar to the \textit{MLDataObject} notation the \textit{DataRepository} notation comes in three flavors covering the respective \textit{MLDataObject} types. A \textit{RawDataRepository} represents the storage for \textit{RawDataObjects}, a \textit{FeatureSetRepository} represents the storage for \textit{FeatureSetObjects} and a \textit{DatasetRepository} represents the storage for \textit{DatasetObjects}. The repository types have the same notation as the BPMN 2.0.2 \textit{DataStore} and are differentiated by the same icons used for the different data objects. The icons are however placed in the lower left corner of the graphical element, indicating that they are a \textit{DataRepository} of the corresponding \textit{RepositoryType}.
\end{itemize}

\subsubsection{LogObject and LogStore}
\begin{figure}
    \centering
    \includegraphics[width=1\linewidth]{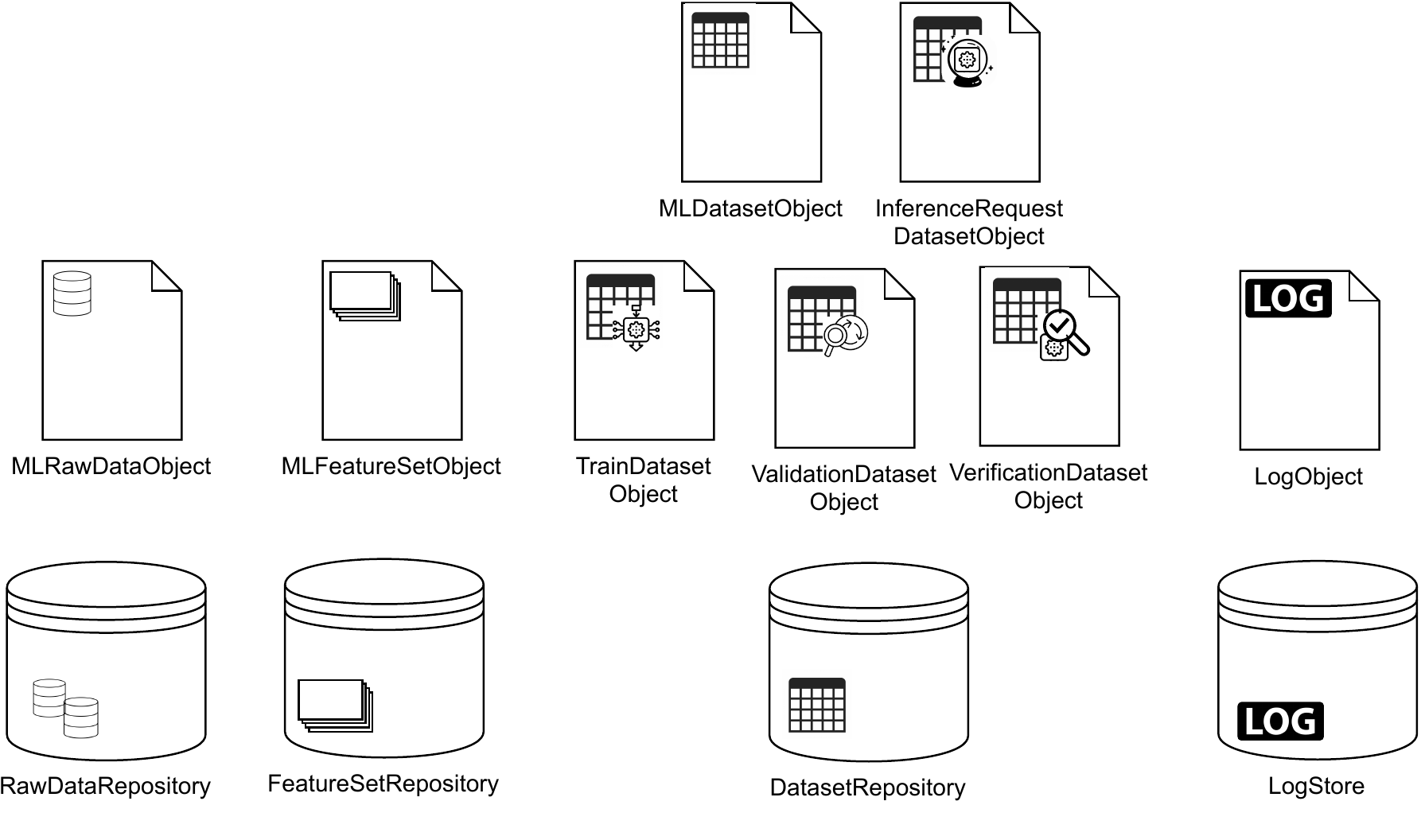}
    \caption[BPMN4sML notation of DataRepositories, MLDataObjects, LogStore and LogObject]%
    {BPMN4sML notation of DataRepositories, MLDataObjects, LogStore and LogObject \par \small }
    \label{fig:Data_and_Log}
\end{figure}

Subsequently, the notation and usage of the \textit{LogObject} and the \textit{LogStore} are described. The notations of the two metamodel elements are shown in Figure~\ref{fig:Data_and_Log}. 
\begin{itemize}
\item A \textbf{LogObject} is an extended \textit{DataObject} that represents the logs of activities and other services executed within and outside of the ML lifecycle phases. It has the same notation as the BPMN 2.0.2 \textit{DataObject}. As addition, a log icon is placed in the upper left corner of the graphical element, indicating that it is a \textit{LogObject}. 
\item A \textbf{LogStore} is an extended \textit{DataStore} that represents a storage for \textit{LogObjects}. Its specific purpose is the provision of a universal access point to write logging information to which other services or activities may read and process. It has the same notation as the BPMN 2.0.2 \textit{DataStore}. As addition, a log icon is placed in the lower left corner of the graphical element, indicating that it is a \textit{LogStore}. 
\end{itemize}

\subsection{Activities}
\label{activities}
As explained in section~\ref{BPMN}, activities represent operations in a BPMN process diagram. Of interest to this study is the \textit{Task} type through which an atomic activity can be defined describing a specific action executed within the process flow~\cite[p.154ff]{OMGBPMN}. The OMG standard provides seven sub-classes to the Task type, namely \textit{Service}, \textit{Send}, \textit{Receive}, \textit{User}, \textit{Manual}, \textit{Business Rule} and \textit{Script}. Each of the sub-types constraint the generic \textit{Task} type to define the operation more precisely addressing various scenarios. In the context of machine learning workflows two sub-classes are potential candidates for extension - the \textit{ScriptTask} and the \textit{ServiceTask}. While the former task allows to specify a script which can be equated with the business logic of a serverless function it is constrained to the execution environment of business process engines and thus not appropriate for the FaaS-based nature of the modeled ML workflows. In contrast, a \textit{ServiceTask} does not constrain the execution environment. While it does not fully reflect the needs for the context of this research, it can be extended to do so. Consequently, within BPMN4sML new task types as sub-classes of \textit{ServiceTask} are defined that express the extension for the \textit{FlowObject} \textit{Activity} in context of the previously described metamodel in Figure~\ref{fig:BPMN4sMLTasks}. Each task comes with its own notation and follows a set of guidelines that are elaborated in the remaining part of this section. For explanation purposes, selected examples describing the application of the extension are given with some of them being visually depicted. The examples are workflow fragments highlighted by using \textit{Link} \textit{Intermediate} events that are of type \textit{Catch} or \textit{Throw}  This is valid for all of the following examples. Note that all tasks can produce \textit{MetadataObjects} and \textit{LogObjects} which is therefore omitted in the task description to highlight task-specific differences. Furthermore, the respective produced and sourced artefacts implicitly imply a read or write operation to the various data store solutions.

\begin{figure}
    \centering
    \includegraphics[width=1\linewidth]{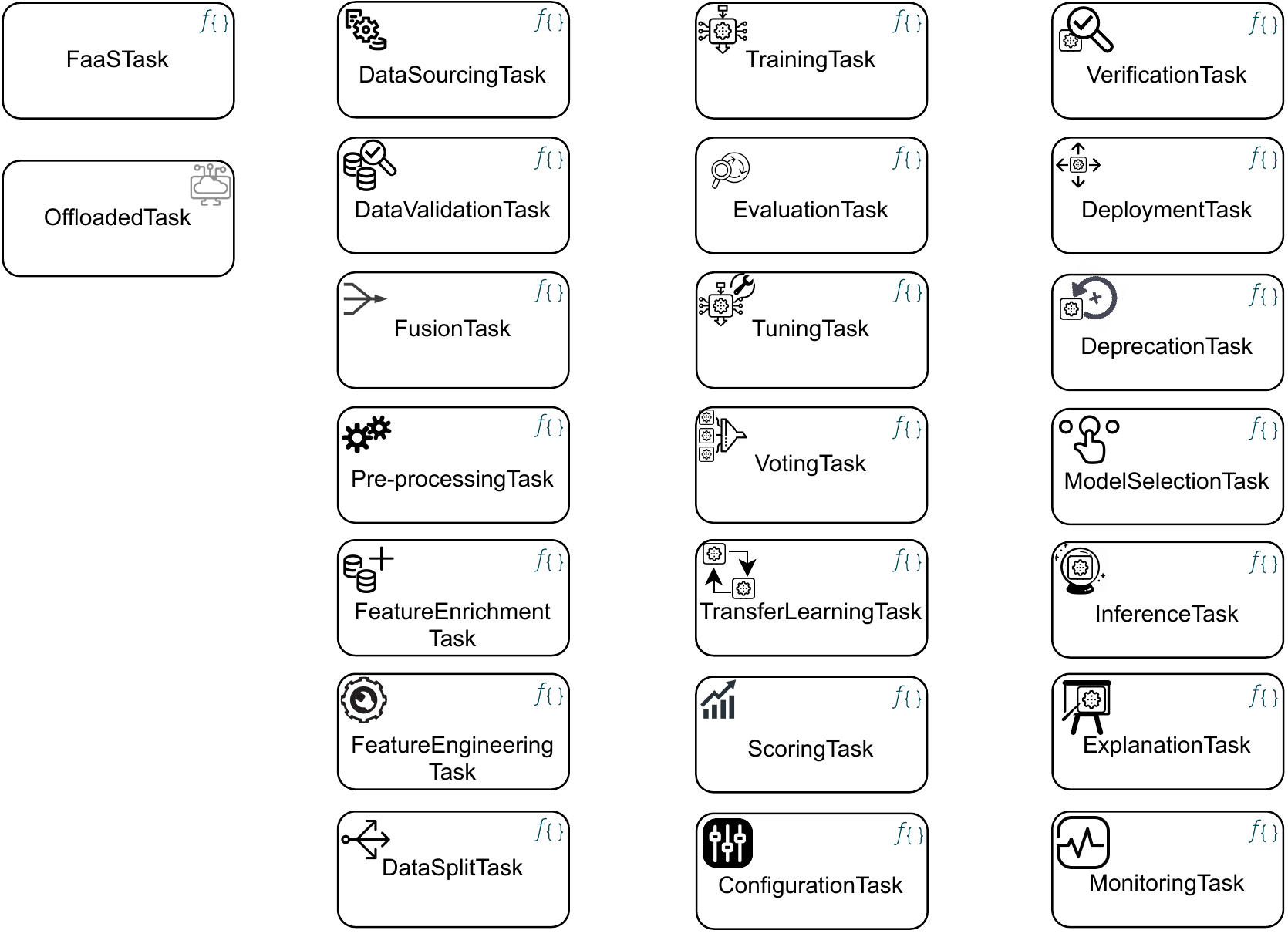}
    \caption[BPMN4sML notation of Tasks]%
    {BPMN4sML notation of Tasks\par \small  Note that the Tasks inheriting from FaaSTask or OffloadedTask are only visualized with inheritance from \textit{FaaSTask} to save space. In case of an offloaded task, the upper right \textit{FaaSIcon} of the shape is exchanged for the icon of the \textit{OffloadedTask}.}
    \label{fig:Tasks}
\end{figure}

\subsubsection{FaaSTask}
A  \textbf{FaaSTask} is a Task that represents a serverless function which is run on a cloud and executes some logic. It can interact with \textit{DataObjects} and \textit{DataStores} and has typical FaaS constraints, i.e. short running time, fixed memory, statelessness. A \textit{FaaSTask} is triggered by an event (e.g. a simple invocation, not to be confused with a BPMN event) which serves as data input and is assumed to be in JSON format. Its content may be identifiers to relevant data artefacts that the \textit{FaaSTask} operates on. On a BPMN modeling level the triggering event may either be explicitly declared via a BPMN event or implicitly by modeling arcs, e.g. sequence flow between activities. After executing its logic, the \textit{FaaSTask} is completed. It can return information in JSON format, i.e. data outputs, and thereby produce events in the serverless context. The selection in favor of JSON has been made as FaaS products (e.g. by Google Cloud, Azure, AWS) and other programming languages support its interpretation. A \textit{FaaSTask} object shares the same shape as the \textit{Task}, which is a rectangle that has rounded corners. However, there is a FaaS icon in the upper right corner of the shape that indicates that the Task is a \textit{FaaSTask}, see Figure~\ref{fig:Tasks}. 
\begin{itemize}
\item \textbf{Example:} After a tuning job, a new tuning result is produced carrying information about hyperparameter sets, their values and respective performance scores that the corresponding models achieved. The result is propagated to a serverless function, i.e. \textit{FaaSTask}, triggering its invocation. The function identifies the best score and the associated hyperparameter values and \textit{MLModelObject} to reference it in the next steps. This ML workflow fragment is illustrated within Figure~\ref{fig:OffloadedTuningExample}.
\end{itemize}

\subsubsection{OffloadedTask}
An \textbf{OffloadedTask} is a Task that represents an operation which is run outside of a serverless function on an optimized hardware stack or on an edge device. The offloaded task executes some logic. The logic can be a pre-defined configuration or handed to it for instance through a \textit{CodeObject}. 
Similar to a \textit{FaaSTask}, in cloud-based scenarios the offloaded task may directly be triggered by an event (from the serverless context) which serves as data input and return a data output, thereby producing an event in the serverless system (not to confuse with BPMN events). An \textit{OffloadedTask} can but does not have to be resource constrained as a \textit{FaaSTask} is. After executing its logic, the \textit{OffloadedTask} is completed. An \textit{OffloadedTask} object shares the same shape as the \textit{Task}, which is a rectangle that has rounded corners. However, there is a cloud-edge icon in the upper right corner of the shape that indicates that the Task is an \textit{OffloadedTask}, see Figure~\ref{fig:Tasks}. An illustrative example is given in Figure~\ref{fig:OffloadedTuningExample}.

\subsubsection{JobConfigurationTask}
\begin{figure}
    \centering
    \includegraphics[width=1\linewidth]{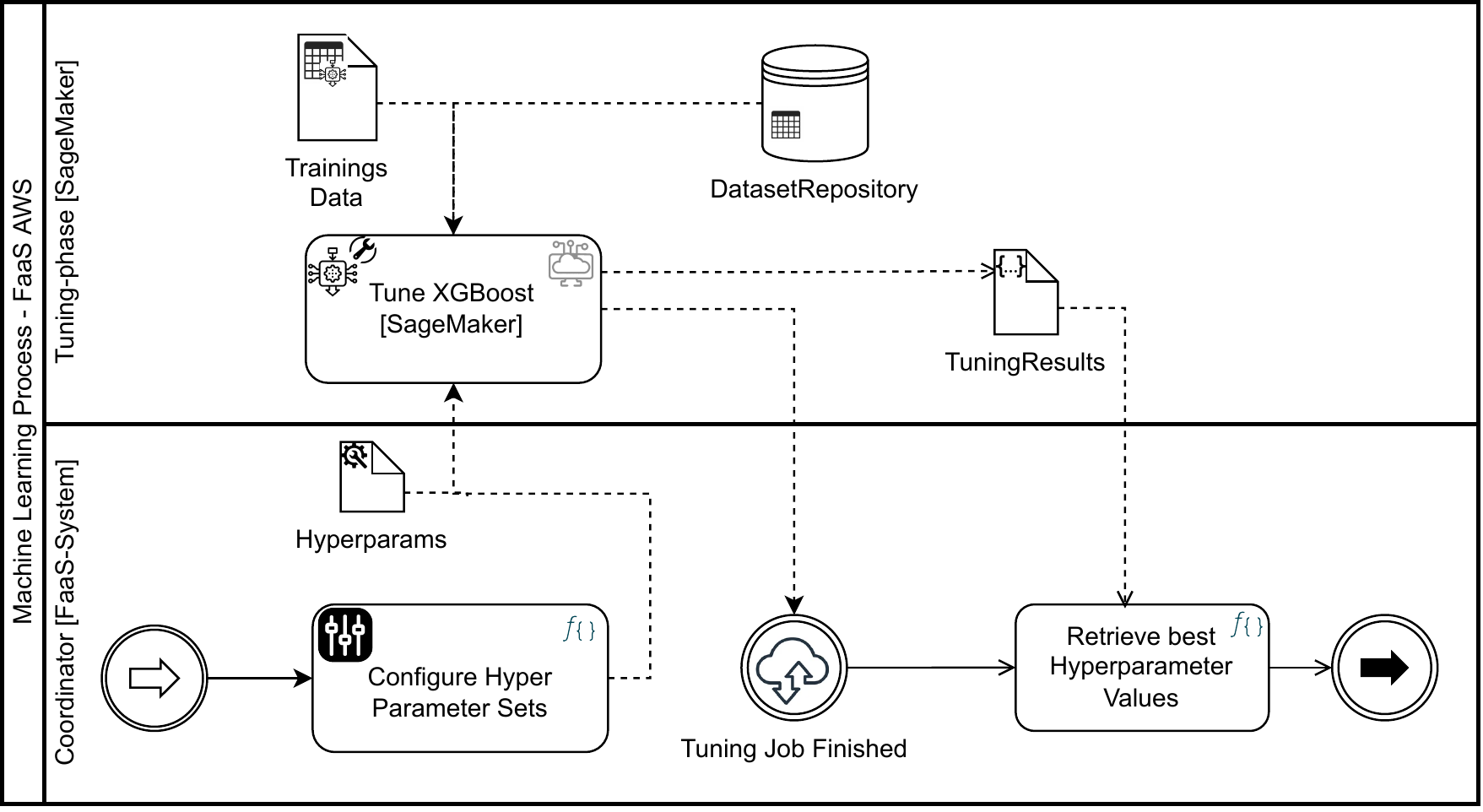}
    \caption[BPMN4sML workflow fragment of ConfigurationTask, offloaded TuningTask, JobOffloadingEvent and FaaSTask]%
    {BPMN4sML workflow fragment of ConfigurationTask, Offloaded TuningTask, JobOffloading Event and FaaSTask\par \small  }
    \label{fig:OffloadedTuningExample}
\end{figure}

In the context of BPMN4sML, a \textbf{JobConfigurationTask} represents a serverless function used to prepare an offloaded task. This can manifest in the generation of a \textit{CodeObject} that specifies a script to run during the offloaded task, in the generation of a \textit{LearningConfiguration}, e.g. to specify a tuning set-up, or in the identification of devices on which the offloaded job shall be executed. The generated artefacts may be directly propagated as files or as identifiers to those files (stored in a \textit{MetadataRepository}). A \textit{JobConfigurationTask} may precede a task that inherits from \textit{OffloadedTask}. Alternatively, it may define the entire operation through the \textit{CodeObject}. A \textit{JobConfigurationTask} object shares the same shape as the \textit{Task}, which is a rectangle that has rounded corners. However, there is a configuration icon in the upper left corner of the shape that indicates that the Task is a \textit{JobConfigurationTask}, see Figure~\ref{fig:Tasks}. 
\begin{itemize}
\item \textbf{Example: } This example describes a use case for a \textit{JobConfigurationTask}, an offloaded \textit{TuningTask} and a \textit{JobOffloadEvent}. 
For a XGBoost algorithm, a complex tuning job defined by a large GridSearch and 10-fold nested cross-validation needs to be executed to identify an appropriate set of hyperparameters. To mitigate computational resource constraints, the tuning job is directly run on AWS SageMaker which serves as a machine learning platform. The tuning strategy may be directly specified within the configuration script in JSON format as part of the \textit{OffloadedTask}. Alternatively for this use case, a \textit{JobConfigurationTask} defines certain aspects of the tuning configuration, such as which hyperparameter sets to try out, dynamically prior to executing the offloaded task. The configuration is then propagated to the \textit{TuningTask}. A \textit{JobOffloadEvent} is triggered once the tuning job finishes and the process continues. In this scenario, the produced \textit{TuningResult} is not lost due to the offloaded tuning task not being stateless. Subsequent tasks may therefore access and analyse the \textit{TuningResult} without it being explicitly written to a \textit{MetadataRepository}. This workflow fragment is illustrated in Figure~\ref{fig:OffloadedTuningExample}.
\end{itemize}

Note that the following tasks can be realized as \textit{FaaSTask} or as \textit{OffloadedTask}. Their notation allows to differentiate between the two by having either the \textit{FaaSTask} or the \textit{OffloadedTask} icon in the upper right corner of the element.

\subsubsection{DataSourcingTask}
A \textbf{DataSourcingTask} represents an operation to retrieve raw data from a raw data provider, potentially transform it and write it to a \textit{RawDataRepository}. The data collected and written to the repository is a \textit{MLRawDataObject}. The task can propagate the identifier to the new \textit{MLRawDataObject}. In case the sourced data is already processed it can be written as a \textit{MLFeatureSetObject} to a \textit{FeatureSetRepository}. This task often is the initial step towards a machine learning workflow. A \textit{DataSourcingTask} object shares the same shape as the \textit{Task}, which is a rectangle that has rounded corners. However, there is a sourcing (ETL) icon in the upper left corner of the shape that indicates that the Task is a \textit{DataSourcingTask}, see Figure~\ref{fig:Tasks}. 
\begin{itemize}
\item \textbf{Example:} For a hypothetical stock trading bot the latest stock information needs to be regularly sourced. Every evening, a \textit{DataSourcingTask} is executed to retrieve the stock data of that day in order to update a machine learning model for the next day.
\end{itemize}

\subsubsection{DataValidationTask}
A \textbf{DataValidationTask} is a Task that examines if anomalies start occurring in a \textit{MLFeatureSetObject} or \textit{MLDatasetObject}. Anomalies may be the disappearance of features or a strong change in the data distribution. The new feature sets or datasets are compared against existing schemata and known data statistics which can be sourced from the \textit{MetadataRepository} to realize the validation activity.
A \textit{DataValidationTask} object shares the same shape as the \textit{Task}, which is a rectangle that has rounded corners. However, there is an additional icon representing data validation in the upper left corner of the shape that indicates that the Task is a \textit{DataValidationTask}, see Figure~\ref{fig:Tasks}. 
\begin{itemize}
\item \textbf{Example:} Elaborating on the previous example, prior to considering the retrieved stock data as input for model training, it is ensured that no drastic change in the data statistics occurred - for instance in case of a crash of a specific stock or a part of the stock market any machine learning model may not be applicable anymore. 
\end{itemize}

\subsubsection{DataFusionTask}
A \textbf{DataFusionTask} is a Task that fuses data or features originating from different data providers but specifying the same phenomenon. It can operate on \textit{MLRawDataObjects} as well as \textit{MLFeatureSetObjects}. It must be connected to at least two \textit{MLDataObjects} that are to be fused and produces exactly one \textit{MLDataObject} representing the now fused data artefact.
A \textit{DataFusionTask} object shares the same shape as the \textit{Task}, which is a rectangle that has rounded corners. However, there is an additional icon representing data fusion in the upper left corner of the shape that indicates that the Task is a \textit{DataFusionTask}, see Figure~\ref{fig:Tasks}. 
\begin{itemize}
\item \textbf{Example:} A radar and a camera capture separate observations of the same car driving through a street. To create a unified \textit{MLFeatureSetObject} the two data artefacts need to be fused by correctly associating the observations with each other. The unified \textit{MLFeatureSetObject} is subsequently written to the \textit{FeatureRepository}. 
\end{itemize}

\subsubsection{PreprocessingTask}
A \textbf{PreprocessingTask} is a Task that performs various data processing operations on a \textit{MLDataObject} such as data cleaning or imputation of missing observations. It can operate on exactly one \textit{MLDataObject} and writes the processed \textit{MLDataObject} to the respective \textit{FeatureRepository} or \textit{DatasetRepository}.
A \textit{PreprocessingTask} object shares the same shape as the \textit{Task}, which is a rectangle that has rounded corners. However, there is an additional icon representing data processing in the upper left corner of the shape that indicates that the Task is a \textit{PreprocessingTask}, see Figure~\ref{fig:Tasks}. An illustrative example is given in Figure~\ref{fig:PreprocessingTrainingExample}.

\subsubsection{FeatureEngineeringTask}
A \textbf{FeatureEngineeringTask} is a Task that creates structural changes on a \textit{MLFeatureSetObject} or a \textit{MLDatasetObject}. For instance this can be the creation of new features through the numerical combination of existing ones (e.g. creating group averages) or selection of a subset of features relevant to the model learning activities. It can operate on exactly one \textit{MLDataObject} and writes the processed \textit{MLDataObject} to the respective \textit{FeatureRepository} or \textit{DatasetRepository}. A \textit{FeatureEngineeringTask} object shares the same shape as the \textit{Task}, which is a rectangle that has rounded corners. However, there is an additional icon representing feature engineering in the upper left corner of the shape that indicates that the Task is a \textit{FeatureEngineeringTask}, see Figure~\ref{fig:Tasks}. 

\subsubsection{FeatureEnrichmentTask}
A \textbf{FeatureEnrichmentTask} is a Task that adds several \textit{MLFeatureSetObjects} or \textit{MLDatasetObjects} into one overall \textit{MLDatasetObject} to improve the predictive signal. The enriched data artefact is then written to the respective \textit{DataRepository}. It differs from the previously introduced \textit{PreprocessingTask} and \textit{FeatureEngineeringTask} as it requires several \textit{MLDataObjects}. In contrast to the \textit{FusionTask}, a \textit{FeatureEnrichmentTask} operates on features and datasets of related but different observations. A \textit{FeatureEnrichmentTask} object shares the same shape as the \textit{Task}, which is a rectangle that has rounded corners. However, there is an additional icon representing feature enrichment in the upper left corner of the shape that indicates that the Task is a \textit{FeatureEnrichmentTask}, see Figure~\ref{fig:Tasks}. 
\begin{itemize}
\item \textbf{Example:} A ML model within a recommender system of a fashion online shop predicts potential products that a site visitor is interested in based on already visited products. If the site visitor is a regular user who can be identified, the dataset used to train the ML model with may be enriched with a feature on the shopping history of users to create a more customized model.
\end{itemize}

\subsubsection{DataSplitTask}
A \textbf{DataSplitTask} is a Task that can split a \textit{MLDatasetObject} into several \textit{MLDataObjects} such as a \textit{TrainDatasetObject}, \textit{VerificationDatasetObject} and a \textit{ValidationDatasetObject}. Not all three splits must be realized and can be further customized. The data artefacts are read from and written to a \textit{DatasetRepository} which the task has a unique connection to. A \textit{DataSplitTask} object shares the same shape as the \textit{Task}, which is a rectangle that has rounded corners. However, there is an additional icon representing a data split in the upper left corner of the shape that indicates that the Task is a \textit{DataSplitTask}, see Figure~\ref{fig:Tasks}. An illustrative example is given in Figure~\ref{fig:PreprocessingTrainingExample}.

\subsubsection{TrainingTask}
A \textbf{TrainingTask} is a Task that trains a machine learning algorithm to produce a \textit{MLModelObject} which may be written to the \textit{ModelRegistry}. The \textit{TrainingTask} requires exactly one connection to a \textit{TrainDatasetObject} which can be sourced from the \textit{DatasetRepository} and an optional one to a \textit{TrainingConfiguration} sourced from a \textit{MetadataRepository} (or directly handed to the task as JSON). A \textit{TrainingTask} object shares the same shape as the \textit{Task}, which is a rectangle that has rounded corners. However, there is an additional icon representing ML model training in the upper left corner of the shape that indicates that the Task is a \textit{TrainingTask}, see Figure~\ref{fig:Tasks}. An illustrative example is given in Figure~\ref{fig:PreprocessingTrainingExample}.

\subsubsection{ScoringTask}
\begin{figure}
    \centering
    \includegraphics[width=1\linewidth]{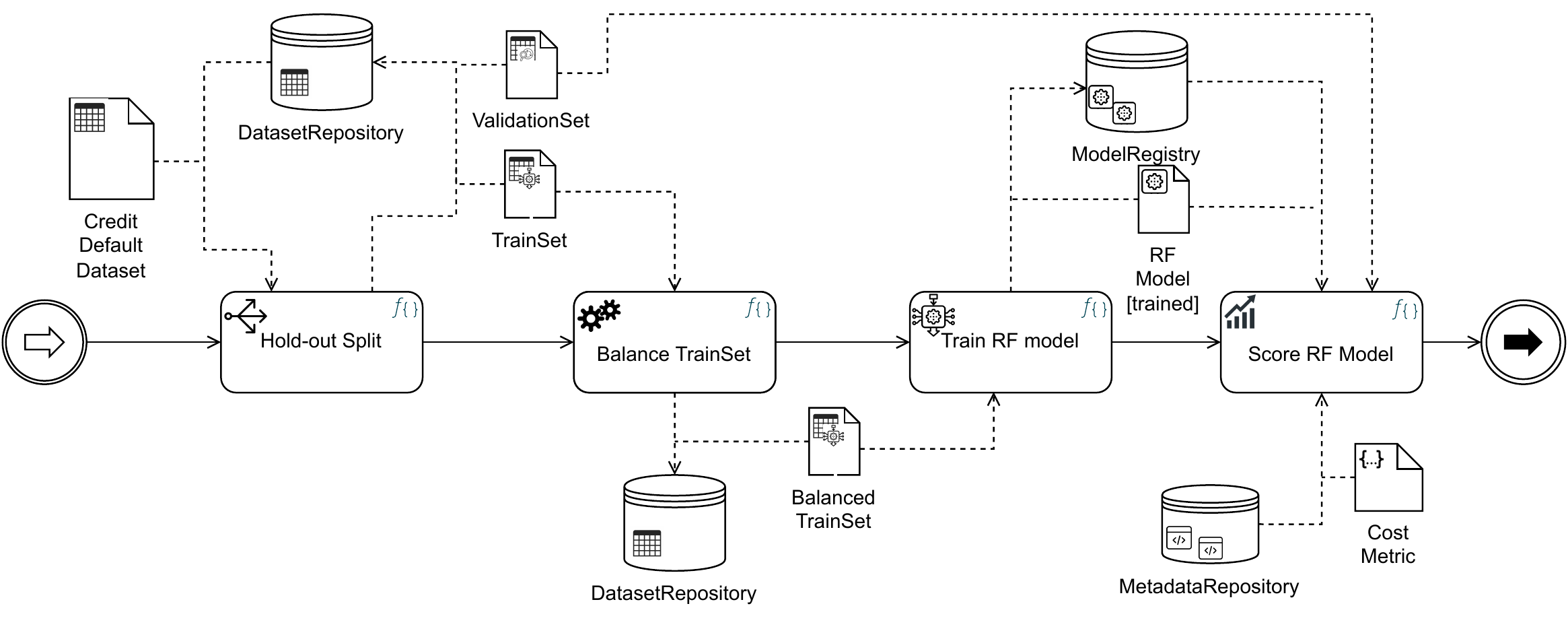}
     \caption[BPMN4sML workflow fragment of DataSplitTask, PreprocessingTask, TrainingTask and ScoringTask alongside respective data artefacts]%
    {PMN4sML workflow fragment of DataSplitTask, PreprocessingTask, TrainingTask and ScoringTask alongside respective data artefacts\par \small  Note that accessing a data artefact implies a read or write operation to the respective repository as all tasks depicted are FaaS-based tasks.}
    \label{fig:PreprocessingTrainingExample}
\end{figure}

A \textbf{ScoringTask} is a Task that scores a single trained \textit{MLModelObject} by means of a scoring metric and a \textit{ValidationDatasetObject}. The scoring metric can be hard-coded as part of the task or sourced from a \textit{Document} of type \textit{RequirementDocument}. The \textit{ScoringTask} requires exactly one connection to a \textit{ValidationDatasetObject} sourced from a \textit{DatasetRepository}. A \textit{ScoringTask} produces a performance score for a \textit{MLModelObject}. Alternatively, if the learning job continuous to not meet the requirements, a \textit{DeficitReport} may be produced. A \textit{ScoringTask} object shares the same shape as the \textit{Task}, which is a rectangle that has rounded corners. However, there is an additional icon representing model scoring in the upper left corner of the shape that indicates that the Task is a \textit{ScoringTask}, see Figure~\ref{fig:Tasks}.
\begin{itemize}
\item \textbf{Example:} This example includes a \textit{DataSplitTask}, \textit{PreprocessingTask}, \textit{TrainingTask} and \textit{ScoringTask}, see Figure~\ref{fig:PreprocessingTrainingExample}. In a simplified two-class classification problem, a bank wants to predict if a loan applicant is likely to default.  Information about the applicant is used as \textit{TrainDatasetObject}. A \textit{DataSplitTask} further creates a new \textit{TrainingDatasetObject} and a \textit{ValidationDatasetObject}. Prior to learning, the training dataset needs to be balanced via a \textit{PreprocessingTask} since the loan application prediction is an imbalanced classification problem, i.e. only a few customers actually default. Afterwards, the \textit{TrainingTask} learns a RandomForest algorithm with pre-specified hyperparameters on the training dataset which is sourced from the \textit{DatasetRepository}. Once trained, the \textit{ScoringTask} scores the RandomForest against the validation dataset using a cost metric defined in the \textit{RequirementDocument}. For this simplified example, the model has to perform better than a dummy model, i.e. better than 50\% of all predictions need to be correct. Once scored, a performance score and a potential deficit report can be generated. 
\end{itemize}

\subsubsection{EvaluationTask}
An \textbf{EvaluationTask} is a Task that evaluates the potential fit of an algorithm for a ML problem. The task requires exactly one connection to a \textit{TrainDatasetObject}. The resampling strategy and evaluation parameters can either be hard-coded as part of the task or sourced from an \textit{EvaluationConfiguration}. An \textit{EvaluationTask} produces an \textit{EvaluationResult}, e.g. a robust performance estimate of the algorithm, and optionally writes the ML models produced during resampling to a \textit{ModelRegistry} (or in case of an offloaded job, stores the models in that environment). Alternatively, if the evaluation indicates that the algorithm underperforms, a \textit{DeficitReport} may be produced. An \textit{EvaluationTask} object shares the same shape as the \textit{Task}, which is a rectangle that has rounded corners. However, there is an additional icon representing model evaluation in the upper left corner of the shape that indicates that the Task is an \textit{EvaluationTask}, see Figure~\ref{fig:Tasks}.
\begin{itemize}
\item \textbf{Example:} Various algorithms are considered to solve a given ML problem. For each algorithm an \textit{EvaluationTask} is executed to robustly estimate the fit of that algorithm for the ML task. The most promising algorithm is trained once more on the entirety of the training dataset before ultimately verifying it.
\end{itemize}

\subsubsection{TuningTask}
A \textbf{TuningTask} is a Task that tunes a machine learning algorithm. The tasks requires exactly one connection to a \textit{TrainDatasetObject} which can be sourced from the \textit{DatasetRepository} and an optional one to a \textit{TuningConfiguration} sourced from a \textit{MetadataRepository} (or directly handed to the task as JSON). A \textit{TuningTask} produces a \textit{TuningResult} that can be stored in a \textit{MetadataRepository} if necessary and optionally writes the tuned model(s) to a \textit{ModelRegistry} (or in case of an offloaded job, stores the model(s) in that environment). A \textit{TuningTask} object shares the same shape as the \textit{Task}, which is a rectangle that has rounded corners. However, there is an additional icon representing ML model tuning in the upper left corner of the shape that indicates that the Task is a \textit{TuningTask}, see Figure~\ref{fig:Tasks}. An illustrative  example is given in Figure~\ref{fig:OffloadedTuningExample}.

\subsubsection{TransferLearningTask}
A \textbf{TransferLearningTask} is a Task that re-learns an existing \textit{MLModelObject} sourced from the \textit{ModelRegistry} on a new but related \textit{TrainingDatasetObject} sourced from the \textit{DatasetRepository}. The task produces a new \textit{MLModelObject} which can be written to the \textit{ModelRegistry}. 
A \textit{TransferLearningTask} object shares the same shape as the \textit{Task}, which is a rectangle that has rounded corners. However, there is an additional icon representing model transfer in the upper left corner of the shape that indicates that the Task is a \textit{TransferLearningTask}, see Figure~\ref{fig:Tasks}. 
\begin{itemize}
\item \textbf{Example:} A city wants to integrate a machine learning model in their smart traffic control system. The model directly extracts the letters on a license plate of speeding cars to automatically identify and fine the owner. To avoid spending computational resources on training a sophisticated model from scratch an existing one shall be leveraged. For this task, convolutional neural networks are ideal solution candidates and consequently a \textit{TransferLearningTask} is picked to model the respective ML workflow. A pre-trained deep convolutional neural network is sourced from the ModelRegistry. Its first few layers are frozen whereas the last layers can be updated to associate the patterns with the numbers of the license plate. Once updated, the new \textit{MLModelObject} is stored in the model registry from where other services and components can access it.
\end{itemize}

\subsubsection{VotingTask}
A \textbf{VotingTask} is a Task to conduct a consensus operation with. The task requires the model \textit{InferenceResults} of at least two \textit{MLModelObjects}. It weighs each inference result according to a pre-defined schema or method (e.g. a majority vote) and produces a new \textit{InferenceResult}. A \textit{VotingTask} object shares the same shape as the \textit{Task}, which is a rectangle that has rounded corners. However, there is an additional icon representing a consensus operation in the upper left corner of the shape that indicates that the Task is a \textit{VotingTask}, see Figure~\ref{fig:Tasks}. 

\subsubsection{VerificationTask}
A \textbf{VerificationTask} is a Task that verifies if a trained \textit{MLModelObject} complies with all constraints of a \textit{RequirementDocument}. The constraints can either be embedded within the task or handed to it via the \textit{RequirementDocument}. It differs from the \textit{EvaluationTask} as it sources a never before seen \textit{VerificationDatasetObject} to test the model with. It can produce a \textit{VerificationResult} in case the constraints are not met. Otherwise, it may modify the status of a \textit{MLModelObject} to \textit{verified}. 
A \textit{VerificationTask} object shares the same shape as the \textit{Task}, which is a rectangle that has rounded corners. However, there is an additional icon representing a verification operation in the upper left corner of the shape that indicates that the Task is a \textit{VerificationTask}, see Figure~\ref{fig:Tasks}. 

\subsubsection{DeploymentTask}
A \textbf{DeploymentTask} is a Task that deploys a \textit{MLModelObject} in the specified environment to make it accessible for inference jobs. It may modify the status of a \textit{MLModelObject} to \textit{deployed}. The deployment operation may send the \textit{MLModelObject} as a file to an endpoint or write it to a specific directory within the \textit{ModelRegistry} and provide its identifiable path.
A \textit{DeploymentTask} object shares the same shape as the \textit{Task}, which is a rectangle that has rounded corners. However, there is an additional icon representing a deployment operation in the upper left corner of the shape that indicates that the Task is a \textit{DeploymentTask}, see Figure~\ref{fig:Tasks}. 

\subsubsection{DeprecationTask}
A \textbf{DeprecationTask} is a Task that retires a deployed \textit{MLModelObject} and removes it from its accessible endpoint. It may modify the status of a \textit{MLModelObject} to \textit{deprecated}. Usually, a \textit{DeprecationTask} is the answer to the presence of a new better performing model. The deprecated \textit{MLModelObject} can be archived in the \textit{ModelRegistry}.
A \textit{DeprecationTask} object shares the same shape as the \textit{Task}, which is a rectangle that has rounded corners. However, there is an additional icon representing a deprecation operation in the upper left corner of the shape that indicates that the Task is a \textit{DeprecationTask}, see Figure~\ref{fig:Tasks}.

\subsubsection{InferenceTask}
An \textbf{InferenceTask} is a Task that generates a prediction, i.e. \textit{InferenceResult}, by accessing a deployed \textit{MLModelObject} and using it to run its \textit{prediction} capability on the \textit{InferenceRequestDataset}. 
In FaaS-based systems the \textit{MLModelObject} is loaded from the \textit{ModelRegistry} into the serverless function to compute the prediction. In case of offloaded jobs, the \textit{MLModelObject} may not have to be loaded and can be directly accessed, i.e. no connection to a \textit{ModelRegistry} is required.
An \textit{InferenceTask} object shares the same shape as the \textit{Task}, which is a rectangle that has rounded corners. However, there is an additional icon representing a inference operation in the upper left corner of the shape that indicates that the Task is an \textit{InferenceTask}, see Figure~\ref{fig:Tasks}.

\subsubsection{ModelSelectionTask}
A \textbf{ModelSelectionTask} is a Task that chooses between several \textit{MLModelObjects} based on some criteria in order to select the most appropriate one to fulfil the prediction task. It then propagates the request to the corresponding ML model endpoint or in case of a purely FaaS-based solution, it may propagate the request alongside the selected \textit{MLModelObject} identifier to the next task.  
A \textit{ModelSelectionTask} object shares the same shape as the \textit{Task}, which is a rectangle that has rounded corners. However, there is an additional icon representing a selection operation in the upper left corner of the shape that indicates that the Task is a \textit{ModelSelectionTask}, see Figure~\ref{fig:Tasks}.
\begin{itemize}
\item \textbf{Example:} A complex ML anomaly detection system provides specialized machine learning models for certain anomaly categories. The categories have been formed by means of an unsupervised clustering algorithm. Prior to executing the \textit{InferenceTask}, the \textit{InferenceRequest} is assigned to a cluster. The \textit{ModelSelectionTask} then identifies the corresponding \textit{MLModelObject} (or endpoint) and propagates the \textit{InferenceRequestDataset} accordingly after which the \textit{InferenceTask} is executed.
\end{itemize}

\subsubsection{ExplanationTask}
An \textbf{ExplanationTask} is a Task that justifies, i.e. explains, a decision for a specific prediction, i.e. \textit{InferenceResult}. It thus focuses on local explainability of a \textit{MLModelObject}. Depending on the implemented method the task may source the \textit{MLModelObject} that generated the \textit{InferenceResult} alongside the \textit{InferenceRequestDataset}. The \textit{ExplanationTask} produces a \textit{Document} of type \textit{ModelExplanation}. 
An \textit{ExplanationTask} object shares the same shape as the \textit{Task}, which is a rectangle that has rounded corners. However, there is an additional icon representing a model explanation operation in the upper left corner of the shape that indicates that the Task is an \textit{ExplanationTask}, see Figure~\ref{fig:Tasks}.

\subsubsection{MonitoringTask}
An \textbf{MonitoringTask} is a Task that can investigate various \textit{LogObjects} and \textit{MetadataObjects} which are sourced from the \textit{LogRepository} and \textit{MetadataRepository}. Depending on its implementation, the task may analyse data and model statistics to identify for instance prospective data drift or concept drift. It can produce new \textit{Model \& Data Statistics}. Additionally, it is the main activity to trigger subsequent events relating to critical ML workflow situations, i.e. data or concept drift. 
A \textit{MonitoringTask} object shares the same shape as the \textit{Task}, which is a rectangle that has rounded corners. However, there is an additional icon representing a monitoring operation in the upper left corner of the shape that indicates that the Task is a \textit{MonitoringTask}, see Figure~\ref{fig:Tasks}.

\subsection{Events}
\begin{figure}
    \centering
    \includegraphics[width=0.7\linewidth]{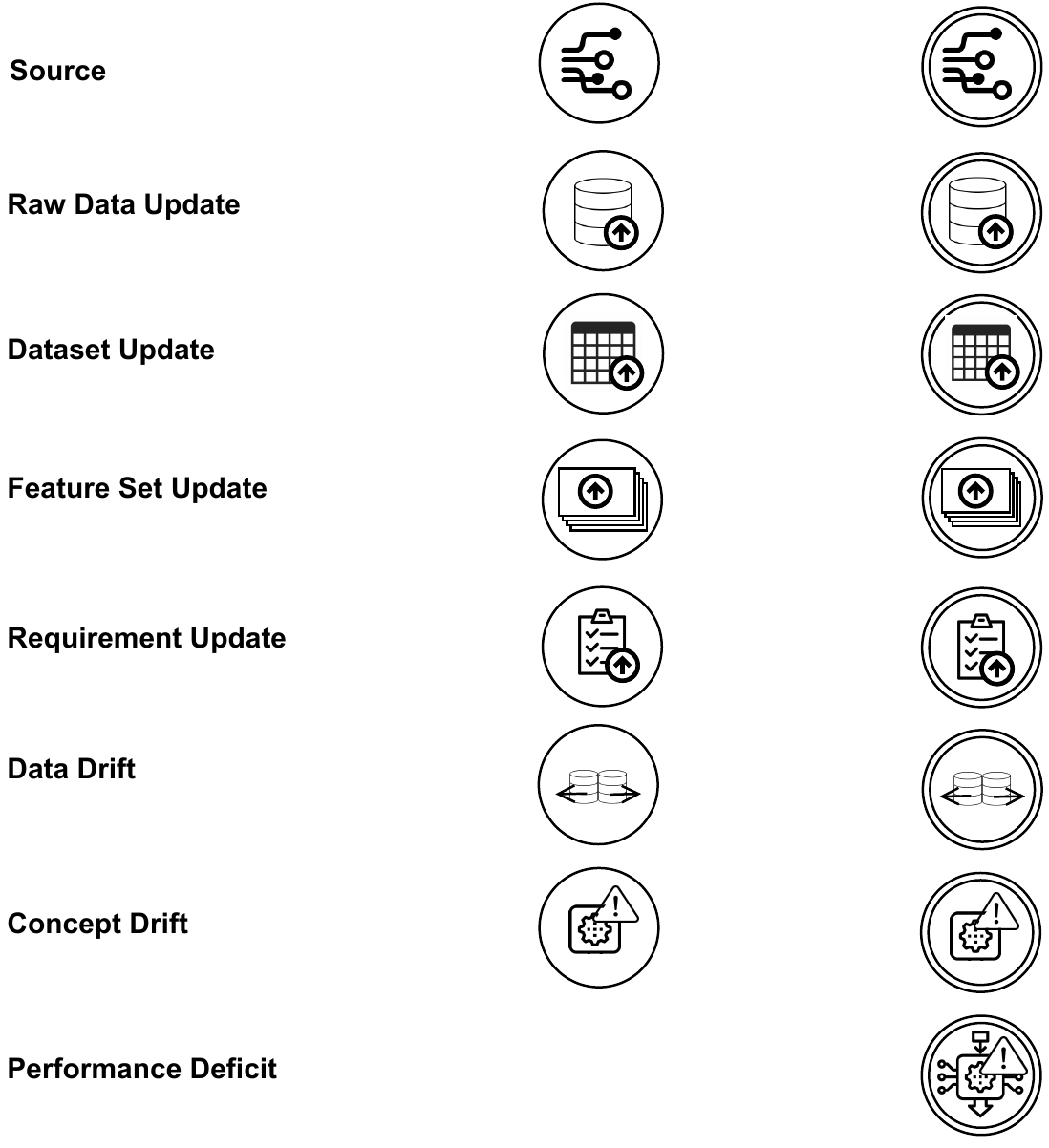}
    \caption[BPMN4sML notation of Events]%
    {BPMN4sML notation of Events\par \small }
    \label{fig:Events1}
\end{figure}

\begin{figure}
    \centering
    \includegraphics[width=0.7\linewidth]{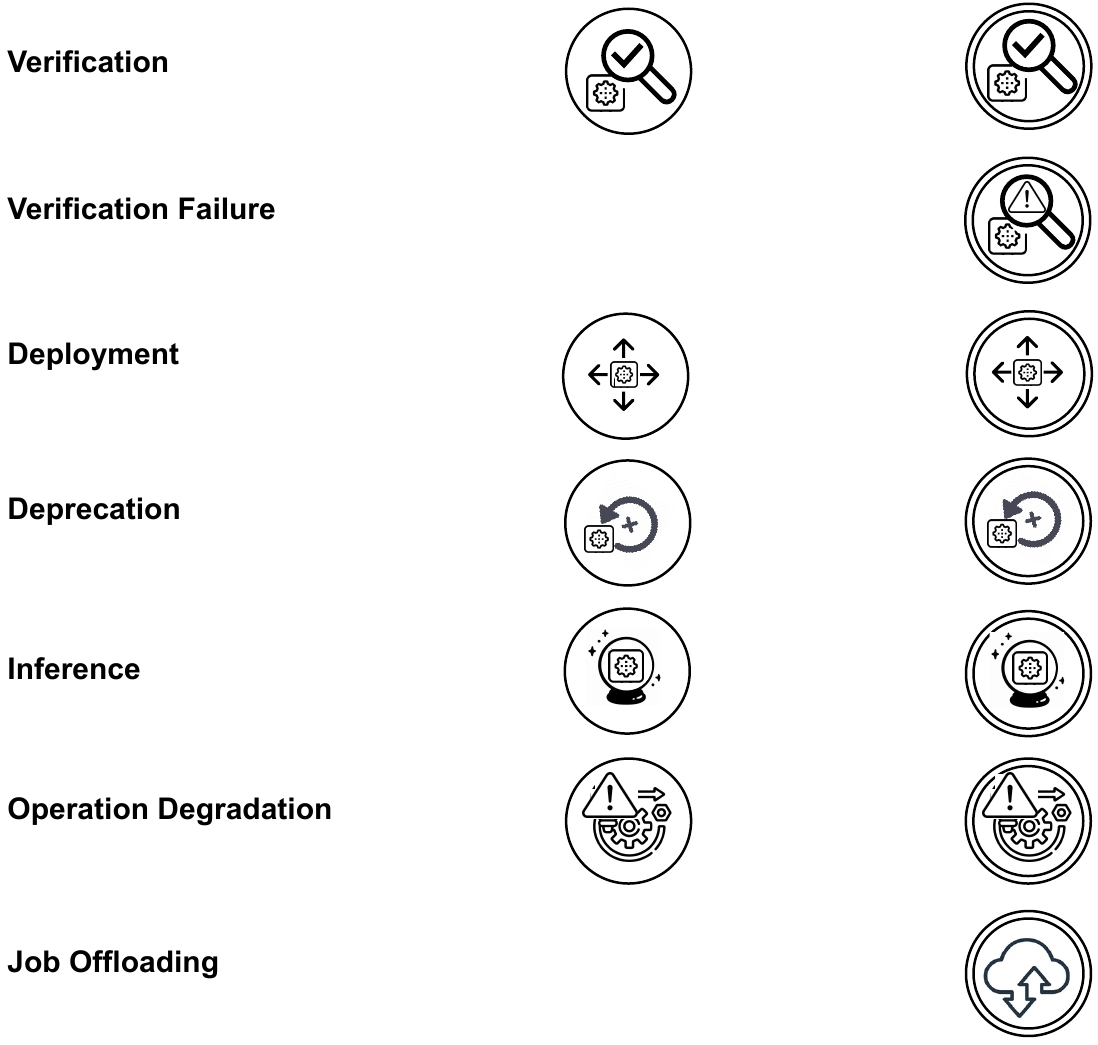}
    \caption[BPMN4sML notation of Events (Cont'd)]%
    {BPMN4sML notation of Events  (Cont'd)\par \small }
    \label{fig:Events2}
\end{figure}

As explained in section~\ref{BPMN} events can be differentiated into \textit{Start}, \textit{Intermediate} and \textit{End} events. They can further be classified into \textit{Catch} and \textit{Throw} events, the former catching a trigger and the latter throwing it. \textit{Start} and \textit{Intermediate Catch} events are of type \textit{Catch} whereas \textit{Intermediate Throw} and \textit{End} events are of type \textit{Throw}~\cite[p.233]{OMGBPMN}. The OMG standard comes with twelve events defined via their \textit{EventDefinition} with exception of a \textit{NoneEvent}~\cite[p.238ff]{OMGBPMN}. The proposed BPMN4sML extension adds fifteen new events to the BPMN 2.0.2 standard (see Figure~\ref{fig:BPMN4sMLEvents}). The corresponding notation is depicted in Figures~\ref{fig:Events1} and~\ref{fig:Events2}.

\subsubsection{DataSourceEvent}
The \textbf{DataSourceEvent} is an \textit{Event} triggered by the presence of new raw data detected in the environment of the raw data provider. It is fired upon detection of the raw data. This event can be of type \textit{Start Event} and \textit{Intermediate Catch}. 
\begin{itemize}
\item \textbf{Example:} In an online mobile game a new match has just concluded. Data on each team, their actions taken, the final score as well as system information become available, triggering the \textit{DataSourceEvent}. Subsequently, the \textit{DataSourcingTask} is automatically run to ingest the raw data into the \textit{RawDataRepository}.
\end{itemize}

\subsubsection{RawDataUpdateEvent}
The \textbf{RawDataUpdateEvent} is an \textit{Event} triggered by an update of a \textit{RawDataObject} in the \textit{RawDataRepository}. It is fired as soon as the new raw data file is written to the storage and its identifier becomes available. This event can be of type \textit{Start Event} and \textit{Intermediate Catch}.

\subsubsection{FeatureSetUpdateEvent}
The \textbf{FeatureSetUpdateEvent} is an \textit{Event} triggered by an update of a \textit{FeatureSetObject} in the \textit{FeatureSetRepository}. It is fired as soon as the new feature set is written to the storage and its identifier becomes available. This event can be of type \textit{Start Event} and \textit{Intermediate Catch}.

\subsubsection{DatasetUpdateEvent}
\begin{figure}
    \centering
    \includegraphics[width=0.7\linewidth]{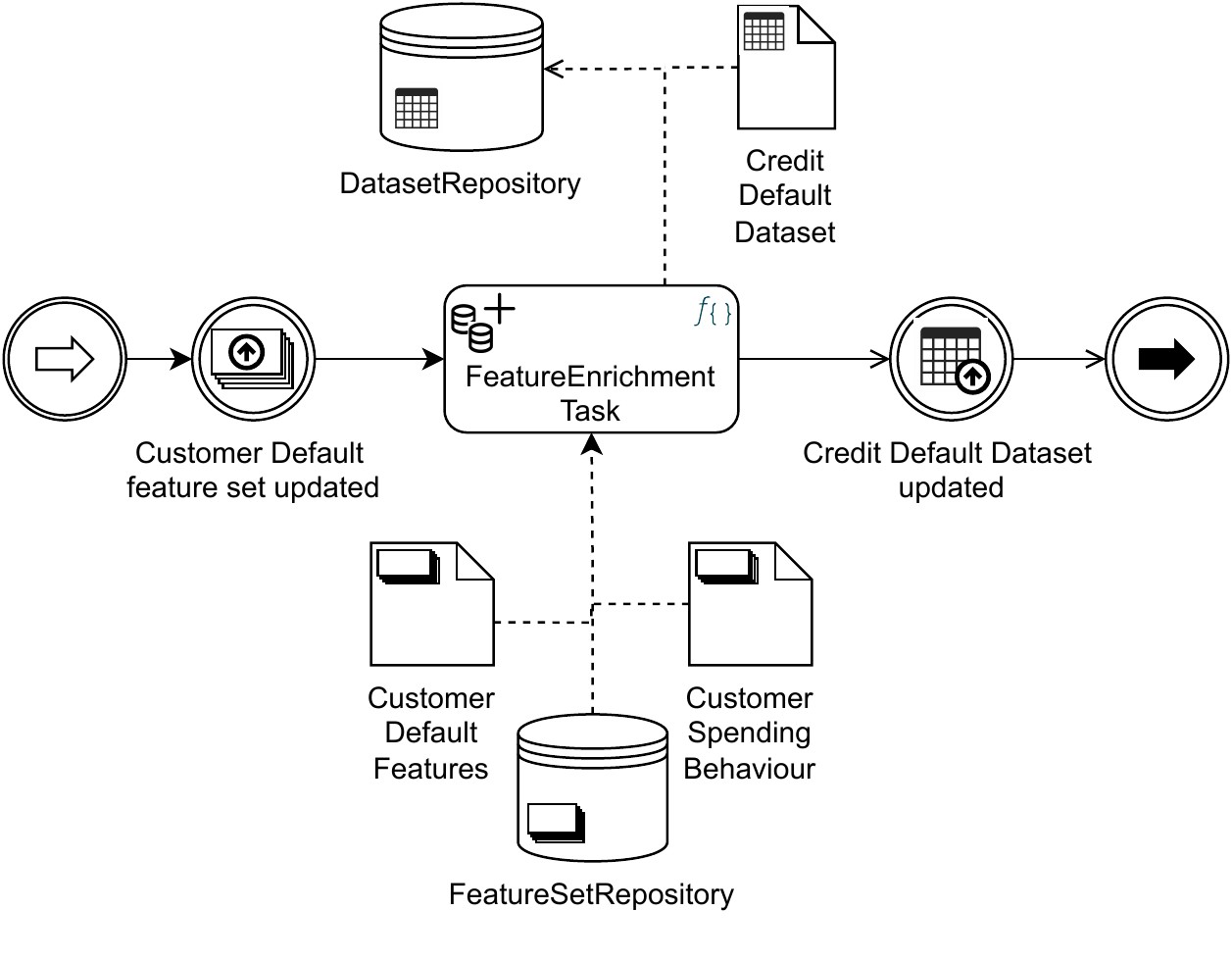}
    \caption[BPMN4sML workflow fragment of FeatureSetUpdateEvent, DatasetUpdateEvent, FeatureEnrichmentTask and corresponding data artefacts]%
    {BPMN4sML workflow fragment of FeatureSetUpdateEvent, DatasetUpdateEvent, FeatureEnrichmentTask and corresponding data artefacts\par \small }
    \label{fig:FeatureDataSetUpdateEvent}
\end{figure}

The \textbf{DatasetUpdateEvent} is an \textit{Event} triggered by an update of a \textit{DatasetObject} in the \textit{DatasetRepository}. It is fired as soon as the new dataset is written to the storage and its identifier becomes available. This event can be of type \textit{Start Event} and \textit{Intermediate Catch}. 
\begin{itemize}
\item \textbf{Example:} This example showcases the \textit{FeatureSetUpdateEvent} and \textit{DatasetUpdateEvent}, see Figure~\ref{fig:FeatureDataSetUpdateEvent}. Its logic can be transferred to the \textit{RawDataUpdateEvent} as well. Referring back to the loan application example, an updated feature set with the latest information on customers that received a loan and paid it back fully or defaulted becomes available, triggering the \textit{FeatureSetUpdateEvent} which initializes \textit{FeatureEnrichmentTask} that combines this feature set with another one on spending behaviour of these customers and writes it as a dataset to the \textit{DatasetRepository} triggering the \textit{DatasetUpdateEvent}. It informs that a new model can be trained.
\end{itemize}

\subsubsection{RequirementUpdateEvent}
The \textbf{RequirementUpdateEvent} is an \textit{Event} triggered by a change or update of a \textit{RequirementDocument} in the \textit{MetadataRepository}. It is fired as soon as the new requirement document presents itself in the process environment. This event can be of type \textit{Start Event} and \textit{Intermediate Catch}.

\subsubsection{DataDriftEvent}
\begin{figure}
    \centering
    \includegraphics[width=0.6\linewidth]{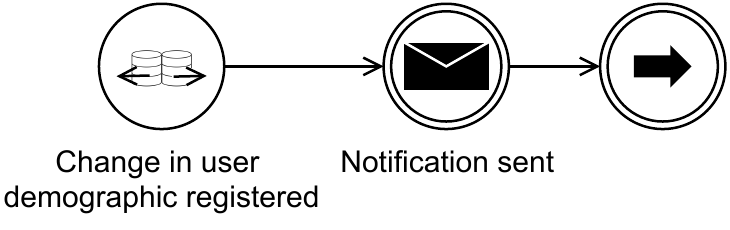}
    \caption[BPMN4sML workflow fragment of a DataDriftEvent of type Start]%
    {BPMN4sML workflow fragment of DataDriftEvent of type Start\par \small}
    \label{fig:DataDriftEvent}
\end{figure}

The \textbf{DataDriftEvent} is an \textit{Event} triggered by a detected increasing skew between the \textit{MLDatasetObject} used for learning and verifying the model and the \textit{InferenceRequestDatasetObject}, i.e. the data that the deployed ML solution receives for prediction. Typically, in consequence of a data drift the deployed model needs to be re-learned (i.e. trained or tuned) on the latest available \textit{TrainDatasetObject}. Possible intervention of a domain expert is necessary to update the data preparation tasks. The \textit{DataDriftEvent} can be of type \textit{Start} as well as of type \textit{Intermediate} (\textit{throw} and \textit{catch}). 
\begin{itemize}
\item An audio streaming media service provider trained a machine learning model to suggest songs to users based on songs that other users, who liked similar songs, listened to. After expanding its user base, a change in the user demographic occurs and also more senior users access the services. The song recommendation is no longer accounting for the new demographic. A \textit{DataDriftEvent} catches this change. Subsequently information are sent to a domain-expert to facilitate their intervention. Note that the event is of type \textit{Start} as the data drift initialises the process, see Figure~\ref{fig:DataDriftEvent}.
\end{itemize}

\subsubsection{ConceptDriftEvent}
The \textbf{ConceptDriftEvent} is an \textit{Event} triggered by a detected changing relationship between the explanatory variables and the target variable used for a \textit{MLModelObject} in a supervised learning setting. Typically, in consequence of a concept drift the referenced \textit{TrainDatasetObject} needs to be updated and the model needs to be re-learned (i.e. trained or tuned). The \textit{ConceptDriftEvent} can be of type \textit{Start} as well as of type \textit{Intermediate} (\textit{throw} and \textit{catch}). 
\begin{itemize}
\item \textbf{Example:} A monitoring service detects a concept drift which subsequently triggers an intermediate throwing \textit{ConceptDriftEvent}. The corresponding catch event registers the concept drift and initializes subsequent activities. 
\end{itemize}

\subsubsection{PerformanceDeficitEvent}
\begin{figure}
    \centering
    \includegraphics[width=0.8\linewidth]{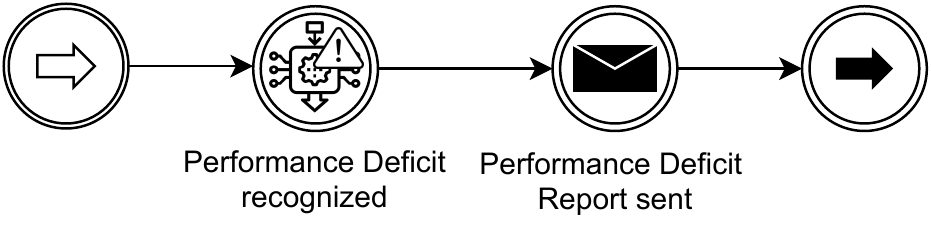}
    \caption[BPMN4sML workflow fragment of an Intermediate PerformanceDeficitEvent]%
    {BPMN4sML workflow fragment of an Intermediate PerformanceDeficitEvent\par \small }
    \label{fig:PerformanceDeficitEvent}
\end{figure}

The \textbf{PerformanceDeficitEvent} is an \textit{Event} triggered by a generated \textit{PerformanceDeficit} \textit{Document} that informs on what happened during the model learning phase that led to no model meeting the requirement constraints. The event can re-route an automated ML workflow to allow for intervention of a domain expert. The \textit{PerformanceDeficitEvent} is an \textit{Intermediate} event of type \textit{Catch}.
\begin{itemize}
\item \textbf{Example:} After an extensive model tuning operation, the ML pipeline is still unable to create a good-enough ML solution. Consequently, a \textit{PerformanceDeficitReport} is generated which is caught by a \textit{PerformanceDeficitEvent} and propagated to a domain expert to request for intervention. Note that in this case the \textit{PerformanceDeficitReport} does not have to be explicitly connected to the \textit{PerformanceDeficitEvent} as it is fired once the report presents itself in the ML workflow. The snippet containing the \textit{PerformanceDeficitEvent} is illustrated in Figure~\ref{fig:PerformanceDeficitEvent}.
\end{itemize}

\subsubsection{VerificationEvent}
The \textbf{VerificationEvent} is an \textit{Event} triggered by the presence of a newly \textit{trained} \textit{MLDocumentObject} in a \textit{ModelRegistry}. It can initialize a \textit{VerificationTask} or request for a confirmation by a domain expert. The \textit{VerificationEvent} can be a \textit{Start} event or an \textit{Intermediate} event of type \textit{Catch}.
\begin{itemize}
\item \textbf{Example:} A trained RandomForest model is written to the \textit{ModelRegistry} for subsequent access. Once registered a notification is produced informing about the existence of the new trained RF model. A \textit{VerificationEvent} catches the trained model ID and initializes a \textit{VerificationTask} to fire-proof the \textit{MLModelObject} prior to its deployment.
\end{itemize}

\subsubsection{VerificationFailureEvent}
The \textbf{VerificationFailureEvent} is an \textit{Event} triggered by a generated \textit{VerificationResult} \textit{Document} that informs on why a ML model verification by means of a \textit{VerificationTask} was unsuccessful. The event fires as soon as a new \textit{VerificationResult} presents itself in the process environment. The \textit{VerificationFailureEvent} is an \textit{Intermediate} event of type \textit{Catch}.
\begin{itemize}
\item \textbf{Example:} A tuned RandomForest model performed well throughout the ML learning phase and complied with the defined requirements when evaluating it on the \textit{ValidationDatasetObject}. After it is written to \textit{ModelRegistry} and the \textit{VerificationTask} is conducted referencing a \textit{VerificationDatasetObject} containing a data sample on which the RF model no longer performs sufficiently well. Consequently, a \textit{VerificationDocument} is generated and caught by a \textit{VerificationFailureEvent}. The report is then propagated as a \textit{Message} to a domain expert similar to the previous example of the \textit{PerformanceDeficitEvent}.
\end{itemize}

\subsubsection{DeploymentEvent}
The \textbf{DeploymentEvent} is an \textit{Event} that can be triggered by information of the \textit{ModelRegistry} such as the presence of a newly \textit{verified} \textit{MLDocumentObject}. Once the ID of the verified model is available the event is fired. The \textit{DeploymentEvent} can be of type \textit{Start} as well as of type \textit{Intermediate} (\textit{Throw} and \textit{Catch}).

\subsubsection{DeprecationEvent}
The \textbf{DeprecationEvent} is an \textit{Event} that can be triggered as response to information captured throughout the ML lifecycle. Once the identifier of the \textit{deployed} \textit{MLModelObject} is received, the event is fired. It can be used to initialize a \textit{DeprecationTask}. The \textit{DeprecationEvent} can be of type \textit{Start} as well as of type \textit{Intermediate} (\textit{Throw} and \textit{Catch}). 
\begin{itemize}
\item \textbf{Example:} In a running ML system a new XGBoost model is deployed that outperforms the currently operating RandomForest model. A \textit{MonitoringTask} identifies this situation and the corresponding ID of the underperforming model. The ID is propagated through a \textit{DeprecationEvent} of type \textit{Throw} and captured by a \textit{DeprecationEvent} of type \textit{Catch} which initializes the \textit{DeprecationTask}. 
\end{itemize}

\subsubsection{InferenceEvent}
\begin{figure}
    \centering
    \includegraphics[width=0.8\linewidth]{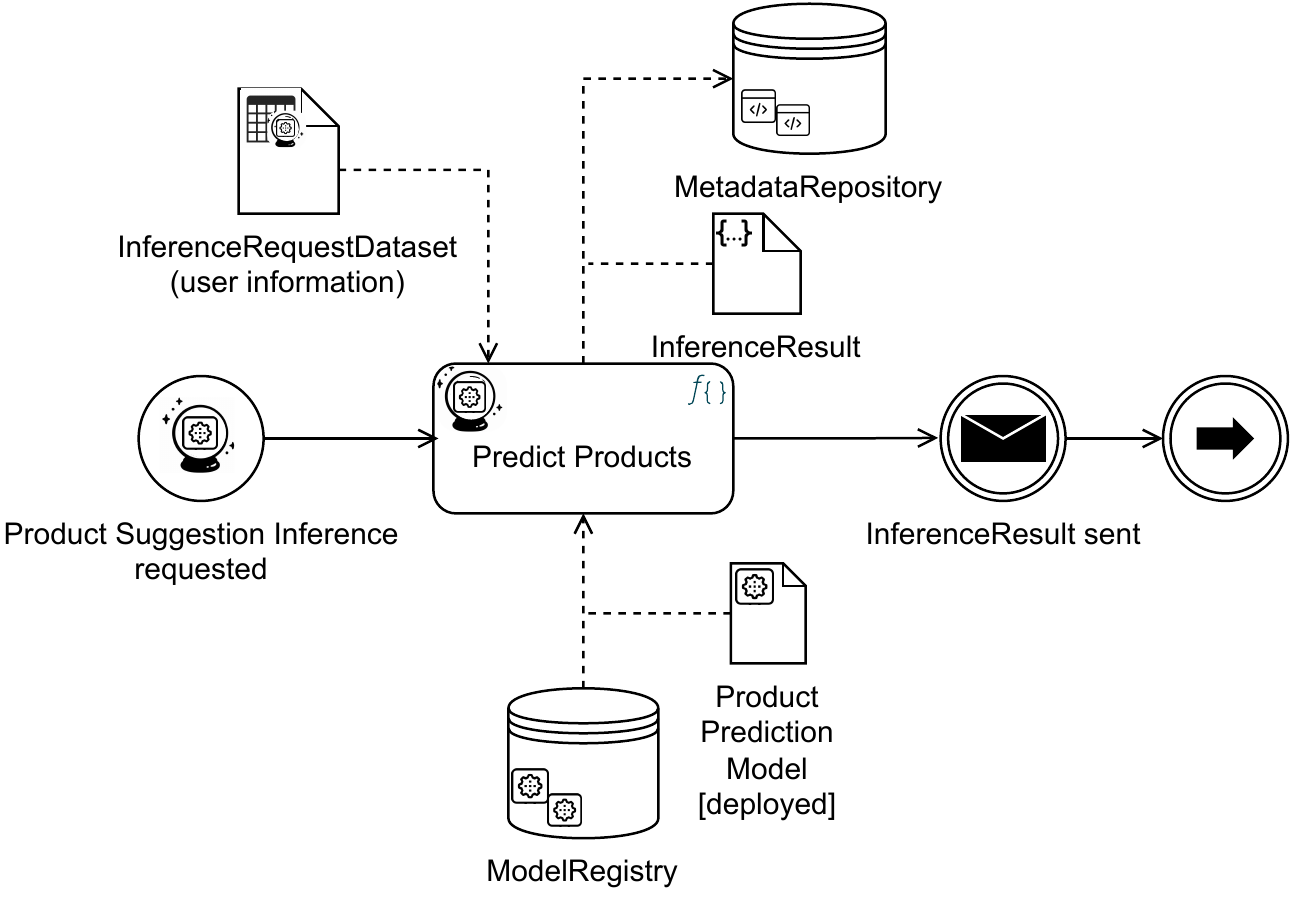}
    \caption[BPMN4sML workflow fragment of an InferenceEvent, InferenceTask and corresponding data artefacts]%
    {BPMN4sML workflow fragment of an InferenceEvent, InferenceTask and corresponding data artefacts\par \small }
    \label{fig:InferenceRequestEvent}
\end{figure}

The \textbf{InferenceEvent} is an \textit{Event} triggered by the presence of a new \textit{InferenceRequestDataSet}. The \textit{InferenceEvent} can be a \textit{Start} event or an \textit{Intermediate} event of type \textit{Catch}.
\begin{itemize}
\item \textbf{Example:} A user arrives on the front page of an online web-shop automatically producing a set of user specific datapoints. The \textit{InferenceEvent} captures the new \textit{InferenceRequestDataSet} and initializes the \textit{InferenceTask} of the recommendation engine to showcase products that the user is most likely to be interested in. Note that an Inference Start Event is used since the prediction process starts with the reception of the \textit{InferenceRequestDataSet}. The example is depicted in Figure~\ref{fig:InferenceRequestEvent}
\end{itemize}

\subsubsection{OperationDegradationEvent}
The \textbf{OperationDegradationEvent} is an \textit{Event} triggered by detected operational under-performance of a \textit{deployed} \textit{MLModelObject}. The event can be of type \textit{Start} as well as of type \textit{Intermediate} (\textit{throw} and \textit{catch}).  
\begin{itemize}
\item \textbf{Example:} An operational ML solution continuously lags behind in responding to inference requests. A \textit{MonitoringTasks} captures this behaviour when analysing the corresponding log files. Subsequently, an intermediate \textit{OperationDegradationEvent} is thrown. 
\end{itemize}

\subsubsection{JobOffloadingEvent}
The \textbf{JobOffloadingEvent} is an \textit{Event} triggered by \textit{JobData} specifying information about an offloaded job. The \textit{JobOffloadingEvent} is an \textit{Intermediate} event of type \textit{catch}.

%% file: chapters/6.Implementation.tex
The proposed BPMN4sML extension enables end-users to design serverless machine learning workflows based on an accepted process modeling standard. This helps in realizing the main objective of standardized modeling, analysis and communication of ML workflows in a technology independent and interoperable manner. To now also facilitate the serverless deployment orchestration of generated BPMN4sML diagrams in a technology-agnostic manner, TOSCA as an OASIS standard can be leveraged. As elaborated upon in section~\ref{TOSCASection}, TOSCA enables creation, automated deployment and management of portable cloud applications and supports declarative deployment modeling~\cite{DeploymentMetamodel}. Further, recent TOSCA extensions form first steps towards support for FaaS-based applications and their choreography~\cite{TOSCAServerless} as well as limited support for function orchestration~\cite{FaaSBPMNOrchestration}. A TOSCA deployment model in form of a service template realizes representation of service components as nodes as well as their relationships and configuration. Respective semantics of both nodes and relationships further define functionality, i. e. attributes, properties, requirements and capabilities and corresponding interfaces. Consequently, TOSCA is an ideal candidate solution that in combination with an orchestrator such as xOpera allows to execute and deploy modeled topologies on specified resources. The missing link between BPMN4sML model diagrams and TOSCA deployment models is a mapping on how to relate the BPMN4sML tasks, artefacts, events and control flow to corresponding TOSCA elements. 

To provide orientation, section~\ref{ProposedMethod} aggregates the entire process as a nominal sequence of activities starting with serverless ML workflow modeling using BPMN4sML and arriving at a deployment model realized by TOSCA.
Following, section~\ref{ApproachOverview} elaborates upon the choices made to realize respective mapping rules towards a potential conceptual model-to-model mapping. Subsequently, section~\ref{MappingRules} presents the identified mapping.

Note that the scope and functionality of the conceptual mapping and deployment modeling is not covering the full and automated conversion from BPMN4sML to TOSCA service templates. First, certain architectural design choices for deployment need to be specified by the modeler. Second, currently availble TOSCA node types and relationship types require extension for different kinds of event flow logic, function chaining and representation of offloaded services (ML platforms). This exceeds the scope and time of this work and thus only allows showcasing the mapping and transformation on a conceptual level and not on a fully implemented one for BPMN4sML models which results in a limitation and opportunity for future research.

\section{Summarized Method Proposal}
\label{ProposedMethod}
\begin{figure}
    \centering
    \includegraphics[width=1\linewidth]{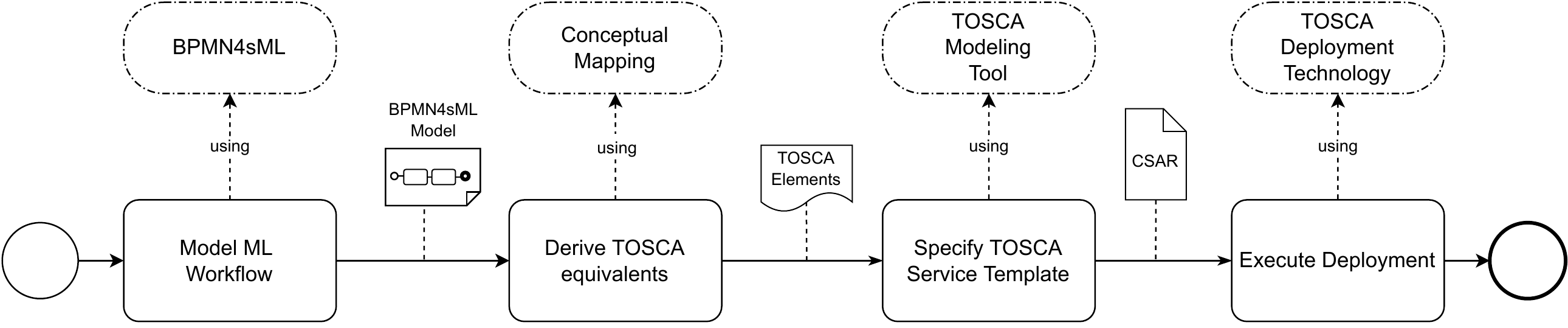}
    \caption[Method for technology independent and interoperable modeling of ML workflows \& their serverless deployment orchestration]%
    {Method for technology independent and interoperable modeling of ML workflows \& their serverless deployment orchestration \par \small }
    \label{fig:Method}
\end{figure}

When consolidating the developed BPMN4sML extension with the possible conceptual mapping to corresponding TOSCA elements, an initial proposal towards a method for the technology independent and interoperable modeling of ML workflows and their serverless deployment orchestration can be derived for ML engineers to draw from. A full (automated) realization of the proposed method exceeds the scope, it is however described conceptually to indicate future research directions. Figure~\ref{fig:Method} explains the process visually.

Initially, a machine learning workflow is specified by applying the BPMN4sML extension resulting in a BPMN4sML model diagram. Any modeling tool supporting BPMN can be used to achieve this step - during this work we refer to \textit{draw.io / diagrams.net} as an open-source tool for the modeling of BPMN workflows. An initial version of the BPMN4sML notation can be loaded into the software to enable modeling.

By applying the conceptual mapping, TOSCA elements are derived for the respective BPMn4sML \textit{Tasks}, repositories and workflow logic. The identified elements conceptually represent the TOSCA Service Template which can be realized and further specified using a TOSCA modeling tool such as Winery~\cite{Winery}. Extracting the Service Template as a CSAR file ultimately enables the deployment of the modeled ML workflow via a TOSCA compliant deployment orchestration technology such as xOpera.

If the workflow logic is to be realized as a function orchestration, a version of the function orchestration generator by Yussupov et al.~\cite{FaaSBPMNOrchestration} can theoretically be integrated to derive the orchestration logic automatically from the BPMN4sML model diagram. However, this requires further extension to allow its application on a full BPMN4sML workflow to among others differentiate between \textit{Tasks} realized as serverless function and other Tasks and artefacts and is therefore left to future research. Instead, if a function orchestration shall be derived from a given BPMN4sML model, the ML engineer may define the function orchestration file themselves when following this initial version of the proposed method.

In addition to integrating a function orchestration generator, the proposed sequence of activities can be further automated in future research. The conceptual mapping can be codified by referencing the BPMN4sML diagram in a XML format and extending, thereby directly deriving the topology template. If also properties of the respective TOSCA node and relationship types are specified during this process, a complete Service Template can be derived and eventually deployed.

\section{Conceptual Mapping Approach Overview}
\label{ApproachOverview}
To derive a conceptual mapping for model to model (M2M) transformations various approaches have been put forward such as model transformation languages~\cite{M2MATL}, extrapolation from given examples~\cite{M2MbyExample} or automated semantic and syntactic checking measurements to derive meaning and relations~\cite{M2MAutomated}. 

Fundamentally, to ensure that a proper model to model transformation is possible one requires a source metamodel as well as a target metamodel that define the concepts, semantics and syntactic rules, i.e. the languages' modeling primitives~\cite{M2MSystematicTransformation}. In case of this work, the BPMN4sML metamodel represents the source and the TOSCA metamodel the target which are to be related. 
Mappings that are built on the metamodel level of both languages can then be applied to the model level to transform actual model instances from source to target~\cite{M2MSystematicTransformation}. 

To arrive at a model to model transformation one may build conceptual mappings and subsequently define transformation rules that select which mapping to apply for a given case by either analysing and relating the respective elements of each modeling language manually or through pre-defined detecting measurements~\cite{M2MAutomated}. Wang et al. propose an automated detection approach based on semantic and syntactic checks to derive model mappings that are built among the models' elements and the group of properties they contain~\cite{M2MSystematicTransformation}. Between models each possible pair and combination is iterated over so that the association (pair of elements) with the highest identified match (semantic \& syntactic value) can be considered a mapping. Adopting such an approach or the one brought forward by Jouault et al~\cite{M2MATL}  exceeds the bounds of this thesis. Instead the former, i.e. a manual analysis, between elements and properties of each metamodel, is applied to generate a conceptual mapping. To do so, the syntax, describing the language's rules and relationships between available elements, as well as the semantics that establish their meaning in the domain context are considered.


\section{Conceptual Mapping from BPMN4sML to TOSCA}
\label{MappingRules}
As explained, BPMN4sML processes (ML workflows) represent model instances conforming to the BPMN4sML metamodels previously introduced and now considered as source. TOSCA topologies represent the target model and conform to the TOSCA metamodel~\cite{TOSCAMetamodel}. Depending on implementation needs and design choices, different TOSCA realizations can be generated from the same source model (ML workflow). 

Ideally, the ML engineer has a certain freedom of choice when transforming a BPMN4sML workflow to TOSCA. For instance, one may wish to translate a workflow of \textit{FaaSTasks} (or even \textit{OffloadedTasks} to ML platforms and related services) into one or several function orchestrations. Depending on cloud provider also other services can be included in the orchestration's definition. This requires to derive a function orchestration file from the modeled BPMN4sML workflow. The format and type of function orchestrations depend heavily on the cloud service provider. Further, function orchestrations necessitate the representation of the orchestration engine also on the level of the TOSCA topology template as respective TOSCA node types. A recent approach has been brought forward by Yussupov et al.~\cite{FaaSBPMNOrchestration} which presents both a first method to derive function orchestration files from limited BPMN fragments as well a modeling a limited number of the corresponding TOSCA node types. 

When instead the same workflow shall be realized as a function choreography, a different TOSCA topology template needs to be identified that accounts for the workflow logic by properly associating event providers with corresponding event consumers. Additionally, TOSCA allows practitioners to design their own node types and relationship templates. Consequently, various solutions to represent event-driven logic have been developed. Wurster et al.~\cite{TOSCAServerless} account for 
the modeling of events and event flow by defining event producers and consumers in node templates and their relationship as TOSCA relationship template. This can take the shape of a cloud service that produces an event and a serverless function or other cloud service that consumes it. The event flow is then modeled by means of derivations from the normative TOSCA relationship type \textit{connectsTo} between the two nodes. An example can be a BPMN4sML \textit{FeatureSetRepository}, realized as an \textit{AWS S3} storage, that produces a \textit{FeatureSetUpdateEvent} when a new file is added or updated, realized as a TOSCA relationship of the normative type \textit{connectsTo} and specified as a \textit{S3 Event} relationship via a \textit{PUT} property which in turn connects to the invoked \textit{FaaSTask} (TOSCA node), as for instance a BPMN4sML \textit{PreprocessingTask}. However as the authors elaborate, such an approach limits the specification of more complex workflow logic with chained functions. One may resolve this by leveraging a notification or queuing service to better connect functions. This however requires a different definition of node types and relationship types, i.e. a TOSCA extension, and is thus left to future research.

Another realization of event-driven logic can be the definition of an event consumer as a new TOSCA node type that directly incorporates the event-driven characteristics, i.e. specifies the event provider. An example can be found in TOSCA blueprints published as part of the RADON project~\cite{RADON} which define for instance a \textit{GoogleCloudBucketTriggeredFunction}~\cite{RadonParticles}, i.e. a Google Cloud serverless function which gets invoked upon changes in a specified Google Cloud bucket, or an \textit{AzureHttpTriggeredFunction}. This however may also limit the extend of modeling function chains on a TOSCA level and requires potential extension of node and relationship types (e.g. by possibly defining a \textit{FunctionTriggeredFunction} Node Type).

Further, both, function orchestrations via orchestration engines and function choreographies via event-flows, may be combined within the TOSCA realization of one BPMN4sML model. For instance, chaining a set of tasks of a sub-process may be simplified by an orchestration. Its result could then produce an event that triggers functions, services or other orchestrations representing subsequent tasks in the BPMN4sML workflow. Modeling such a logic on a TOSCA level also requires extension. Moreover, offloaded tasks that are for instance realized as services provided by a machine learning platform such as AWS SageMaker or Azure Machine Learning need to be incorporated as part of the workflow. Certain function orchestration engines provide direct interfaces to services of the same cloud provider, for instance Amazon Web Services or Google Cloud Workflow. Otherwise, the offloaded task needs to be included in an event-driven fashion within the deployment model. 

To abstract from these possibilities and allow for an initial solution candidate, a conceptual mapping shall first and foremost relate the high-level BPMN4sML elements (mainly tasks and data artefacts) to their abstract TOSCA element counterparts. The conceptual mapping is proposed as one of potentially several different possible mapping realizations. A focus is placed on the utilization of function orchestrations where possible to represent process logic and workflow patterns as TOSCA only provides limited support to model function choreographies. Table~\ref{tab:conceptualMapping} describes the identified explicit and implicit mappings between the metamodels. In case a mapping implicates a potential TOSCA extension it is explicitly mentioned - however solely to indicate future research directions to build on top of this proposition as a new TOSCA extension exceeds this work's scope. Note that specifications such as including TOSCA requirements and capabilities of TOSCA Node Types are not explicitly mentioned but can be considered to further structure semantics, i.e. connections between nodes. Further note that BPMN4sML data artefacts such as \textit{MLModelObjects} are exclusively used for explanation purposes in a BPMN4sML model and do not have a counterpart in respective TOSCA models.

\begingroup
\footnotesize  
\setlength\tabcolsep{3pt}
\setcellgapes{2pt}
\makegapedcells
\setlength\LTcapwidth{\dimexpr\linewidth-3pt}
   \begin{xltabular}{\linewidth}{>{\bfseries}P{1cm} L L L}
\caption{Explicit and implicit mappings between BPMN4sML and TOSCA metamodels}
\label{tab:conceptualMapping} \\
    \Xhline{1pt}
\thead{Nr.}	& \thead{BPMN4sML}  & \thead{TOSCA \\ (explicit mapping)}& \thead{TOSCA \\ (implicit mapping)} \\
\endfirsthead
\Xhline{1pt}
\thead{Nr.}	&	\thead{BPMN4sML}  & \thead{TOSCA \\ (explicit mapping)}& \thead{TOSCA \\ (implicit mapping)}  
\endhead
    \Xhline{0.8pt} 
1	&	FaaSTask (and inherinting elements such as VotingTask etc.)	&	TOSCA Node Type (e.g. technology-specific AwsLambdaFunction Node Type) 	&		\\	\hline
2	&	FaaSTask \textit{name/configuration} ; OffloadedTask \textit{script}	&	TOSCA Node Properties (translates to any of the available node properties - for Function Node Type e.g. runtime, memory, timeout and function provider specific properties etc.)	&		\\	\hline
3	&	FaaSTask \textit{platform} / OffloadedTask \textit{offloadingTechnology}	&		&	TOSCA Node Type + Relationship Type to the actual Node that the BPMN4sML task represents (e.g. cloud provider / edge Node as Node Type + hostedOn Relationship Type to the hosted Node)	\\	\hline
4	&	FaaSTask \textit{script}	&	TOSCA Node Property (specifying a path to a zip file attached as TOSCA Implementation Artefact)   	&		\\	\hline
5	&	Instance of FaaSTask, OffloadedTask (or subclasses) 	&	TOSCA Node Template with values for node properties	&		\\	\hline
6	&	OffloadedTask (\textit{MLPlatform})	&	TOSCA Node Type (translates to different Types depending on implementation such as 1) edge Node Type for application component on edge; or 2) ML cloud service Node Type for ML platform e.g. AwsSageMaker - note that cloud service specific ML platform Node Types currently have no TOSCA counterpart and require extension) 	&		\\	\hline
7	&	DataRepository ; MetadataRepository ; LogStore ; ModelRegistry	&	TOSCA Node Type (e.g. AWS DynamoDB Table, AWS S3 Bucket, Google Cloud Bucket etc.); Repositories and Registries can potentially be realized as one node with different directories  	&		\\	\hline
8	&	\textit{Platform} (DataRepository ; MetadataRepository ; LogStore ; ModelRegistry)	&		&	TOSCA Node Type + Relationship Type (hostedOn) to the actual Node representing the BPMN4sML data artefact - similar to mapping 4	\\	\hline
9	&	BPMN DataAssociation (connecting Tasks with DataStores)	&	TOSCA Relationship Type (connectsTo)	&		\\	\hline
10	&	BPMN4sML Event	&	Conditional semantic equivalence with TOSCA Relationship Type or TOSCA Node Type (depends on actual implementation of event flow and type of BPMN4sML event - difference between function orchestration and event-driven; e.g. \textit{DatasetUpdateEvent} can equal \textit{connectsTo} Relationship Type with S3 trigger specification or Node Type such as \textit{S3TriggeredFunction}) - see Wurster et al.~\cite{TOSCAServerless} and Radon Particles	&	Potentially requires additional TOSCA Node Type to realize TOSCA Relationship Type (e.g. Node for Bucket or Notification service) \\	\hline
11	&	BPMN4sML process (connecting Tasks with Tasks + workflow patterns)	&	TOSCA Node Type (e.g. for function orchestration realized as abstract Workflow Node Type or as technology-specific AwsSFOrchestration or AzureOrchestrating Function Node Type); Function Orchestration further requires an orchestration file that can be derived from a BPMN process model as shown by Yussupov et al.~\cite{FaaSBPMNOrchestration}; Alternative: specify workflow logic as event-driven via Relationship Types or Node Types but requires TOSCA extension	&	In case of function orchestration: TOSCA Function Orchestration Node Type connects via TOSCA Relationship Type to the orchestrated Nodes, can require further Node Types for realization of Function Orchestration Node Type depending on technology provider - see Yussupov et al.~\cite{FaaSBPMNOrchestration}; Also necessitates a hostedOn Relationship Type to the cloud platform represented by a Node Type (similar to mapping 4)	\\ \Xhline{1.2pt}
\end{xltabular}
\endgroup

\begin{figure}[!b]
    \centering
    \includegraphics[width=1\linewidth]{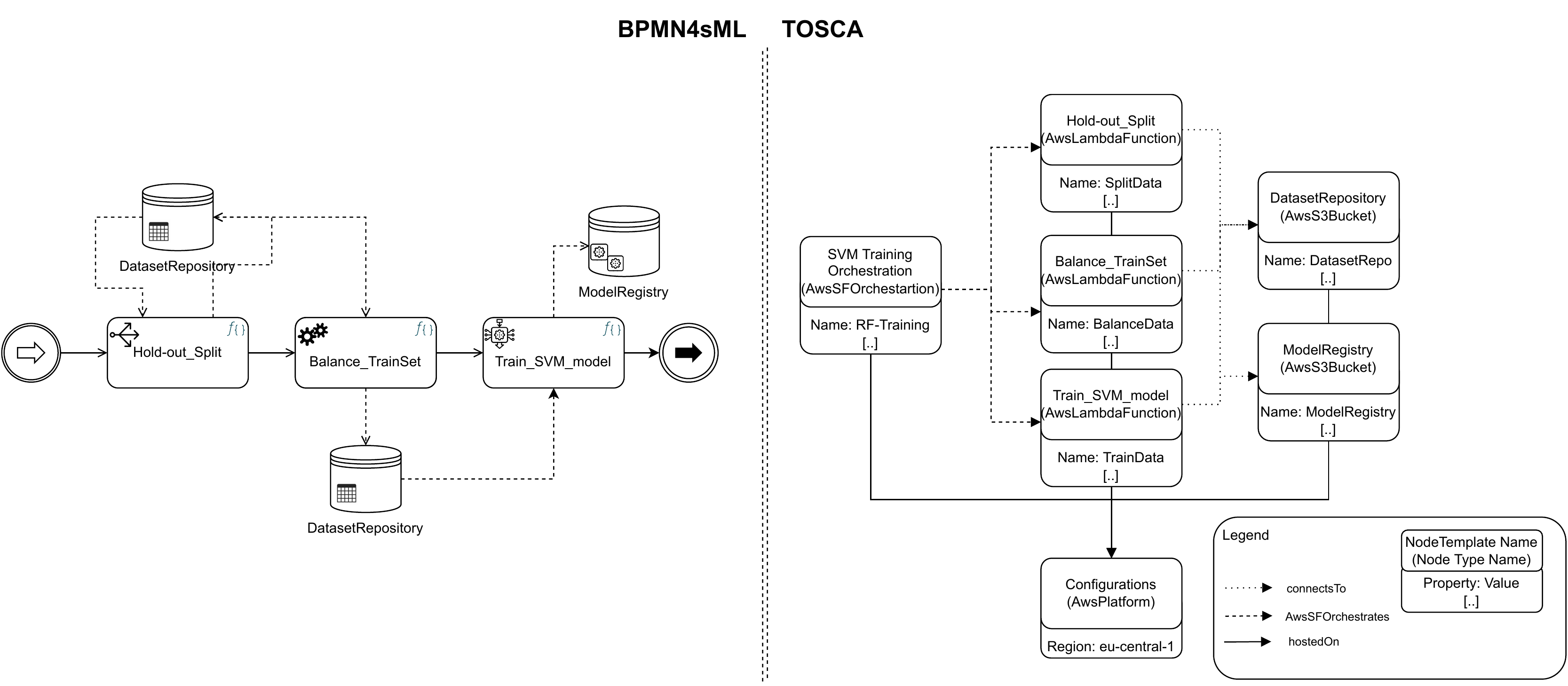}
    \caption[Mapping example of a BPMN4sML workflow fragment (left) to a TOSCA deployment model (right)]%
    {Mapping example of a BPMN4sML workflow fragment (left) to a TOSCA deployment model (right)}
    \label{fig:implementation_example}
\end{figure}

To illustrate the proposed mapping, we consider an abstract ML scenario. The BPMN4sML workflow fragment depicts a chain of tasks connecting the data management phase with the mode learning phase. A dataset is split into a training and validation set, the training set is further preprocessed to balance the data and a random forest model is learned. For this illustration the workflow is represented as a simple sequence of BPMN4sML Tasks that are realized as serverless functions and make use of a data repository and a model registry. To highlight that the conceptual mapping does not necessitate any other data artefacts such as the modeling of a training dataset or ML model, we omit it in the diagram. The BPMN4sML workflow fragment and the correspond TOSCA deployment model are shown in Figure~\ref{fig:implementation_example}.

Next, the mapping towards respective TOSCA nodes and relationships is illustrated. For this case, we want to implement the BPMN4sML model as a function orchestration on Amazon Web Services, making use of StepFunctions. Accordingly, each Task corresponds to a TOSCA Node Type of type \textit{AwsLambdaFunction}. Further, the dataset repository and model registry correspond to a TOSCA Node Type of type \textit{AwsS3Bucket}. The whole workflow fragment logic is captured within the StepFunction definition which is not modeled in this example but is attached as deployment artefact to the TOSCA Node Type of type \textit{AwsSFOrchestration}. The Relationship Types correspond to the BPMN4sML connection arcs. An \textit{AwsSFOrchestrates} TOSCA Relationship Type connects the \textit{AwsSFOrchestration} node to each task node. This nodes can be connected via TOSCA \textit{Requirements} and \textit{Capabilities}. The {AwsSFOrchestration} node implements a requirement as \textit{Orchestrator} whereas the \textit{AwsLambdaFunction} nodes have the corresponding capability \textit{Orchestrated}. 
Further, the access of the BPMN4sML tasks to the respective data repository and model registry corresponds to a TOSCA Relationship Type of type \textit{connectsTo}. Similar requirements and capabilities are implemented for the respective Node Types. Finally, given that the workflow model shall be deployed on AWS, an additional TOSCA Node Type as \textit{AwsPlatform} needs to be implicitly created that implements the capability \textit{Container} and can be connected to the corresponding requirement of all other nodes.

%% file: chapters/7.IllustrativeUseCase.tex
To further validate the proposed BPMN4sML extension artefact and the conceptual mapping towards deployment, two illustrative use cases are referenced.
A modeling example inspired by Alipour et al.~\cite{UseCaseOnlineML} is considered that encompasses a simple yet convincing ML application, showcasing the functionality and feasibility of BPMN4sML to describe ML workflows. The authors present a self-managing service architecture for cloud data services that leverages serverless functions and machine learning models to predict future workload of said data services. The prediction informs decision-making for better resource-provisioning. Section~\ref{UseCase1} elaborates on the setting of the use case and presents the corresponding BPMN4sML diagram.

Next to validating the potential of representing serverless ML workflows through BPMN4sML, a minimalistic example illustrating the application of the established conceptual mapping to derive a TOSCA service template is demonstrated in section~\ref{UseCase2}. The example references a simplified instance of a home credit default risk prediction challenge published on Kaggle by the Home Credit Group~\cite{HomeCreditGroup}. Note that at the moment of writing, the respective available Ansible roles corresponding to the available TOSCA node Types and Relationship Types provided by Yussupov et al.~\cite{FaaSBPMNOrchestration} that this work draws from are not fully aligned with the updates on serverless functions made by the respective cloud providers and result in erroneous deployment orchestration. Updating the Ansible Roles as well as TOSCA Node Types exceeds the scope and is left to future research.

\section{Illustrative Example 1 : The Case of Self-managing Service Architectures}
\label{UseCase1}

\subsection{ML Application Architecture}
\begin{figure}
    \centering
    \includegraphics[width=1\linewidth]{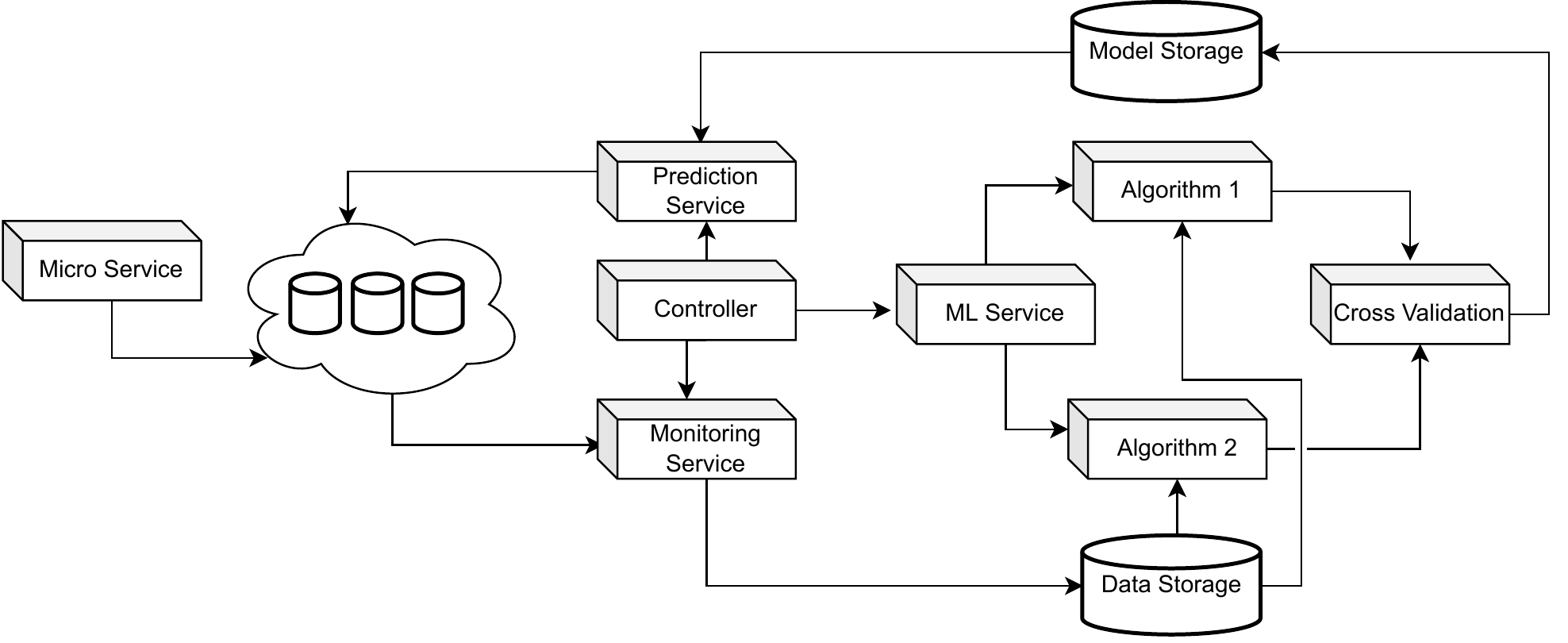}
    \caption[Illustrative example 1 - Conceptual service architecture for online machine learning]%
    {Illustrative example 1 - Conceptual service architecture for online machine learning \par \small Example adjusted and drawn from Alipour et al.~\cite{UseCaseOnlineML}. }
    \label{fig:MLOnlineExample}
\end{figure}

The referenced high-level prediction service architecture involves six components, see Figure~\ref{fig:MLOnlineExample}: 
\begin{itemize}
\item External monitoring - creates log files on metrics on resource demand from data service and writes them to a storage; translates to an external monitoring service that tracks metrics and generates log files and a \textit{DataSourcingTask} to source these files and write them to a storage
\item Controller - periodically invokes the different services (Monitoring to source data, ML services for model updates, invocation of Prediction service); translates to the BPMN4sML model diagram
\item Storage -  stores sourced monitoring data that serve as ML model training samples; translates to \textit{DatasetRepository} 
\item ML service - wrapper for offloaded training jobs on ML platform. Note that the proposed implementation leverages Amazon Machine Learning which has been replaced by AWS SageMaker to provide easier access and direct endpoints; translates to offloaded \textit{TrainingTask} and an optional \textit{OffloadingJobEvent} depending on service provider (i.e. if ML platform allows direct integration with serverless functions or communication via notification services)
\item Cross-validation service - estimates a robust perfomance score of each algorithm through resampling, i.e. cross-validation, and writes learned models to a model storage (similar logic as ML services by referencing a ML platform); roughly translates to an offloaded \textit{EvaluationTask} that runs a resampling strategy to produce a robust performance estimation and a \textit{ModelRegistry}
\item Prediction service - performs inference on real-time workloads to predict future workloads in scheduled intervals by selecting the appropriate model and applying it; translates to an optional \textit{ModelSelectionTask} and an \textit{InferenceTask}.
\end{itemize}

\subsection{Representation as a BPMN4sML Workflow}
\begin{sidewaysfigure}
    \centering
    \includegraphics[width=0.8\linewidth]{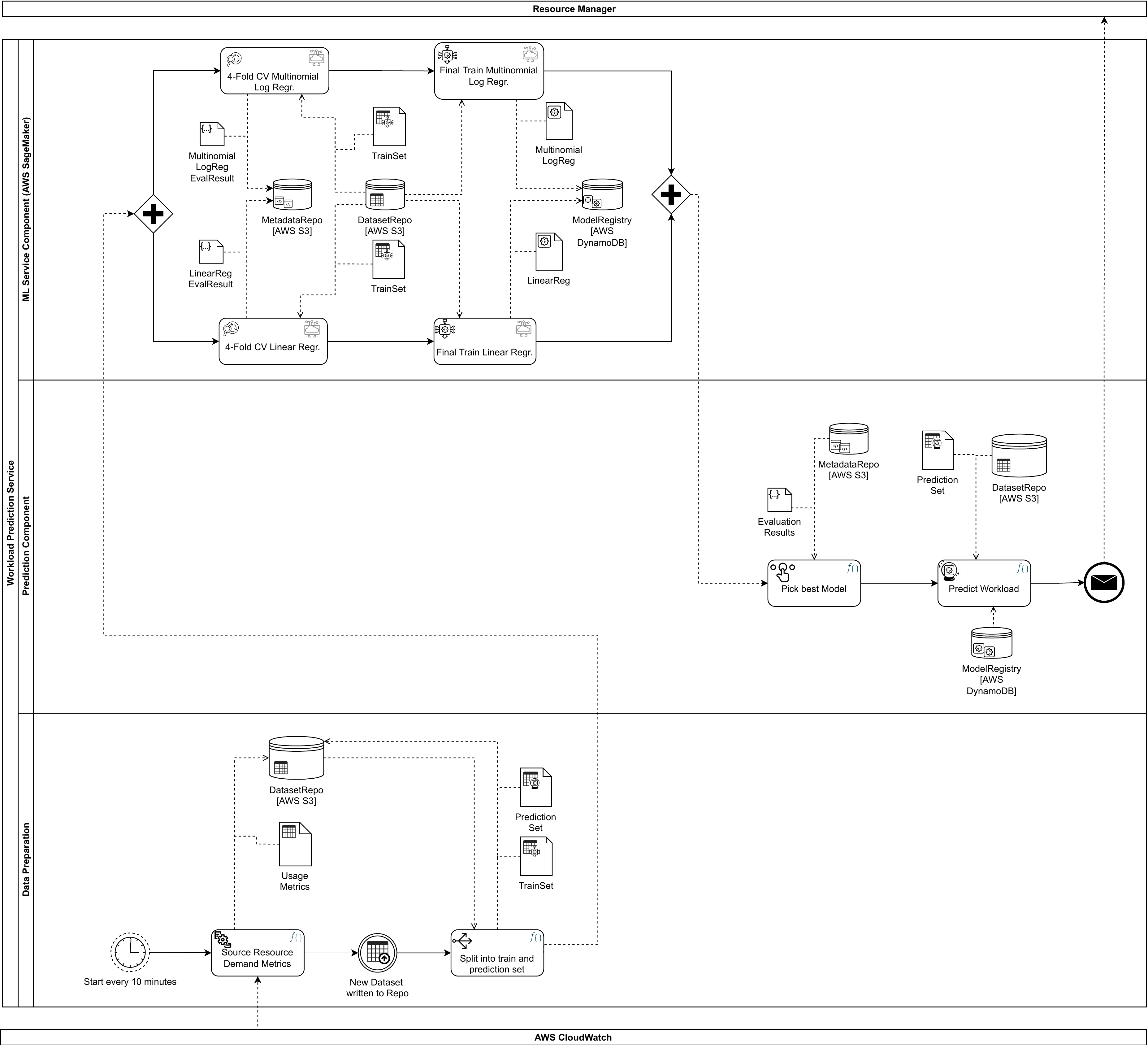}
    \caption[Illustrative example 1 - Referenced use case depicted as BPMN4sML model]%
    {Illustrative example 1 - Referenced use case depicted as BPMN4sML model \par \small }
    \label{fig:BPMN4sMLOnlineMLExample}
\end{sidewaysfigure}

For this modeling example, the referenced general service architecture is specified according to the explanation by Alipour et al.~\cite{UseCaseOnlineML}. Two ML models, a multinomial logistic regression and a linear regression, are trained and compared. 
The scheduling interval for sourcing the data and retraining the models is set to 10 minutes and the platform of choice is Amazon Web Services. The first half of the monitoring service to create log files with is AWS CloudWatch similar to the authors' choice and kept outside of the actual ML workflow. Some functionality is adapted to highlight the ML workflow fragment, to account for developments on ML platforms and to clarify the training and evaluation routine - namely, the training and cross validation service are merged into one overall evaluation service that runs a 4-fold cross validation. To account for proper machine learning practice, a final model is created based on the entire training dataset that was used beforehand for the cross-validation in order to make use of the largest dataset possible. Note that with the use of AWS SageMaker no explicit \textit{JobOffloadingEvent} needs to be specified as SageMaker can itself communicate with endpoints, produce (serverless) events and propagate results. Also note that in this use case a separate model verification phase and deployment phase are omitted. Instead, the final models are directly made accessible via the model registry which is also modeled in the transformed BPMN4sML workflow, see Figure~\ref{fig:BPMN4sMLOnlineMLExample}. Further note that the original implementation does not specify a data preprocessing operation apart from the data sourcing task as it is assumed that the sourced dataset comes in a cleaned and digestable format. In the transformed BPMN4sML model, a \textit{DataSplitTask} is added to highlight the different datasets - one used to train the models with and one (the logs of the last minutes) to predict the future workload. As the focus lies on the machine learning section, the subsequent provisioning operations are hidden within a closed \textit{resource manager} that adjusts the required resources for the data services based on the latest inference result.
In this modeling example the different lanes depict the services specified by Alipour et al.~\cite{UseCaseOnlineML}, see Figure~\ref{fig:MLOnlineExample}. The \textit{Controller} service is no longer necessary as the modeled workflow replaces its logic. 

\subsection{Evaluation of BPMN4sML's Modeling Capability}
The modeled online ML scenario demonstrates the application of BPMN4sML notations and semantics to design ML workflows. All services, data artefacts and logic described by Alipour et al. can be modeled using the extension. The referenced BPMN4sML elements are summarized in table~\ref{tab:usecase1Eval}.

\begingroup
\footnotesize  
\setlength\tabcolsep{3pt}
\setcellgapes{2pt}
\makegapedcells
\setlength\LTcapwidth{\dimexpr\linewidth-3pt}
    \begin{xltabular}{\linewidth}{L L L }
\caption{Summary of referenced BPMN4sML elements for modeling of illustrative use case 1}
\label{tab:usecase1Eval} \\
    \Xhline{1pt}
\thead{BPMN4sML Element}  & \thead{Represented Functionality}& \thead{Connected Elements} \\
\endfirsthead
\Xhline{1pt}
\thead{BPMN4sML Element}  & \thead{Represented Functionality}& \thead{Referenced Data Artefacts} 
\endhead
    \Xhline{0.8pt} 
DataSourceTask (FaaSTask)	&	Partial Monitoring Component - reads raw data from AWS CloudWatch and writes it as DatasetObject to a DatasetRepository	&	RawDataSetObject, DatasetObject, DatasetRepository \\ \hline
DatasetUpdateEvent	& Partial Controller Component - initiates the subsequent activities once a new dataset has been written to storage	& No direct connection but triggered upon update of DatasetRepository\\ \hline
DataSplitTask (FaaSTask)	&	Added functionality for clarification purposes	&	DatasetRepository, TrainDatasetObject, InferenceRequestDatasetObject \\ \hline
EvaluationTask (OffloadedTask)	&	ML Service Component - estimates a robust performance score of the considered Learners to allow for later selection	& EvaluationResult, MetadataRepository, DatasetRepository, TrainDatasetObject \\ \hline
TrainingTask (OffloadedTask)	&	ML Service Component - trains the final model on the full TrainDatasetObject	&	DatasetRepository, ModelRegistry, TrainDataset, MLModelObject \\ \hline
ModelSelectionTask (FaaSTask)	&	Inference Service Component - compares the estimated performance scores and chooses the best fitting model for this cycle of predictions	&	MetadataRepository, EvaluationResults \\ \hline
InferenceTask (FaaSTask)	&	Inference Service Component - takes the best available model and predicts future workload	&	DatasetRepository, InferenceRequestDatasetObject, ModelRegistry, implicitly MLModelobject \\

\Xhline{1.2pt}
    \end{xltabular}
\endgroup

In comparison, considering the referenced ML workflow, certain challenges arise if it was to be represented solely by conventional BPMN 2.0 notation. The available BPMN 2.0 Task types do not allow for differentiation of activities executed by serverless functions or via offloading. As previously explained, the ScriptTask is inappropriate and the ServiceTask cannot fully represent serverless characteristics. Further, instead of modeling activities via generic BPMN tasks, the semantic of the BPMN4sML specific tasks helps in clarifying their responsibility within the ML workflow. Next, the modeling of domain-specific data artefacts can hardly be achieved by abstract DataObjects which therefore necessitates new elements such as EvaluationResults, different DatasetObjects  and MLModelObjects. Also the new repositories and registries, i.e. ModelRegistry, MetadataRepository, DatasetRepository, help in explicitly associating the modeled elements with their exact counterpart in a ML workflow. Reviewing this subset of requirements based on the modeled scenario, it becomes clear that 1) a modeling language extension is needed and 2) BPMN4sML is a step in the right direction towards realizing an extension that covers the entirety of the ML domain. 

Nevertheless, while modeling the workflow it is noticeable that the more complex the ML workflow gets, the more cluttered the BPMN4sML model may become. Especially for ML workflows, complex set-ups involving a vast number of potential models and data artefacts are norm rather than exception. The modeler can mediate this issue for instance by not explicitly modeling all data artefacts as was done in this scenario but only the required repositories. Also, as common for BPMN modeling, (closed) sub-processes can be leveraged to design a ML workflow over several levels of complexity. Note that not all proposed BPMN4sML elements can be covered in an illustrative use case of this work. It is however assumed that based on the preceding requirement analysis and argumentation as well as modeled workflow snippets and textual examples, all elements' existence is justified and can potentially be leveraged for modeling in other scenarios.




\section{Illustrative Example 2 : The Case of Home Credit Default Risk Prediction}
\label{UseCase2}

\subsection{Simplified BPMN4sML Machine Learning Pipeline}
\begin{figure}
    \centering
    \includegraphics[width=1\linewidth]{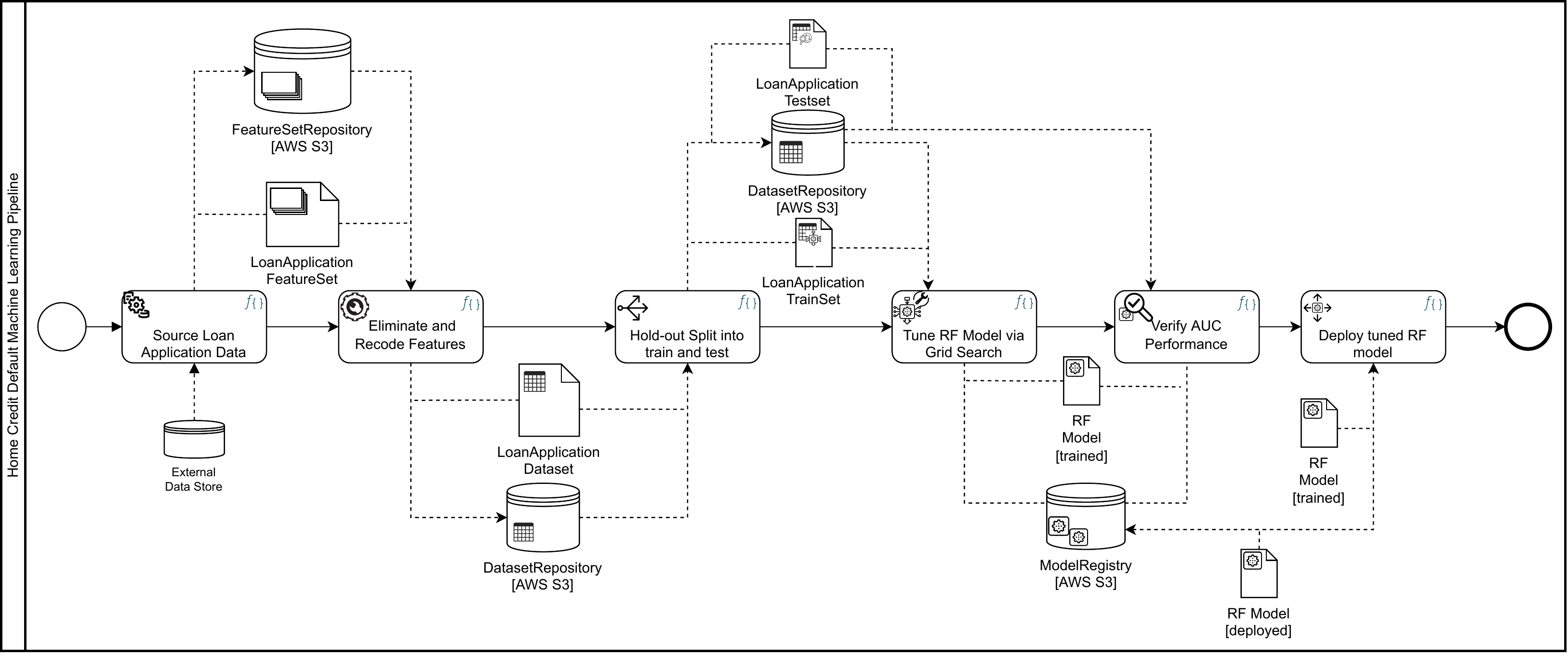}
    \caption[Illustrative Example 2 - simplified BPMN4sML model for home credit default machine learning pipeline]%
    {Illustrative Example 2 - simplified BPMN4sML model for home credit default machine learning pipeline\par \small }
    \label{fig:usecase2BPMN4sML}
\end{figure}

In this illustrative example we limit ourselves to a machine learning pipeline that is already agreed upon, that is to say that it assumes a chosen set of steps to follow and no extensive data analysis or experimentation needs to be performed, reason being to show the functionality of converting a BPMN4sML model to a TOSCA deployment model. The illustrative example is based on the home credit default risk prediction challenge published in 2018 on Kaggle by the Home Credit Group~\cite{HomeCreditGroup}. The goal is to find a well performing machine learning solution to predict the likelihood of a home credit applicant going into default. The machine learning task is a two-class classification problem. For this example a subset of $61000$ observations of the available \textit{application data} is considered that depicts information on the applicants and their background within 122 features. Figure~\ref{fig:usecase2BPMN4sML} describes the simplified machine learning pipeline via BPMN4sML.

Six tasks realized as serverless functions process the application data, create a machine learning model, verify it and deploy it. The workflow starts by sourcing the dataset from an external data store - for this example the data store is a public Amazon S3 bucket. The data comes in a processed format and can be written as a feature set to a feature set repository. Note that the feature set already contains the target variable.

Following, a feature engineering task further re-structures the feature set by eliminating and formatting a set of columns. In this case, 22 columns are eliminated based on a lack of signal identified by either having too many missing observations or being uninformative as a feature. Moreover, categorical features are dummy encoded. The updated feature set is then written as a dataset to a dataset repository.
Subsequently, a data split task divides the dataset into a training and test set by performing a simple 80:20 holdout split. The new datasets are again written to the dataset repository.
Next, a Random Forest (RF) learner is tuned on the training dataset via a small GridSearch testing various hyperparameter values for the number of trees per model and their respective depths. The tuning routine is defined as a 5-fold cross validation and applies ROC-AUC as a scoring metric. The best performing Random Forest model is directly written to the model registry.
Afterwards, a verification task verifies the RF model performance on the available test data. Again, ROC-AUC is measured.
Finally, the deployment task writes the verified model as a deployed model to a new access point, i.e. directory of the model registry, through which other applications and services can retrieve it.
To simplify this workflow, some considerations were made. We omit an extensive data preparation and tuning routine. Further, the verification task does not consider potential verification failure that might trigger a verification failure event or request for intervention. Overall the entire workflow is kept light so that it can be realized solely on serverless functions as it would otherwise necessitate further TOSCA extension to represent ML platforms by cloud providers as TOSCA Node Types.


\subsection{BPMN4sML to TOSCA via Conceptual Mapping}
\begin{figure}
    \centering
    \includegraphics[width=1\linewidth]{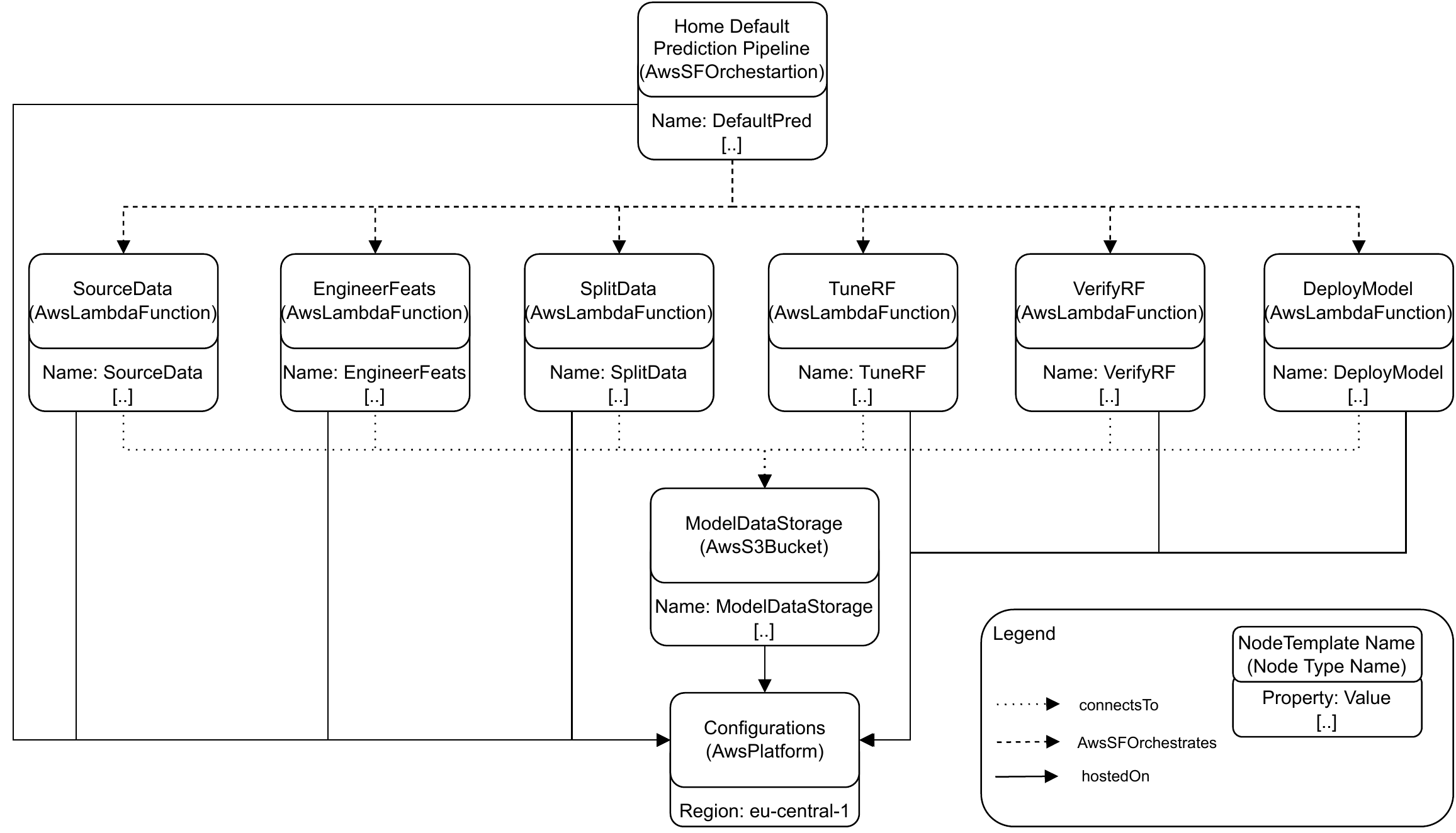}
    \caption[Illustrative Example 2 - TOSCA Topology Template of BPMN4sML home credit default ML pipeline]%
    {Illustrative Example 2 - TOSCA Topology Template of BPMN4sML home credit default ML pipeline \par \small }
    \label{fig:usecase2TOSCA}
\end{figure}

To convert the described BPMN4sML model, the conceptual mapping identified in section~\ref{MappingRules} is applied which results in the TOSCA Topology Template presented in Figure~\ref{fig:usecase2TOSCA}. All BPMN4sML tasks are modeled as serverless functions. Moreover the workflow shall be realized as a function orchestration and hosted on AWS cloud. Consequently, the conceptual mapping dictates three corresponding TOSCA Node Types, namely 1) \textit{AwsSFOrchestration}, 2) \textit{AwsLambdaFunction} and 3) \textit{AwsPlatform}. Further, two TOSCA Relationship Types are required, 1) \textit{AwsOrchestrates} and 2) \textit{hostedOn}. The orchestrator node represents the Amazon StepFunction engine which coordinates the Amazon Lambda functions. Next, the BPMN4sML model depicts a \textit{FeatureSetRepository}, \textit{DatasetRepository} and \textit{ModelRegistry}. These data artefacts correspond to a TOSCA Node Type and a \textit{connectsTo} Relationship Type. For simplification the TOSCA model, the different data repositories and the model registry are aggregated as one Amazon S3 Bucket, i.e. \textit{AwsS3Bucket} Node Type, and translated into different directories on this S3 bucket which itself is again \textit{hostedOn} the \textit{AwsPlatform} node. Alternatively, one \textit{AwsS3Bucket} can be modeled for each repository and the model registry.

To derive a complete service template and compile it as a CSAR file, the TOSCA Topology Template can now be specified via Winery to for instance define the property values of the \textit{AwsLambdaFunctions} (allocated memory, runtime, function layers, timeout threshold etc.) and to attach the respective task logic as a zipped code file artefact and the orchestration logic as orchestration file artefact. The Topology Template modeled in Winery is available in appendix~\ref{fig:usecase2Winery}.

As noted in the introduction of this chapter, at the moment of writing the available TOSCA Node Types, Relationship Types and corresponding Ansible roles~\footnote{An Ansible role is a recipe to define an installation process of an application or component on a specific platform.} that enable the orchestration of serverless functions via TOSCA deployment models lag behind the development and update of the actual cloud provider's functionality. As such, currently when the displayed TOSCA Topology Template is packaged as a Service Template into a CSAR file, the subsequent deployment orchestration via xOpera runs into errors. Specifically, serverless functions realized as AWS Lambdas incorporate a \textit{state}, e.g. active, that is not currently considered during deployment. Consequently, as soon as a TOSCA \textit{AwsLambdaFunction} is to be configured without verifying if it is already \textit{active}, deployment fails. As previously stated, extending TOSCA functionality and Ansible roles exceeds the scope and time available for this thesis which is therefore left for future research.

\begin{figure}
    \centering
    \includegraphics[width=1\linewidth]{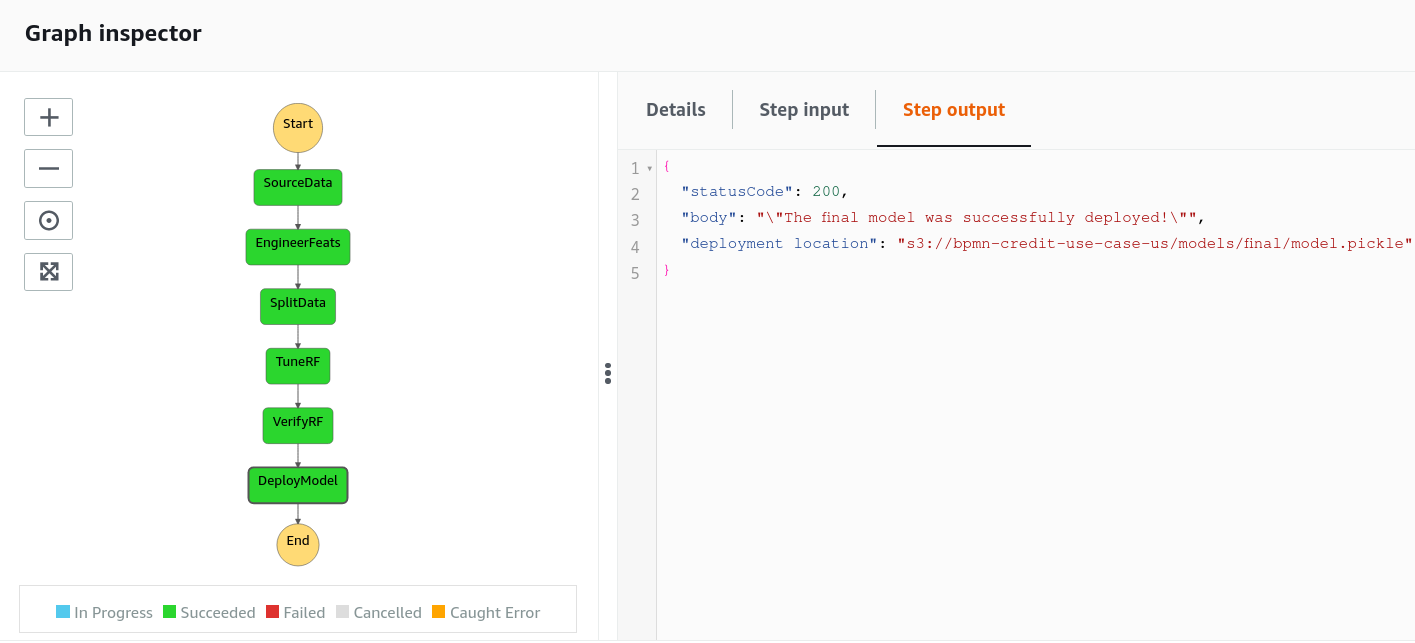}
    \caption[Illustrative Example 2 - direct execution of the modeled workflow on AWS via Step Functions]%
    {Illustrative Example 2 - direct execution of the modeled workflow on AWS via Step Functions \par \small }
    \label{fig:AWSSFExample}
\end{figure}

As a substitute for displaying the deployment of the modeled BPMN4sML workflow, the pipeline is directly implemented on the AWS cloud while referencing equivalent logic, functionality and properties as discussed before. The workflow is then orchestrated by a StepFunction. A successful execution of the ML pipeline is shown in Figure~\ref{fig:AWSSFExample}. The workflow is initialized manually and each task is represented by a Lambda function. Upon successful completion, the location of the deployed model is returned. Note that for this example we do not consider complex branching logic or error-handling, it can however be included in the workflow definition and orchestrated by the StepFunction engine.



\subsection{Evaluation of Current Extent of Conceptual Mapping}
The modeled workflow and showcased conceptual mapping allow successful conversion of a BPMN4sML model diagram into a TOSCA Topology Template. All modeled BPMN4sML elements can be translated to TOSCA equivalents validating the potential of the proposed method discussed in section~\ref{ProposedMethod}. Moreover by relying on BPMN and TOSCA as standards, the goal to allow technology independent and interoperable modeling is realized. The same workflow depicted in Figure~\ref{fig:usecase2BPMN4sML} can be implemented on any other cloud provider without having to change the workflow model diagram given that the provider offers a similar set of services. With regards to the TOSCA Topology Template the Nodes and Relationships corresponding to the ML workflow model will then depict the respective node and relationship types of the chosen cloud provider. Certain differences can occur as the implementation of function orchestrations differs among providers, see Yussupov et al.~\cite{FaaSBPMNOrchestration} for a discussion on this topic. 

Nevertheless, while this illustrative example helps in validating the conceptual mapping identified in section~\ref{MappingRules}, conditions apply. First and foremost, as previously stated, at the moment of writing, automated orchestration of the TOSCA Service Template via xOpera is not fully supported due to the discrepancy between functionality provided by TOSCA Node Types and Ansible roles and the functionality required to align with the latest cloud provider versions of serverless functions and function orchestrations.

Next, only a subset of the BPMN4sML extension elements are covered in this example. No complex workflow patterns such as branching or looping could be investigated. Similarly, events such as the modeling of an update of a repository triggering a subsequent task are not included in the illustrative use case due to time constraints. Such scenarios are however covered by the conceptual mapping and can in TOSCA either be depicted as an event-flow or as a function orchestration.
Other complex workflow logic such as error handling has not been directly addressed. BPMN4sML allows modeling such scenarios by leveraging BPMN native constructs. They are part of the overall workflow logic and can therefore be incorporated implicitly within a function orchestration in a corresponding TOSCA deployment model. Further investigation is however required to fully validate such mappings.

Overall, modeling a serverless machine learning workflow in a technology-agnostic manner is facilitated by BPMN4sML and supported by TOSCA. Given that the mentioned TOSCA elements are extended, a full mapping is possible which in turn also enables deployment orchestration of the modeled BPMN4sML diagram.

%% file: chapters/8.discussion.tex
This study proposes three artefacts as significant contributions, two core artefacts and a second-order one. 
First, it presents a metamodel extension to BPMN based on a preceding extensive requirement analysis. Second, it brings forward constructs corresponding to the metamodel and incorporating the necessary modeling elements, extending the BPMN notation and respective semantic. Together, the two artefacts form the proposed BPMN4sML extension addressing the requirements of machine learning workflows as well as characteristics stemming from their serverless deployment. 
Third, this thesis establishes a conceptual mapping between BPMN4sML and TOSCA elements with a focus on serverless deployment orchestrations. The artefacts are consolidated as a method describing a process for technology independent and interoperable modeling of (serverless) ML workflows via BPMN4sML and their mapping to TOSCA for subsequent deployment.
In this chapter, the findings of the requirement analysis and artefacts are discussed and interpreted in context of the research questions.

\section{Machine Learning and Serverless Requirements Affecting Conceptual Modeling}
The first set of sub-questions \textbf{1.1, 1.2, 1.3} is resolved by means of a literature review in chapter~\ref{chapter:related_work}. The second set of sub-questions \textbf{2.1, 2.2, 2.3} builds on this knowledge base, extends and analyses it to identify requirements in chapter~\ref{chapter:requirement_analyses}.

\subsubsection{Machine Learning}
Different interpretations exist on what constitutes a machine learning workflow~\cite{AssuringMLLifecycle,MicrosoftMLWorkflow,AssetML}. We establish basics on machine learning as a domain and organize machine learning processes along the ML lifecycle phases, see sections~\ref{MachineLearning},~\ref{MLLifecycle}. We consider \textit{requirement analysis}, \textit{data management}, \textit{model learning}, \textit{model verification}, \textit{model deployment} and \textit{model monitoring and inference} as phases relevant to the derivation of specific requirements for the conceptual modeling of ML, see section~\ref{MLrequirements}. Each phase is characterized by a set of activities shared across ML processes. Nonetheless, activities do not pertain solely to one phase. Further, each phase produces and consumes data artefacts such as machine learning models, data sets and other code, configuration, result or information objects. Besides, events and decision points can influence the respective ML lifecycle phase, making the whole processes potentially iterative. 

We further observe that machine learning is a vast and quickly developing field of research. Thus, the scope of this work constrains the extend to which requirements specific to ML methodologies can be considered. For example, while particularities of federated learning are identified, they are not further considered in the creation of the extension artefact. However, future integration of the outlined concepts is supported.

\subsubsection{Serverless Computing}
Realizing a machine learning workflow with serverless computing can be achieved mainly via two options, 1) function composition and 2) function orchestration, see section~\ref{FaaS}. Moreover, the inherent characteristics of serverless computing - 1) utilization-based billing, 2) limited control, 3) event-driven - and of Function-as-a-Service - 1) statelessness, 2) fixed-memory, 3) short running time - require job offloading in the case of a more compute-intensive ML activity, see section~\ref{FAASCDReq}.

This study concentrates on an approach for generic modeling of ML workflows and their serverless deployment orchestration. Accordingly, creation of frameworks to manage for instance resource provisioning and mitigate the serverless / FaaS characteristics in new ways as shown by Carreira et al.~\cite{ServerlessCirrus} or Jiang et al.~\cite{DemystifyingSML} is not pursued. Instead, we refer to a combination of BPMN and TOSCA to realize our objective.

\subsubsection{Modeling}
Making machine learning more accessible to involved stakeholders within an organization is an ongoing challenge. 
By reviewing process modeling fundamentals as well as potential modeling languages and basic workflow concepts in sections~\ref{PMCM},~\ref{BPMN}, we identify the Business Process Model and Notation as the de-facto standard and reference it henceforth. BPMN and BPMN extensions however do not offer the required constructs to fully represent machine learning workflows as we explain in section~\ref{relatedWork}. Similarly, existing literature on standard-based deployment of ML workflows is scarce.

The absence of a fitting modeling language and notation poses large obstacles for non-technical stakeholder to get involved and for technical stakeholders to communicate and design workflows in a consistent manner~\cite{AiBMInnovation,McKinseyMLOperationalization}. Further, integration of a ML process into the overall process infrastructure and environment of an organization is difficult~\cite{ConceptualModelingML}. Also technical stakeholders such as data scientists and ML engineers are confronted with an abundance of potential tools and offerings, making the modeling and serverless deployment orchestration of ML workflows in a technology independent and interoperable manner a challenge. Therefore, abstraction is necessary to be able to represent ML concepts irrespective of chosen provider or technology.

Synthesizing  the literature review and requirement analysis, we identify 57 requirements needed to enable fundamental conceptual modeling of (serverless) ML workflows. Ten identified requirements can be related directly to existing BPMN concepts or concepts proposed by BPMN extensions. Representing the remaining ones requires extension.

\section{BPMN4sML Suitability for (Serverless)  ML Workflow Modeling }
\label{DiscussBPMN4sML}
The core sub-questions \textbf{3.1 (a, b)} are resolved in chapter~\ref{chapter:results} by referencing the established set of elements that are required to enable modeling of (serverless) ML workflows and incorporating them in a BPMN extension as Business Process Model and Notation for serverless Machine Learning.

This study serves as both, a specification document for BPMN4sML as well as a modeling guide. We propose the extension to start addressing the various cases in the field of machine learning which supersede capabilities of the conventional BPMN 2.0 standard and existing extensions. We develop BPMN4sML as an extension by addition while adhering to the process adopted by many publications directed towards extending BPMN, see for instance~\cite{BPMNExclinical,uBPMN,bpmnt}. Further, our phrasing pattern corresponds to the one used by the Object Management Group to better assist BPMN practitioners with understanding the idea and concepts behind BPMN4sML.  

To the BPMN metamodel we add 1) new \textit{Task} elements, 2) new \textit{Data Artefacts} and 3) new \textit{Events} to conceptually capture the requirements and formalize them in context of the standard, see section~\ref{metamodelExt}. Class properties are considered with respect to ML specificities and essential serverless implementation characteristics.
Further, corresponding constructs are provided as a BPMN notation extension which allow to visually represent the new elements of the metamodel extension, see section~\ref{NotationExt}. In doing so, we lay the foundation to tackle technology independent and interoperable modeling of (serverless) ML workflows.

Specifically, BPMN4sML facilitates modeling of machine learning tasks, event streams, data objects and repositories across the entire ML lifecycle. It supports the large part of supervised and unsupervised machine learning and accounts for various methodologies such as ensemble or transfer learning.\\ 
Further, several ways of implementing and deploying the workflow are considered (serverless, offloaded, hybrid), enabling modelers to differentiate both conceptually and visually between entirely FaaS-based solutions or ML services leveraging specific machine learning platforms.
Notably, while this work focuses on serverless machine learning workflows, the identified requirements, conceptualization, notation and semantics are generalizable (or can be easily generalized) to ML workflows outside of the domain of serverless computing. Moreover, the identified ML concepts and notation are applicable to largely represent machine learning workflows as they are advocated by main cloud providers such as Google Cloud~\cite{googleML,GoogleMLOPsDoc}, Amazon Web Services~\cite{AWSmlconcepts,AWSmlarchitecting,AWSmltaming} and Microsoft Azure~\cite{AzureMLOpsref,MicrosoftMLWorkflow} by incorporating a similar (or more extensive) list of modeling elements.

With BPMN4sML, we further answer to the call for research by Lukyanenko et al.\cite{ConceptualModelingML}. 
When modeling ML workflows, ML engineers, data scientists and other process analysts can minimize ambiguity by referring to the most expressive core modeling elements which best portray the machine learning workflow instead of overloading the model diagrams with redundant text annotations to describe an element. Consequently, when implementing a model, ideas can be conveyed precisely, reducing need for clarifications and thereby also time and concomitant costs. 

Leveraging common process modeling, practitioners can increase business understanding with respect to ML projects. Moreover, by visualizing a ML workflow as a process model, transparency and comprehensibility can be improved. Each task, data artefact and decision can be unambiguously presented. As a result, BPMN4sML also contributes to the field of explainable artificial intelligence by providing overview and insight into the mechanics of a ML solution. 

Nonetheless, the current version of BPMN4sML does not allow modeling of the entire domain of machine learning. Extending it further with respect to specific ML methodologies is necessary. 
Moreover, this work focused on the conceptualization of the artefact. As a result, schema descriptions corresponding to the metamodels are still required and will have to be added in the future.
As with any new modeling extension, BPMN4sML requires machine learning practitioners and stakeholders to first learn and understand the notation in order to reap its benefits. Thus, communicating directly with such domain experts and gathering feedback still is paramount to a successful adoption.

As shown in our first illustrative scenario in section~\ref{UseCase1}, using BPMN4sML one can describe end-to-end machine learning workflows in a standardized and coherent manner. It is however noteworthy that ML workflows in general operate on a large set of different data artefacts. Further, by focusing on serverless deployment most tasks require a connection to one or several of the introduced repositories. Hence, a modeler may need to continuously balance the expressiveness of a BPMN4sML model with element overload in order to only depict the core set of required data artefacts. BPMN4sML supports this by offering a concise notation and specific semantics.

\section{TOSCA Suitability for Serverless ML Workflow Deployment}
\label{discussTOSCA}
We resolve the remaining sub-question, RQ \textbf{4.1}, in chapter~\ref{chapter:implementation} by leveraging the new extension artefact, BPMN4sML, and deriving a conceptual mapping between the BPMN4sML metamodel and the TOSCA metamodel to enable the conversion of BPMN4sML workflow models to TOSCA deployment models. Additionally, a consolidating sequence of activities is brought forward as an initial proposal of an end-to-end method to model ML workflows via BPMN4sML and their corresponding serverless deployment orchestration via TOSCA.

Referencing TOSCA as a standard helps in adhering to technology independence and interoperability with respect to serverless deployment orchestration. Moreover, it allows us to leverage an extensive body of existing work on declarative deployment modeling
Drawing from our proposed method can ultimately enable ML engineers to speed up the time it takes from ideating ML solutions to deploying them and, by extension, shortens the time-to-market.
Nevertheless, to realize this objective further work is necessary in automating the BPMN4sML to TOSCA conversion. In section~\ref{ProposedMethod} we elaborate that this relates to three pillars - 1) expressing BPMN4sML models as XML documents, 2) potentially incorporating functionality to directly derive workflow logic as a function orchestration file and 3) accordingly extend TOSCA Node and Relationship Types as well as corresponding Ansible roles. Realizing this was constrained by the scope and time of this thesis.

Our illustrative scenario 2 in section~\ref{UseCase2} showcases the potential to map a BPMN4sML model to a TOSCA deployment model. It however also highlights the need for more work on the side of TOSCA to fully support deployment of serverless workflows. In particular with respect to serverless computing, current TOSCA solutions are limited. Yussupov et al.~\cite{FaaSBPMNOrchestration} and Wurster et al.~\cite{TOSCAServerless} take first steps in this direction. Nevertheless, the latter one's proposition does not support serverless workflows and the former one's necessitates further extension and maintenance to restore its functionality. Similarly, TOSCA propositions as part of the RADON project~\cite{RadonParticles} are not directed towards serverless workflow orchestration. Further, to fully realize the proposed method, new TOSCA Node and Relationship Types are required for representing offloading technologies such as machine learning platforms.
Alternatively, also other tools, albeit not necessarily standards, can be experimented with to circumvent current TOSCA shortcomings. Terraform~\cite{Terraform} may be another solution towards cloud infrastructure automation. Similarly, the Serverless Workflow Specification may aide in better realizing function orchestrations.

\section{Threats to Validity}
As with any research, certain factors can influence the extent to which our findings are consistent internally, i.e. as created artefacts, and generalizable externally, i.e. to be applied to other related areas.
In the preceding sections we already reflected on the artefacts, thus in this section only specific relevant aspects are highlighted. 

An extensive literature review is conducted to identify relevant concepts and requirements based on which the BPMN extension is developed. Nonetheless, certain aspects of it can be further improved. We derive domain concepts as well as extension requirements in an argumentative manner based on the identified literature and industry tools. To further validate these aspects, interviews or surveys with other experts and domain practitioners are in order such that results can be compared against our established set of requirements.

Similarly, machine learning as a domain is evolving at an unprecedented pace and incorporates several large sub-domains. Accounting for all and deriving respective concepts and requirements is not feasible within this thesis scope and requires a much larger study set-up. We reduce this influence on our extension by referring to 1) extensively studied ML domains such as supervised learning and 2) incorporating existing reference architectures and surveys. Nonetheless, as we do not explicitly incorporate ML methodologies such as active learning, multi-task learning or similar upcoming ML derivatives, the current BPMN4sML extension is limited in its extent to generalize to the entire ML domain. 

Moreover, no explicit user tests are administered. As BPMN4sML is an artefact meant for practical application, gathering evidence on its fit to solve the identified research problem by letting third parties utilize it is necessary. This refers to both, the identified concepts as well as the usability and intuitiveness of the developed notation. User testing is explicitly excluded from this thesis due to time constraints. We reference illustrative use cases to mitigate this aspect. Additionally, we refer to an officially published ML scenario to further indicate the validity of BPMN4sML. During presentation of the artefact we provide supplemental example snippets showcasing that the extension elements can in fact be applied according to their definition.

The metamodels as well as the notation are developed according to the Object Management Group guidelines and in line with the approaches proposed by similar work which indicates proper use of the established theoretical foundation and, by extension, validity. 

With respect to orchestrating the deployment of modeled ML workflows, we reference an existing standard, i.e. TOSCA, which directly proves support for technological independence and interoperability. Nonetheless, more elaborate use cases on different cloud providers are necessary to fully determine the extent to which ML workflows can be orchestrated serverlessly. Moreover, at the moment of writing deployment via TOSCA is still error-prone and requires maintenance and updates for proper funtionality.

%% file: chapters/9.conclusions.tex
This thesis researches how machine learning workflows and their serverless deployment orchestration can be modeled in a technology independent and interoperable manner via existing standards (BPMN and TOSCA). As a result, we propose the specification BPMN4sML\footnote{https://github.com/LamaTe/BPMN4sML} (Business Process Model and Notation for serverless Machine Learning) as an extension to BPMN 2.0 to support machine learning engineers and other process analysts in modeling ML processes. Our extension addresses the various challenges that practitioners face when describing a (serverless) machine learning workflow with the existing standard by:
\begin{itemize}
\item Conceptualizing the machine learning lifecycle and workflow by means of a metamodel.
\item Developing a new, more accurate and meaningful notation and semantic to better represent machine learning concepts and artefacts. 
\item Aiding both technical and managerial personnel to describe, communicate and understand ML workflows in a consistent and transparent manner.
\item Taking a first step towards a standardized depiction of machine learning solutions which can be further extended for upcoming machine learning methodologies. 
\end{itemize}
We realize this by conducting a thorough literature review and extensive requirements analysis of the core machine learning domains as well as the relevant paradigms for serverless computing and process and deployment modeling.
We further propose a conceptual mapping from the introduced BPMN4sML elements to corresponding TOSCA elements showcasing the potential of directly deriving technology specific deployable topology templates from a generic machine learning process model.
We exemplify our findings by means of example workflow snippets and two illustrative use cases inspired by 1) a publication and 2) a real-world Kaggle challenge.
The BPMN4sML notation was created leveraging \textit{draw.io / diagrams.net} and can be imported in form of a first version as XML library\footnote{https://github.com/LamaTe/BPMN4sML/tree/main/Extension}.

\section{Implications for Research and Practice}
This study synthesizes machine learning research and serverless computing with practices in business process and deployment modeling. At the moment of writing, no BPMN extension exists to specifically address machine learning workflow modeling. Hence, BPMN4sML joins other publications on extending the process modeling standard. We operationalize conceptual modeling for the machine learning domain, thereby addressing the call for research by Lukyanenko et al.~\cite{ConceptualModelingML}. We moreover contribute to research on technology independent and interoperable modeling of machine learning workflows and their serverless deployment by combining BPMN and TOSCA. As a result, we contribute to the research for mental frameworks to better describe machine learning processes and, in general, the machine learning lifecycle. 

Via BPMN4sML practitioners can leverage established business process management practices in their organization to transparently and more meaningfully develop, model, communicate and integrate machine learning services in their process ecosystem. Our BPMN4sML extension can help in clarifying logic and concepts behind a machine learning service thereby reducing potential risks created by misunderstandings and improving overall documentation quality.
Similarly, machine learning engineers can benefit from a consistent notation and semantic to more clearly and concisely design machine learning solutions in a technology independent and interoperable manner. Applying BPMN4sML and drawing from the method proposed by this thesis paves the way towards reducing lock-in effects created by provider-specific languages and service offerings. 
Ultimately, the extension can help in communicating machine learning ideas across organizational boundaries and assists in lowering the threshold of adoption for ML by making workflows more accessible to a non-technical audience, i.e. personnel familiar with BPMN.

\section{Future Research Directions}
The identified limitations in chapter~\ref{chapter:discussion} indicate at the same time directions for future research. 
As explained, the BPMN extension proposed by this study is of conceptual nature. Hence, corresponding XML schema documents shall be derived in the future to enable more automated extraction of workflow logic as function orchestration instruction files and conversion to corresponding TOSCA elements. In this context, the function orchestration generator prototype by Yussupov et al.~\cite{FaaSBPMNOrchestration} can be updated, extended and potentially integrated. 
Moreover, by representing BPMN4sML workflow models via their formalized XML schema document, verification and validation techniques can be applied to better guide practitioners during their modeling activities. Suitable industrial products to realize this are for instance Camunda Modeler which moreover supports integration of BPMN extensions.

Furthermore, the identified requirements and respective concepts of federated learning can be integrated in the current version of BPMN4sML. Similarly, a more extensive systematic white and grey literature review can be conducted following the methodology by Petersen et al.~\cite{systematicLiterature} and building up on the existing structure of this study. Likewise, also a more extensive look at machine learning services offered by cloud providers is recommended. In this manner, more exotic ML methodologies can be integrated systematically and a larger range of provider specific concepts can be covered. Another opportunity is to create a fully technology and implementation agnostic machine learning modeling extension, i.e. separating the current combination of serverless computing and machine learning concepts in the proposed metamodel.

To realize serveless deployment orchestration in its entirety, TOSCA elements, specifically TOSCA Node and Relationship Types for serverless computing and event-driven workflows shall be extended as well as corresponding Ansible roles. Alternatively, other cloud infrastructure deployment automation technologies such as Terraform and the Serverless Workflow Specification can be experimented with.

In addition, researchers may study on how to integrate the proposed extension within larger MLOps systems that go beyond the actual machine learning workflow.

Finally, we recommend adapting BPMN4sML in practice to gather feedback which in turn can inform new BPMN4sML development cycles following the design science research methodology process. Similarly, more extensive experimentation and evaluation scenarios as well as expert interviews are advised to scrutinize and improve the artefact further and thereby increase its internal validity.

%% file: appendices/main.tex

\chapter{Winery TOSCA Topology Template}
\begin{figure}[h!]
    \centering
    \includegraphics[width=1\linewidth]{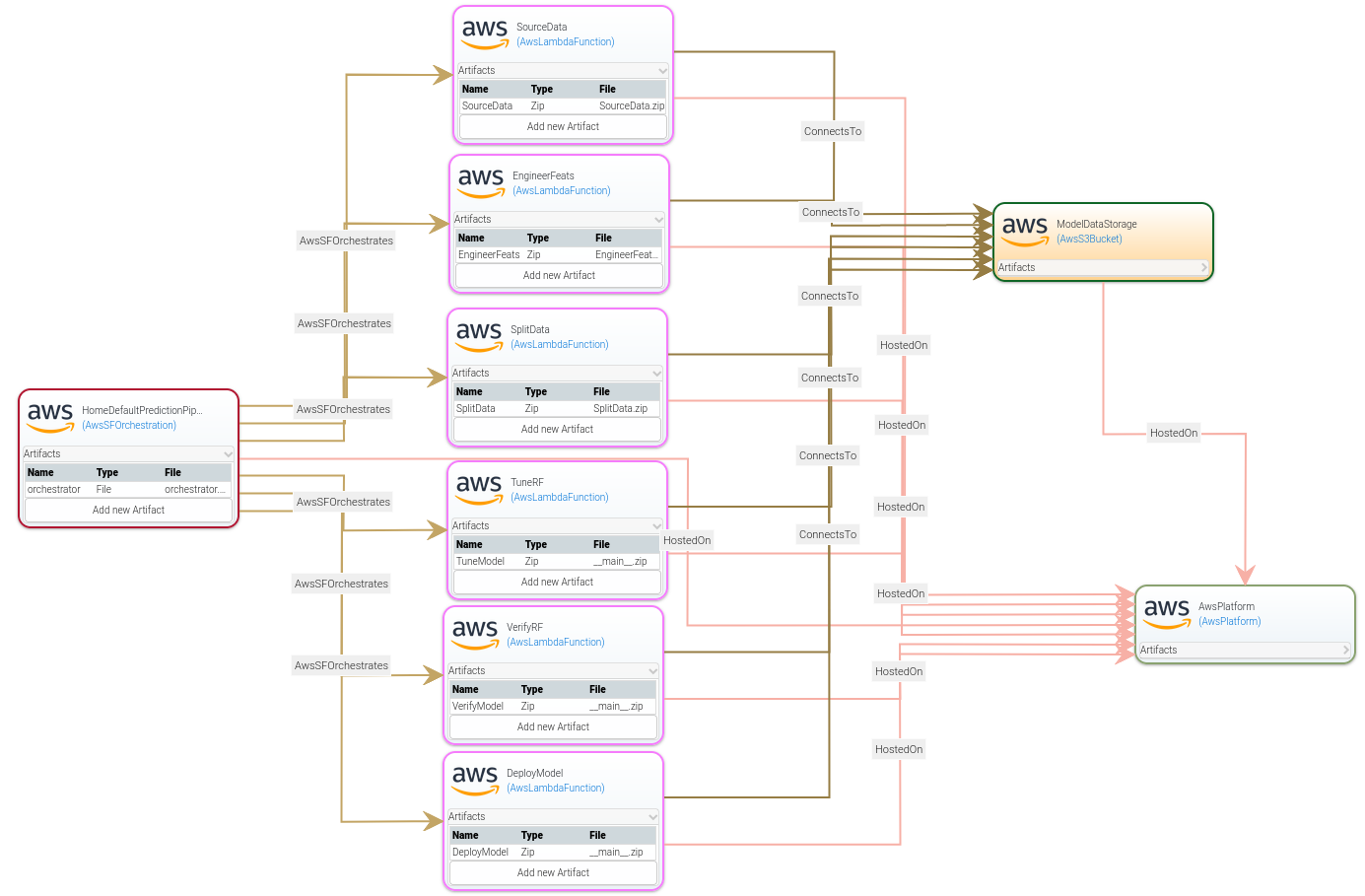}
    \caption[Winery TOSCA Topology Template of home credit default ML pipeline]%
    {Winery TOSCA Topology Template of BPMN4sML home credit default ML pipeline. \par \small To model this Topology Template Winery is started with reference to the TOSCA repository provided by IAAS Serverless Prototyping Lab~\cite{FaaSGit}. Alternatively, Winery can be started with reference to the adapted version of the repository that already contains the displayed Topology Template and is provided as part of this thesis.}
    \label{fig:usecase2Winery}
\end{figure}